\documentclass[a4paper,fleqn,usenatbib]{mnras}

\usepackage{newtxtext,newtxmath}
\usepackage[T1]{fontenc}
\usepackage{ae,aecompl}

\usepackage{graphicx,color,rotating}	% Including figure files
\usepackage{amsmath}	% Advanced maths commands
\usepackage{caption}
\usepackage{subfig, siunitx}
\def\hi{\mbox{H\sc{i}}}

\def\kms{km\,s$^{-1}$}

\def\msun{M$_{\odot}$}

\def\arcsec{$^{\prime \prime}$}
\definecolor{Mygrey}{gray}{0.75}

\newcommand{\ltsimeq}{\raisebox{-0.6ex}{$\,\stackrel{\raisebox{-.2ex}{$\textstyle <$}}{\sim}\,$}}
\newcommand{\gtsimeq}{\raisebox{-0.6ex}{$\,\stackrel{\raisebox{-.2ex}{$\textstyle >$}}{\sim}\,$}}
\newcommand{\farc}{\mbox{\ensuremath{.\!\!^{\prime\prime}}}}

\mathchardef\mhyphen="2D

\title[WISDOM X: ISM morphology in galaxy centres ]{WISDOM Project - X. The morphology of the molecular ISM in galaxy centres and its dependence on galaxy structure} 
\author[Timothy A. Davis et al.]{\parbox{\textwidth}{
Timothy A. Davis,$^{1}$\thanks{E-mail: DavisT@cardiff.ac.uk} Jindra Gensior,$^{2,3}$ Martin Bureau,$^{4}$ Michele Cappellari,$^{4}$ Woorak Choi,$^{5}$ Jacob S. Elford,$^{1}$ J. M. Diederik Kruijssen,$^{3}$ Federico Lelli,$^{6}$ Fu-Heng Liang,$^{4}$ Lijie Liu,$^{7,8}$ Ilaria Ruffa,$^{1}$ Toshiki Saito,$^{9,10}$ Marc Sarzi,$^{11}$ Andreas Schruba$^{12}$ and Thomas G. Williams$^{13}$}
\vspace{0.4cm}\\
\parbox{\textwidth}{
$^{1}$Cardiff Hub for Astrophysics Research \&\ Technology, School of Physics \&\ Astronomy, Cardiff University, Queens Buildings, Cardiff, CF24 3AA, UK\\
$^{2}$Institute for Computational Science, University of Z\"urich, Winterthurerstrasse 190, 8057 Z\"urich\\
$^{3}$Astronomisches Rechen-Institut, Zentrum f\"ur Astronomie der Universität Heidelberg, M\"onchhofstraße 12-14, 69120 Heidelberg, Germany\\
$^{4}$Sub-department of Astrophysics, Department of Physics, University of Oxford, Denys Wilkinson Building, Keble Road, Oxford OX1 3RH, UK\\
$^{5}$Department of Astronomy, Yonsei University, 50 Yonsei-ro, Seodaemun-gu, Seoul 03722, Republic of Korea\\%Yonsei Frontier Lab and Department of Astronomy, Yonsei University, 50 Yonsei-ro, Seodaemun-gu, Seoul 03722, Republic of Korea\\
$^{6}$INAF, Arcetri Astrophysical Observatory, Largo Enrico Fermi 5, I-50125, Florence, Italy\\
$^{7}$Cosmic Dawn Center (DAWN) \\
$^{8}$DTU-Space, Technical University of Denmark, Elektrovej 327, DK-2800 Kgs. Lyngby, Denmark\\
$^{9}$Department of Physics, General Studies, College of Engineering, Nihon University, 1 Nakagawara, Tokusada, Tamuramachi, Koriyama, Fukushima, 963-8642, Japan\\
$^{10}$National Astronomical Observatory of Japan, 2-21-1 Osawa, Mitaka, Tokyo, 181-8588, Japan\\
$^{11}$Armagh Observatory and Planetarium, College Hill, Armagh BT61 9DG, UK\\
$^{12}$Max-Planck-Institut für extraterrestrische Physik, Giessenbachstraße 1, D-85748 Garching, Germany\\
$^{13}$Max Planck Institut für Astronomie, Königstuhl 17, 69117 Heidelberg, Germany
}}
\date{Accepted 2022 March 2. Received 2022 March 2; in original form 2021 December 10}

\pubyear{2022}

\begin{document}
\label{firstpage}
\pagerange{\pageref{firstpage}--\pageref{lastpage}}
\maketitle

% Abstract of the paper
\begin{abstract}
We use high-resolution maps of the molecular interstellar medium (ISM) in the centres of eighty-six nearby galaxies from the millimetre-Wave Interferometric Survey of Dark Object Masses (WISDOM) and Physics at High Angular Resolution in Nearby GalaxieS (PHANGS) surveys to investigate the physical mechanisms setting the morphology of the ISM at molecular cloud scales. 
We show that early-type galaxies tend to have smooth, regular molecular gas morphologies, while the ISM in spiral galaxy bulges is much more asymmetric and clumpy when observed at the same spatial scales. We quantify these differences using non-parametric morphology measures (Asymmetry, Smoothness and Gini), and compare these measurements with those extracted from idealised galaxy simulations.
 We show that the morphology of the molecular ISM changes systematically as a function of various large scale galaxy parameters, including galaxy morphological type, stellar mass, stellar velocity dispersion, effective stellar mass surface density, molecular gas surface density, star formation efficiency and the presence of a bar. We perform a statistical analysis to determine which of these correlated parameters best predicts the morphology of the ISM. We find the effective stellar mass surface (or volume) density to be the strongest predictor of the morphology of the molecular gas, while star formation and bars maybe be important secondary drivers. We find that gas self-gravity is not the dominant process shaping the morphology of the molecular gas in galaxy centres. Instead effects caused by the depth of the potential well such as shear, suppression of stellar spiral density waves and/or inflow affect the ability of the gas to fragment.
\end{abstract}

% Select between one and six entries from the list of approved keywords.
% Don't make up new ones.

%%%%%%%%%%%%%%%%%%%%%%%%%%%%%%%%%%%%%%%%%%%%%%%%%%

%%%%%%%%%%%%%%%%% BODY OF PAPER %%%%%%%%%%%%%%%%%%

\section{Introduction}

The cold gas within the interstellar media (ISM) of galaxies plays a key role in governing their evolution. The atomic material provides the reservoir for future galaxy growth \citep[e.g.][]{2014MNRAS.442.2398P}, while the molecular medium is closely linked to ongoing star formation \citep[e.g.][]{2008AJ....136.2782L}. The properties of this cold gas are thus closely linked to the evolution of the galaxies themselves.

The molecular gas in present-day galaxy discs is formed where the atomic medium becomes dense, either because it has been compressed by external forces (e.g. in a spiral arm density wave or due to turbulence), or because it has become locally unstable to collapse. If a dense region also has a sufficient column density to shield itself from the interstellar radiation field then it will form a giant molecular cloud (GMC). The molecular medium as a whole tends to be dynamically cold, and is found in a thin disc very close to the galaxy midplane, with a scale height of 50-200\,pc \citep[e.g.][]{2013AJ....146..150C,2018ApJ...860...92L,2019MNRAS.484...81P}. 

The morphology of the gas in these molecular discs is governed by mechanisms that clump gas together (e.g. spiral density waves, self-gravity) and dissipative forces (e.g. shear and turbulence) which tend to break clouds apart.
In spiral galaxies the GMCs typically follow the spiral structure of the system \citep[e.g.][]{1980ApJ...239L..53C,2001ApJ...547..792D,2021arXiv210407739L}, and are thought to be in approximate energy equipartition with their surrounding media (e.g. \citealt{1981MNRAS.194..809L,2007ApJ...654..240R,2020ApJ...901L...8S}). 
In recent years the morphology of the molecular medium in the discs of nearby galaxies has been studied on cloud scales (by e.g. the Physics at High Angular Resolution in Nearby GalaxieS (PHANGS) project; \citealt{2021arXiv210407739L}), revealing the importance of stellar structures such as spiral arms in setting its morphology \citep[e.g.][]{2021ApJ...913..113M,2021arXiv210904491Q}. 
How the molecular material behaves on small scales in other galactic environments is, however, not as well understood.

One environment which has received comparatively less attention is the centres of galaxies, where gas properties can be strongly affected by both the deep potential well (creating high pressures, and high shear), and by the presence of bars and/or active galactic nuclei (AGN). The nuclei of spiral galaxies have been revealed to be dynamic environments, where gas streams, resonances and warps are common \citep{2003A&A...407..485G,2005A&A...442..479K,2013MNRAS.429..987L,2019A&A...632A..33A,2020ApJ...901L...8S,2021MNRAS.505.4310C,2021arXiv210410227G}. However, disentangling which effects are due to the AGN/bars, and which are due to the galaxy potential can be difficult. 

Elliptical and lenticular galaxies (also known as early-type galaxies; ETGs) are another environment in which the molecular media has been less extensively studied. In the local Universe $\approx$23\% of these systems are found to host molecular gas reservoirs \citep{2011MNRAS.414..940Y,2019MNRAS.486.1404D}. This gas is typically found in centrally-concentrated discs, which are dynamically cold \citep{2013MNRAS.432.1796A,2013MNRAS.429..534D,2019MNRAS.484.4239R}. ETGs have large bulge components, and thus very centrally-concentrated stellar mass distributions. The circular velocity curves of ETGs therefore rise sharply \citep[e.g.][]{2021arXiv211006033Y}, leading to strong shear rates, which have been suggested to impact the efficiency with which stars form \citep{2014MNRAS.444.3427D}.  ETGs can thus be seen, from a dynamical point of view, as scaled-up versions of the bulges of spiral galaxies, and thus it is instructive to consider them together. 

The first cloud-scale investigations of the ISM of ETGs have revealed some have gas in discrete molecular clouds like spiral galaxies \citep{2015ApJ...803...16U}, while others have very smooth discs \citep{2017MNRAS.468.4675D}, and some show an absence of large molecular clouds \citep{2021arXiv210604327L}. The physical mechanism(s) causing the observed diversity of ISM morphologies in ETGs have not yet been identified. 

In this paper we investigate the mechanisms governing the morphology of the molecular medium in the centres of eighty-six nearby spiral and early-type galaxies. We use Atacama Large Millimeter/submillimeter Array (ALMA) observations from the millimetre-Wave Interferometric Survey of Dark Object Masses (WISDOM\footnote{\url{https://www.wisdom-project.org}}; \citealt{2017MNRAS.468.4663O,2017MNRAS.468.4675D}), which reveals the properties of the molecular ISM in the centre of a diverse range of spiral and early-type galaxies at very high (typically 0\farc1, $\approx$30\,pc) angular resolution.
We complement the WISDOM data with lower-resolution ($\approx$1\arcsec, $\approx$80\,pc) observations on the centres of a larger sample of galaxies from the PHANGS ALMA large programme\footnote{\url{http://www.phangs.org}} \citep{2021arXiv210407739L}. 
These data are also compared with a suite of numerical simulations of isolated galaxies designed to systematically probe a wide range of bulge-to-disc ratios and central densities from \cite{2020MNRAS.495..199G}. We extract non-parametric measurements of the morphologies of the molecular gas in the centres of the observed and simulated galaxies, and investigate how the morphologies of the ISM in galaxy centres vary as a function of galaxy properties.

In Section \ref{data} we briefly discuss the ALMA data reduction and product-creation routines used for the WISDOM data, along with the salient details of the PHANGS dataset. In Section \ref{nonparamorpho} we extract non-parametric measurements of the morphology of the molecular gas, and connect these to galaxy properties in Section \ref{results}. Finally in Sections \ref{discuss} and \ref{conclude} we discuss these results and draw conclusions.

\begin{figure*}
\begin{tabular}{llll}
\begin{turn}{90}\huge \hspace{1.5cm} ETGs \end{turn} & 
\subfloat[NGC0383]{\includegraphics[height=5cm,trim=0cm 0cm 0cm 0cm,clip]{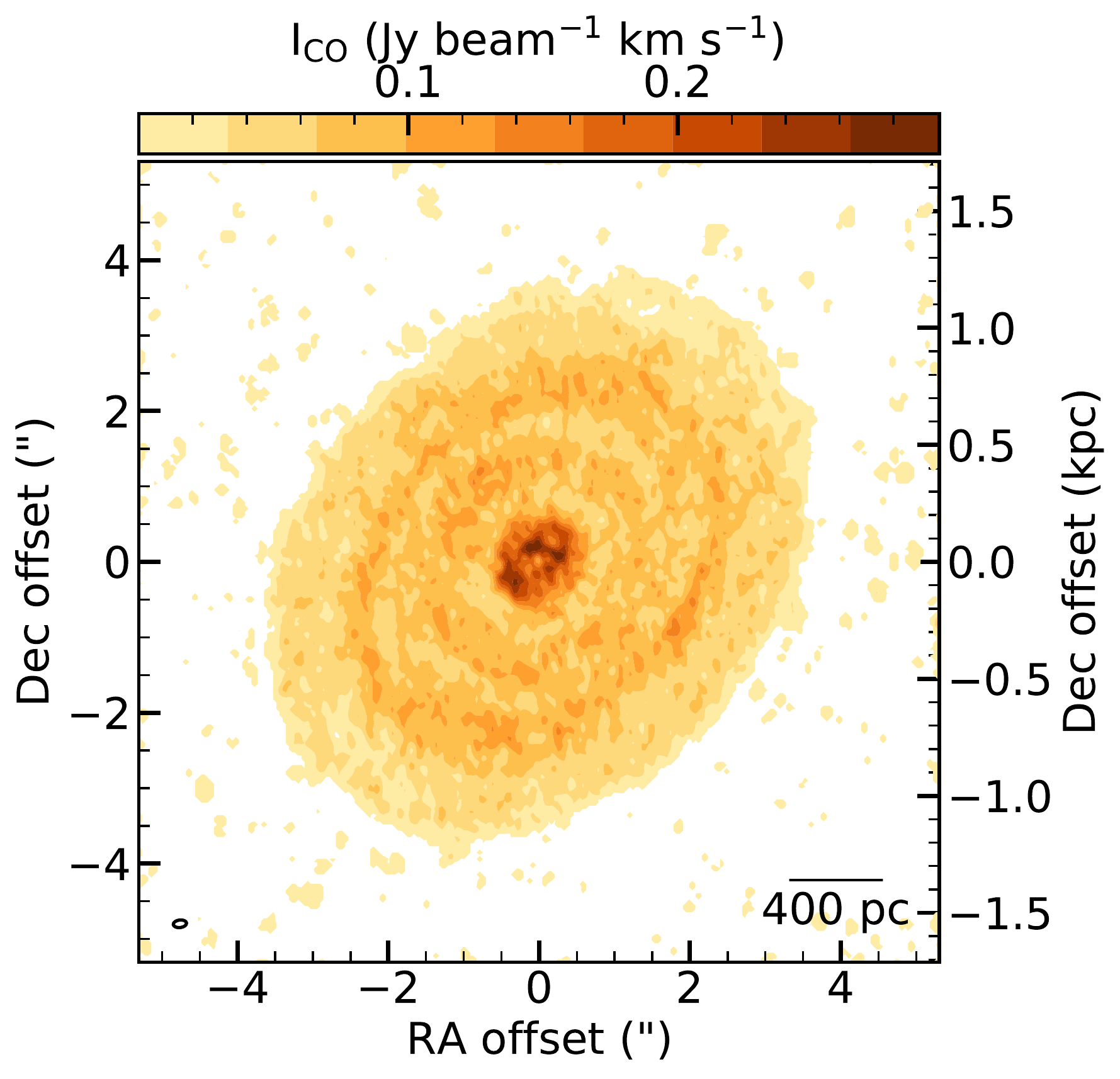}} &
\subfloat[NGC0524]{\includegraphics[height=5cm,trim=0cm 0cm 0cm 0cm,clip]{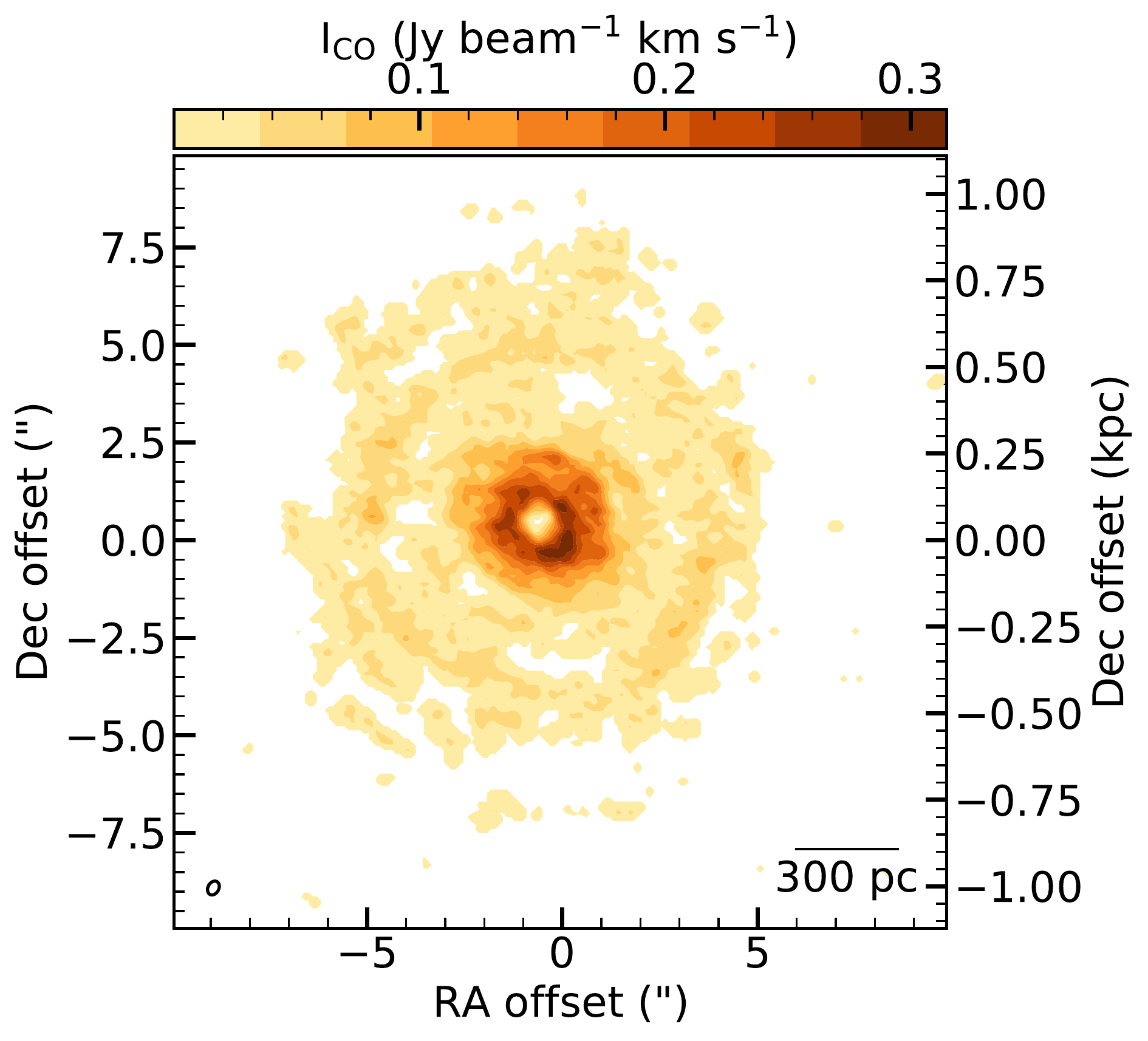}} &
\subfloat[NGC1387]{\includegraphics[height=5cm,trim=0cm 0cm 0cm 0cm,clip]{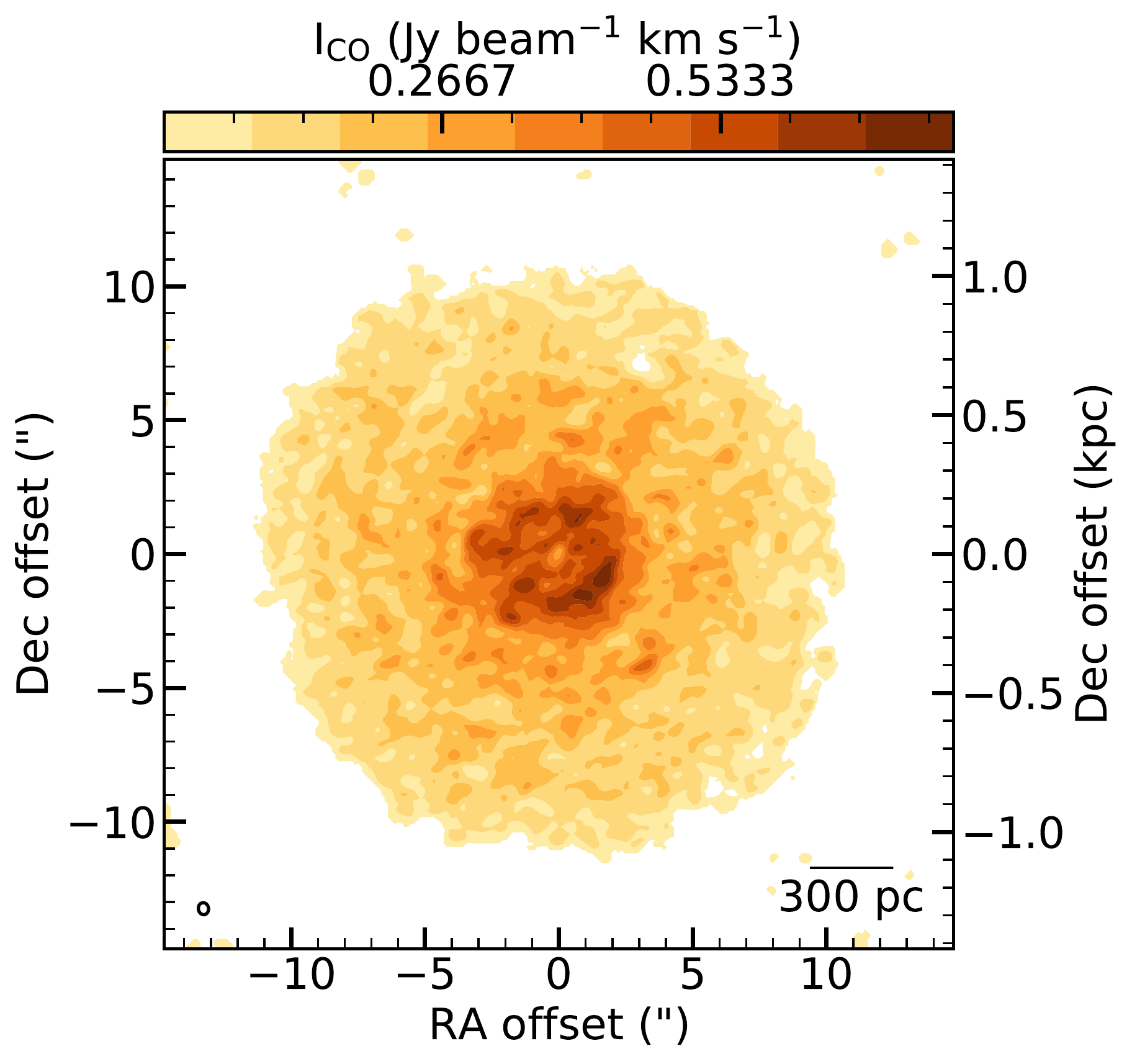}} \\
\hline
\begin{turn}{90}\huge \hspace{1.5cm} Spirals \end{turn} & 
\subfloat[NGC3368]{\includegraphics[height=5cm,trim=0cm 0cm 0cm 0cm,clip]{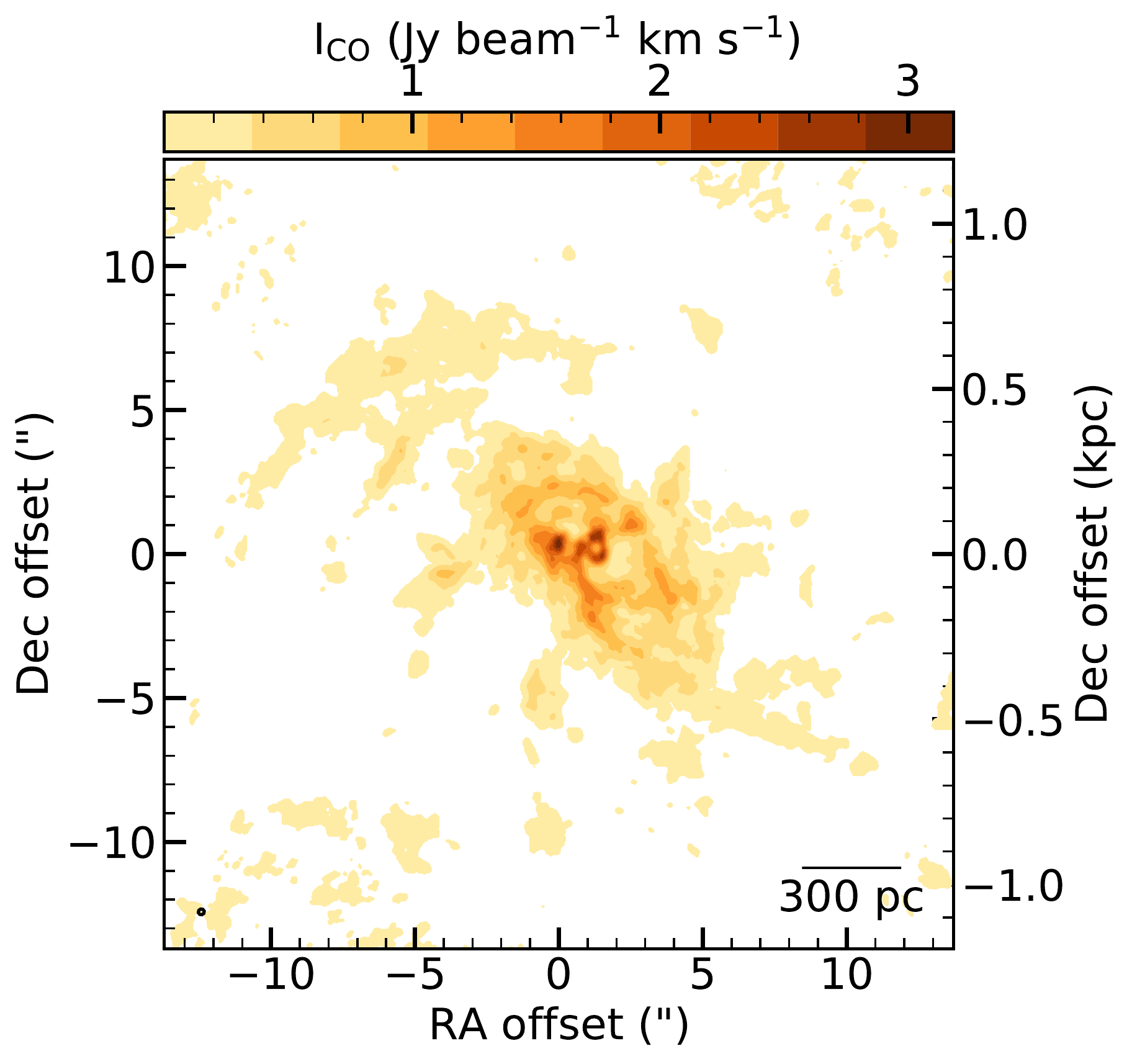}} &
\subfloat[NGC4501]{\includegraphics[height=5cm,trim=0cm 0cm 0cm 0cm,clip]{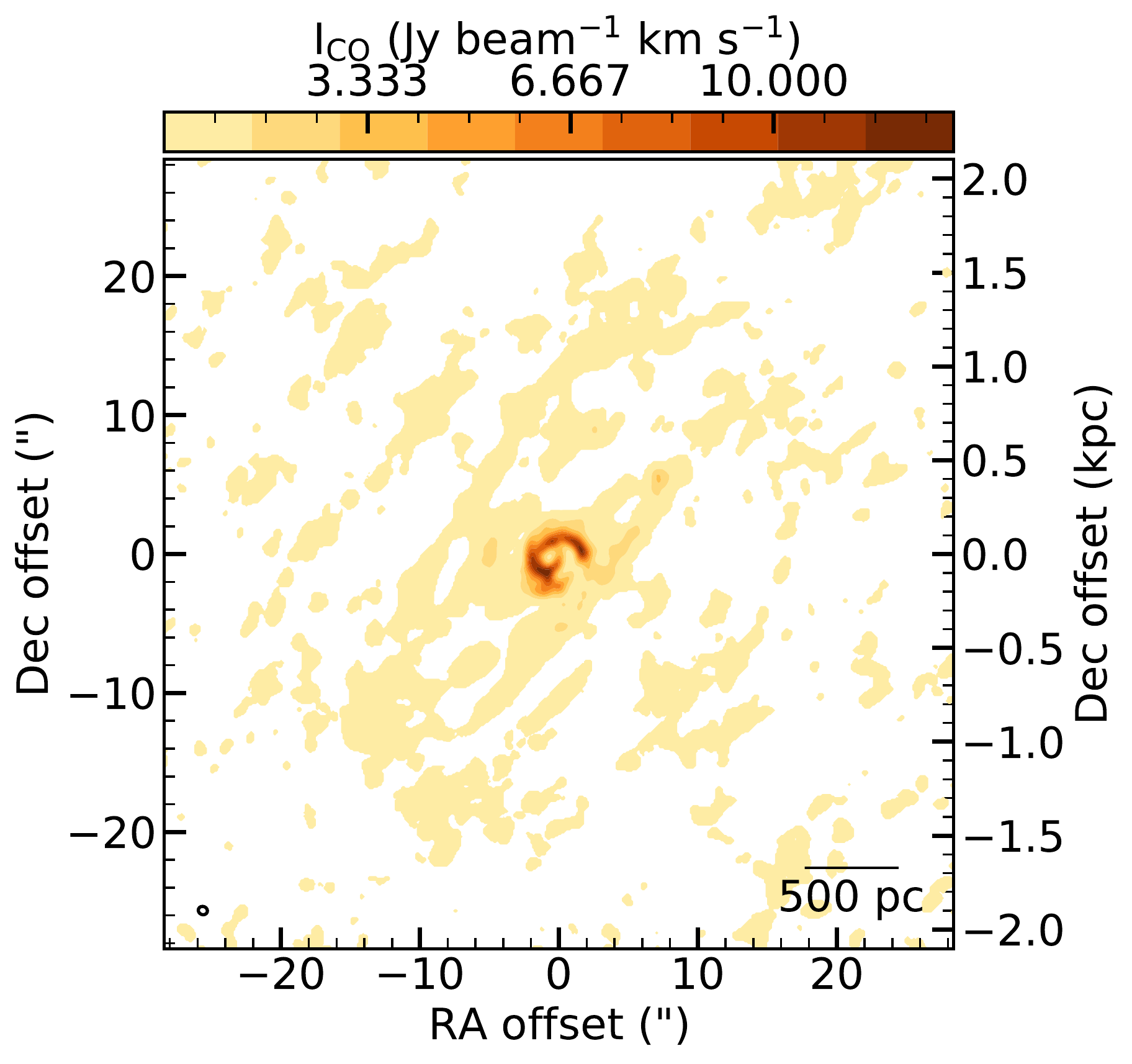}} &
\subfloat[NGC4826]{\includegraphics[height=5cm,trim=0cm 0cm 0cm 0cm,clip]{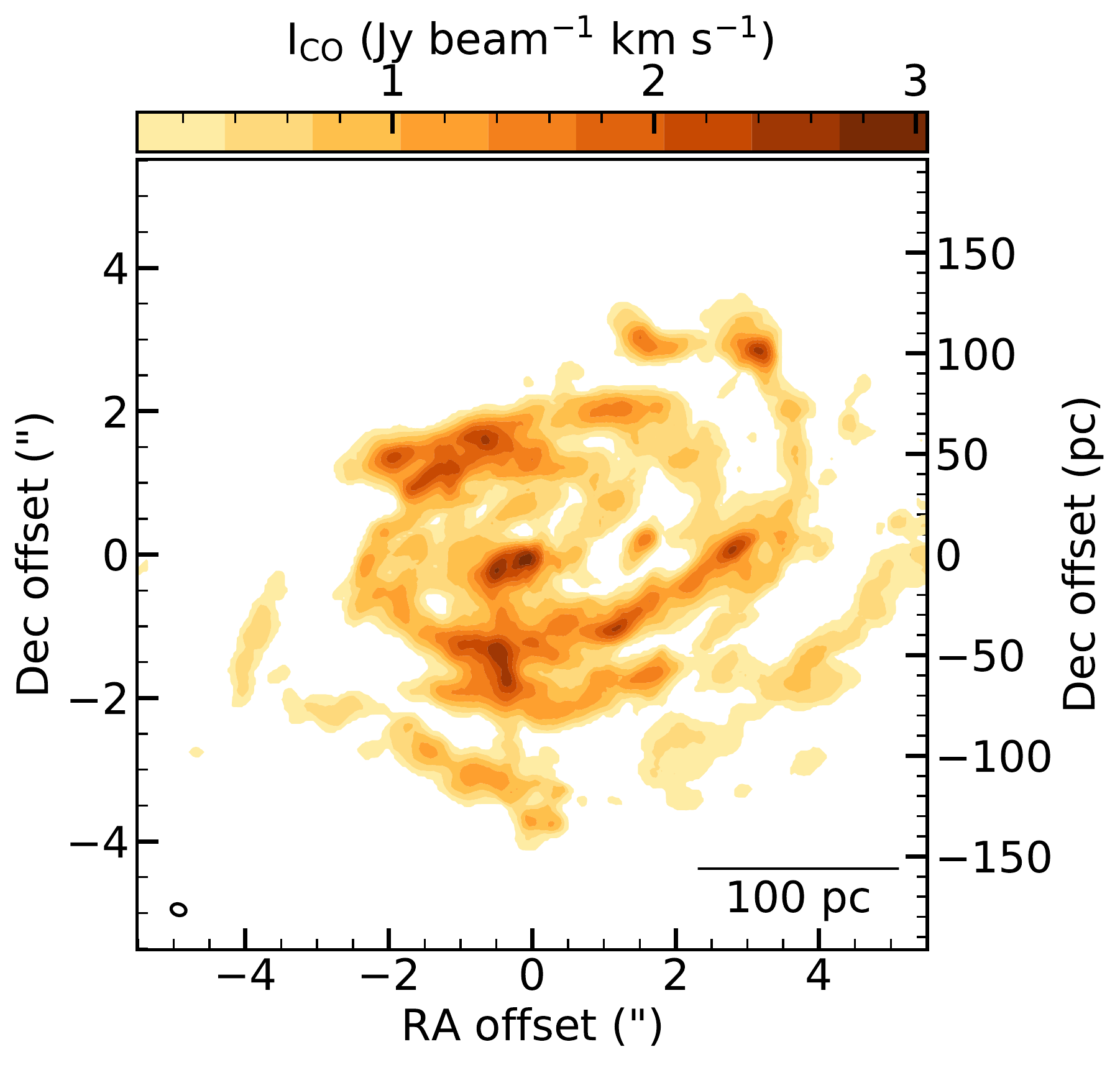}} \\
\end{tabular}
\caption{Integrated-intensity maps of the CO(2-1) or CO(3-2) transition for six galaxies in the WISDOM survey. Galaxies in the top row are all classified as ETGs, while those in the bottom row are spiral galaxies. The synthesised beam is shown as a black ellipse in the bottom left of each panel. The physical resolution reached is similar in each case ($\approx$30\,pc). The morphology of the molecular gas in the ETGs is clearly different from that found in the spirals, suggesting that the properties of the galaxy impact ISM structure on small scales.  } 
\label{mom0s}
\end{figure*}

\section{Data}
\label{data}

In this work we use observations from two surveys: WISDOM and PHANGS. We discuss the WISDOM data in Section \ref{wisdomdata}, and the PHANGS data in Section \ref{phangs_data}.

\subsection{WISDOM galaxies}
\label{wisdomdata}

\subsubsection{Sample selection}
\label{selection_wisdom}
In this work we include 26 galaxies observed as part of the WISDOM project. The data collected by WISDOM were originally obtained with the intent to measure supermassive black hole (SMBH) masses through kinematic modelling of the molecular gas. As such the sample is fairly heterogeneous, containing both nearly quenched ETGs, and star-forming spirals. The sample contains galaxies with both active and inactive nuclei. Those that are active include jetted radio galaxies and radio-quiet Seyfert-like objects.
The main selection criterion for this survey was that the sphere of influence\footnote{The radius of the sphere of influence is given by $R_{\rm SOI}=GM_{\rm BH}/\sigma_*^2$, where $M_{\rm BH}$ is the SMBH mass and $\sigma_*$ the central stellar velocity dispersion.} of the SMBH was spatially resolvable. This typically leads to massive galaxies (that have more massive SMBHs) being observed at somewhat lower spatial resolutions. In addition, where possible, targets were selected to have regular dust lanes in their centres (based on imaging from the {\it Hubble Space Telescope; HST}). This may bias the morphology of the molecular media in these targets, assuming the dust and molecular gas are co-spatial. This issue should be ameliorated somewhat by the inclusion of the larger PHANGS sample which used different selection criteria (see below), and will be discussed further in Section \ref{select_effect}.

\subsubsection{ALMA data}

Here we utilise observations of the CO(2-1) and CO(3-2) lines taken by ALMA as part of the WISDOM project. These observations were carried out between 2013 and 2020 as part of a large number of projects (2013.1.00493.S, 2015.1.00466.S, 2016.1.00419.S, 2016.1.00437.S, 2016.2.00053.S, 2017.1.00277, 2017.1.00391.S, 2017.1.00904.S, 2018.1.00517.S and 2019.1.00363.S). Each object in the sample was observed multiple times using ALMA array configurations with different minimum and maximum baseline lengths, to reach high angular resolutions while ensuring excellent flux recovery.  

Some of these data have been already presented in previous works (see Table \ref{datatable}), where full details of the data reduction and imaging procedures can be found. A brief summary is provided in the following. All datasets were calibrated, combined and imaged using the ALMA pipeline, as provided by the European ALMA Regional Centre staff, and the \textsc{Common Astronomy Software Applications} (\textsc{CASA}) package \citep{2007ASPC..376..127M}. Continuum emission was measured over the full line-free bandwidth, and where detected was subtracted from the data in the $uv$--plane using the \textsc{casa} task \textsc{uvcontsub}. 
The line and continuum data from the combined ALMA datasets were then cleaned and imaged using the \textsc{casa} task \textsc{tclean} and Briggs weighting with a robust parameter of 0.5 (which should provide the best trade-off between sensitivity and angular resolution). The angular resolution reached for each object is indicated in Table \ref{datatable}. {The median surface brightness sensitivity reached by these observations is 0.2~K (or $\approx$14\,\msun\ pc$^{-2}$ with 10\,\kms\ channels and assumptions as per Section \ref{molgasmasses})}. The resulting three-dimensional (RA, Dec, velocity) datacubes were typically produced with channel widths of 10\,\kms, and a pixel size which approximately Nyquist samples the synthesised beam. 

\subsubsection{Moment maps}
In this paper we are interested in the morphology of the ISM, and thus utilise maps of the integrated intensities of the CO lines. These moment-zero maps (shown in Fig. \ref{mom0s} and in online Appendix \ref{allmaps}) were created using the smooth-mask technique \citep[e.g.][]{Dame2011}. Each mask was generated by taking a copy of the cleaned, primary beam-corrected cube and smoothing it, first spatially using a Gaussian of full-width half-maximum (FWHM) equal to 1.5 times that of the synthesised beam, and then spectrally using a boxcar with a width of 4 channels. We then select pixels with an amplitude in the smoothed cube greater than 3 times the root-mean-squared (RMS) noise in that cube. The mask was then applied to the un-smoothed cube to create the moment maps. The exact threshold used to create the moment maps does not have a strong effect on our results, unless it is set unreasonably low (where significant correlated noise at large radii can dominate the non-parametric morphology measurements).

\subsubsection{Molecular gas masses}
\label{molgasmasses}

We estimated the H$_2$ gas mass present within our ALMA field of view, using the standard formalism (see e.g. \citealt{2013ARA&A..51..207B,2021MNRAS.503.5179N})
We assume a typical Milky Way-like CO-to-H$_{2}$ conversion factor of 2$\times$10$^{20}$ cm$^{-2}$ (K \kms)$^{-1}$ \citep{Dickman:1986jz} (equivalent to $\alpha_{\mathrm{CO}}\approx4.36\,$\msun\,(K\,\kms)$^{-1}$\,pc$^{-2}$) and that the $J_{\rm upper}=2$ or $3$ CO lines have a line ratio ${T_{\mathrm{b, CO(2-1)}}/T_{\mathrm{b, CO(1-0)}}}=0.7$ and ${T_{\mathrm{b, CO(3-2)}}/T_{\mathrm{b, CO(1-0)}}}=0.3$ \citep{2021arXiv210911583L}. Derived molecular gas masses are presented in Table \ref{datatable}. 

  Mean molecular gas surface densities within the central one kiloparsec (in radius) of each galaxy were estimated from our moment zero map, using the $\alpha_{\mathrm{CO}}$ and line ratio assumptions described above, and taking the inclination of the source into account. These are also listed in Table \ref{datatable}. 
  
  As is typical with these measurements the systematic uncertainties dominate over random errors. We here assume a fixed 10\% uncertainty in our total H$_2$ masses and surface densities to reflect the uncertainty in the distance estimates we have available.  Further uncertainties due to the CO-to-H$_2$ conversion factor and assumed line-ratio may also be present, and we discuss these additional factors further in later sections.

\subsubsection{Ancillary data}
\label{wisdom_ancillary}
In addition to our ALMA maps, for each WISDOM galaxy we collect a variety of ancillary information from literature sources.

The morphology of each object (as listed in Column 2 of Table \ref{datatable}) is taken from the the NASA Extragalactic Database (NED), or HyperLEDA\footnote{http://leda.univ-lyon1.fr/} \citep{2014A&A...570A..13M}, and confirmed by visual inspection. While optical morphologies are somewhat subjective, the reclassification of any individual source does not change our results. 
The distance to each galaxy is taken from the NED redshift-independent distance catalogue \citep{2017AJ....153...37S}. We adopt the median of the surface-brightness fluctuation distances if they are available, or the median of all available measurements if they are not. We place Virgo and Fornax cluster objects at the mean distance to the cluster (16.5 Mpc and 19.9 Mpc respectively; \citealt{2007ApJ...655..144M,2001ApJ...546..681T}). 

We take the stellar mass of each object ($M_*$) from various sources: ATLAS$^{\rm3D}$ \citep{2013MNRAS.432.1862C}, MASSIVE \citep{2017MNRAS.471.1428V}, the $z$=0 Multiwavelength Galaxy Synthesis (z0MGS) project \citep{2019ApJS..244...24L} and \cite{2017GCN.21707....1C}. Where stellar masses were not available from literature sources we estimated them here from the $K_s$-band magnitude of each galaxy (from the Two Micron All-Sky Survey: 2MASS; \citealt{2003AJ....125..525J}) using Eq. 2 of \cite{2013ApJ...778L...2C}. We assume our stellar masses have an uncertainty of 0.1 dex. The effective radius of each galaxy (R$_e$) was also taken from 2MASS, and we adopt a typical uncertainty on these values of 1\farc5. Star formation rates are taken from \cite{2014MNRAS.444.3427D,2016MNRAS.455..214D} and \cite{2019ApJS..244...24L}, and we assume an uncertainty in these of 0.2 dex. Stellar velocity dispersion measurements were collected from a range of literature sources \citep{2001A&A...374..394V,2007MNRAS.379.1249D,2007MNRAS.378..163H,2009MNRAS.400.2098P,2009ApJS..183....1H,2012A&A...544A.105S,2013MNRAS.432.1709C,2014ApJ...795..158M,2015MNRAS.446.2837S,2015MNRAS.453.1727D,2015ApJS..218...10V,2019MNRAS.483...57P}, and here we adopt a typical uncertainty of 30\,\kms. Given their disparate origins (e.g. from IFU, long-slit and fibre spectroscopic measurements within different apertures), significant scatter should be expected when correlating these with other galaxy properties, but we include them here for completeness. 

From the above parameters we calculated the effective stellar mass surface density ($\mu_*$) as
\begin{equation}
\mu_* \equiv \frac{M_*}{2 \pi R_e^2}.
\end{equation}
 All of the above information is listed in Table \ref{datatable} for each object.

\subsection{PHANGS galaxies}
\label{phangs_data}

In addition to the WISDOM objects (that span a range of visual morphologies), we include CO(2-1) observations of galaxies from the primary data release of the PHANGS \citep{2021arXiv210407739L} survey. 

The PHANGS main galaxy sample includes all 75 systems which meet the following selection criteria:

\begin{itemize}
\item Distances $\ltsimeq$17\,Mpc.
\item Inclinations $<75^{\circ}$
\item Visible to ALMA ($-75^{\circ}< {\rm declination} <+25^{\circ}$)
\item Stellar masses $\gtsimeq$10$^{9.75}$ \msun.
\item Specific star formation rates $>$10$^{-11}$ yr$^{-1}$
\end{itemize}

Given the above cuts, the vast majority of the PHANGS systems are spiral galaxies. Clearly this sample has some limitations, such as not containing low specific star formation rate systems, but crucially it has not been selected based on any measure of {ISM regularity (see Section \ref{selection_wisdom})}.

The PHANGS galaxies were observed with ALMA at lower spatial resolutions than typical for the WISDOM systems. PHANGS reaches a median physical resolution of $\approx$80\,pc in the galaxies included here, compared with a median resolution of 34\,pc for WISDOM.  This is, however, enough to resolve the structures of the molecular ISM in detail. Comparisons of the one source in common (NGC4826) suggest that the morphological parameters we extract here are robust at these resolutions (see also Section \ref{effect_of_res}).  {The PHANGS galaxies were observed with ALMA to a typical surface brightness sensitivity of 0.17~K (or $\approx$11\,\msun\ pc$^{-2}$ with 10\,\kms\ channels and assumptions as per Section \ref{molgasmasses}), very similar to the sensitivity reached by the WISDOM observations, simplifying comparisons between the two}.

To conduct our analyses we use the `strict' moment zero maps contained in the PHANGS data release {(i.e. moment maps created using masks based on emission detected at high
confidence)}, which are most similar to those used in WISDOM. Regenerating the PHANGS moment-zero maps by applying the WISDOM masked-moment technique would not affect our results. For full details of the PHANGS data processing and product creation see \cite{2021arXiv210407665L}. As we concentrate on the structure of the molecular ISM in galaxy centres we extract the innermost 3$\times$3 kpc$^2$ of each PHANGS map (the typical area covered by the WISDOM observations). Quantifying how the morphology of the molecular ISM changes with radius will be explored in a future work. 

We directly use the galaxy properties for the PHANGS galaxies from the data release catalogue as described in \cite{2021arXiv210407739L}. Stellar velocity dispersions were taken from the HyperLEDA catalogue\footnote{http://leda.univ-lyon1.fr/} \citep{2014A&A...570A..13M} or literature sources where available \citep{2006ApJ...642..711L, 2006MNRAS.366..480G, 2009ApJ...690.1031B, 2013A&A...556A..98V, 2015MNRAS.451..859L}. As discussed for the WISDOM stellar velocity dispersions these disparate measurements are extracted from a variety of apertures using various different techniques.  As consistently derived measurements are unavailable we include them here for completeness.  Sixty PHANGS systems have all the data required, and hence can be used in this work, of which 59 are spiral galaxies. Thirty-five of these galaxies have stellar velocity dispersion measurements available, typically the higher stellar mass systems. NGC4826 is included in both surveys. In this case we use the WISDOM data, which have a higher spatial resolution, but very similar morphological parameters are derived from the PHANGS data.%, and as such this would not change our results. 

\begin{table*}

\caption{WISDOM galaxy properties}
\begin{center}
\begin{tabular*}{\textwidth}{@{\extracolsep{\fill}} l l r r r r r r r r r r l l}
\hline
Name & \hspace{-0.25cm}Type & Dist. &  \hspace{-0.25cm}log\,M$_{\rm H_2}$ & log\,${\Sigma_{\rm H_2,1kpc}}$ & log\,M$_*$ & $\sigma_*$ & Re$_{\rm Ks}$ & log\,SFR & log $\mu_*$ & Beam & Beam & Mass Ref. &  Data Ref. \\
 & & \hspace{-0.35cm}(Mpc) & \hspace{-0.25cm}(\msun) & \hspace{-0.25cm}(\msun\ pc$^{-2}$) & (\msun) & \hspace{-0.25cm}(km s$^{-1}$) & \hspace{-0.25cm}(arcsec) & \hspace{-0.25cm}(\msun\ yr$^{-1}$)  & \hspace{-0.25cm}(\msun\ kpc$^{-2}$) & \hspace{-0.25cm}(arcsec) & (pc) & & \\
 (1) & (2) &(3) &(4) & (5) & (6) & (7) &(8) &(9) & (10)&(11)&(12) & (13) & (14)\\
\hline
FRL49 & E${\star}$ & 85.7 & 8.68 & 2.91 & 10.30 & -- & 3 & 0.78 & 9.31 & 0.19 & 77.2 & Lelli+ subm. & Lelli+subm. \\
MRK567 & S & 140.6 & 8.79 & 3.28 & 11.26 & -- & 6 & 1.30 & 9.24 & 0.14 & 93.4 & C17 & -- \\
NGC0383 & E & 66.6 & 9.18 & 2.66 & 11.82 & 239 & 11 & 0.00 & 9.92 & 0.13 & 42.8 & MASSIVE & \cite{2019MNRAS.490..319N} \\
NGC0449 & S & 66.3 & 9.50 & 2.24 & 10.07 & 250 & 7 & 1.19 & 8.60 & 0.66 & 211.2 & z0MGS & -- \\
NGC0524 & E & 23.3 & 7.95 & 1.41 & 11.40 & 220 & 24 & -0.56 & 9.75 & 0.32 & 36.7 & z0MGS & \cite{2019MNRAS.485.4359S} \\
NGC0612 & E & 130.4 & 10.30 & 1.73 & 11.76 & -- & 9 & 0.85 & 9.13 & 0.19 & 122.2 &M$_{K\mathrm{s}}$   & Ruffa+ in prep  \\
NGC0708 & E & 58.3 & 8.48 & 2.04 & 11.75 & 230 & 24 & -0.29 & 9.30 & 0.09 & 24.1 & MASSIVE & \cite{2021MNRAS.503.5179N} \\
NGC1387 & E & 19.9 & 8.33 & 2.04 & 10.67 & 87 &  16 & -0.68 & 9.51 & 0.42 & 40.3 & z0MGS & Boyce+ in prep \\
NGC1574 & E & 19.3 & 7.64 & 2.02 & 10.79 & 180 &21 & -0.91 & 9.41 & 0.17 & 15.4 & z0MGS & Ruffa+ in prep \\
NGC3169 & S & 18.7 & 9.53 & 2.29 & 10.84 & 165 & 86 & 0.29 & 8.26 & 0.60 & 54.0 & z0MGS & -- \\
NGC3368 & S & 18.0 & 9.03 & 2.46 & 10.67 & 102 & 37 & -0.29 & 8.87 & 0.20 & 17.9 & z0MGS & -- \\
NGC3607 & E & 22.2 & 8.42 & 1.86 & 11.34 & 207 & 22 & -0.54 & 9.80 & 0.55 & 59.0 & A3D & -- \\
NGC4061 & E & 94.1 & 9.43 & 2.43 & 11.64 & -- & 8 & -0.71 & 9.71 & 0.13 & 59.2 & MASSIVE & -- \\
NGC4429 & E & 16.5 & 8.00 & 1.60 & 11.17 & 177 & 49 & -0.84 & 9.19 & 0.16 & 12.8 & A3D &  \cite{2018MNRAS.473.3818D} \\
NGC4435 & E & 16.5 & 8.63 & 1.63 & 10.69 & 153 & 29 & -0.84 & 9.18 & 0.24 & 19.1 & A3D & -- \\
NGC4438 & S & 16.5 & 9.56 & 2.38 & 10.75 & 142 & 23 & -0.30 & 9.42 & 0.56 & 45.1 & z0MGS & -- \\
NGC4501 & S & 15.3 & 8.90 & 2.09 & 11.00 & 102 & 58 & 0.43 & 8.94 & 0.63 & 42.6 & z0MGS & -- \\
NGC4697 & E & 11.4 & 7.20 & 0.77 & 11.07 & 169 & 40 & -1.08 & 9.59 & 0.55 & 30.5 & A3D &  \cite{2017MNRAS.468.4675D} \\
NGC4826 & S & 7.4 & 7.89 & 2.59 & 10.20 & 90 & 69 & -0.71 & 8.62 & 0.18 & 6.5 & z0MGS & -- \\
NGC5064 & S & 34.0 & 9.90 & 2.75 & 10.93 & 210 &  18 & 0.11 & 9.19 & 0.06 & 9.9 & z0MGS & Onishi+ in prep \\
NGC5765b & S & 114.0 & 10.08 & 2.96 & 11.21 & -- &  7 & 1.43 & 9.30 & 0.32 & 178.4 &M$_{K\mathrm{s}}$   & -- \\
NGC5806 & S & 21.4 & 8.97 & 1.97 & 10.57 & 110 & 30 & -0.03 & 8.80 & 0.30 & 31.0 & z0MGS & -- \\
NGC6753 & S & 42.0 & 9.62 & 2.72 & 10.78 & 214 & 20 & 0.32 & 8.78 & 0.14 & 28.4 & z0MGS & -- \\
NGC6958 & E & 35.4 & 8.66 & 1.99 & 10.76 & 168 & 12 & -0.58 & 9.35 & 0.13 & 19.0 & z0MGS & Thater+ in prep \\
NGC7052 & E & 51.6 & 9.26 & 2.23 & 11.75 & 266 & 15 & -0.07 & 9.82 & 0.13 & 32.1 & MASSIVE & \cite{2021MNRAS.503.5984S} \\
NGC7172 & E & 33.9 & 9.78 & 2.53 & 10.76 & 180 & 19 & 0.38 & 8.97 & 0.14 & 22.2 & z0MGS & -- \\
\hline
\end{tabular*}
\label{datatable}\vspace{0.01cm}
\parbox[t]{\textwidth}{ \textit{Notes:} Column 1 lists the galaxy name, and column 2 the galaxy type (E for ETG or S for Spiral). FRL49, indicated with a ${\star}$ was classified as an ETG from low-resolution optical imaging, but hosts a starburst at its core, and so does not nicely fit in either the ETG or spiral category. Column 3 contains the distance assumed for each galaxy (taken from the NASA extragalactic database redshift independent distance catalogue where possible \citealt{2017AJ....153...37S}). Column 4 lists the molecular gas mass within our ALMA field of view estimated using a galactic CO-to-H$_2$ conversion factor as described in Section \ref{molgasmasses} (apart from FRL49, where the molecular gas mass is estimated dynamically, see Lelli et al, submitted). Column 5 contains the mean molecular gas surface density within the inner kiloparsec of the galaxy (see Section \ref{molgasmasses} for full details). Column 6 lists the stellar mass of the galaxies (calculated from dynamics where possible, and through photometry proxies otherwise; see Column 13). Column 7 contains stellar velocity dispersion measurements from a range of sources. See the text for full details. Column 8 contains the $K_{\rm s}$-band effective radius estimated for each system from the 2$\mu$m all-sky survey (2MASS; \citealt{2003AJ....125..525J}). Column 9 contains the total star formation rate of each galaxy taken from  \cite{2014MNRAS.444.3427D,2016MNRAS.455..214D} and \cite{2019ApJS..244...24L}. Errors on the quantities in columns 4-9 are described fully in the text. Column 10 contains the stellar mass surface density estimated within the effective radius of each system. Columns 11 and 12 detail the angular and physical resolution reached by our ALMA data, respectively.  Column 13 details the source of the stellar mass measurements: A3D refers to \cite{2013MNRAS.432.1862C}, C17 to \cite{2017GCN.21707....1C}, MASSIVE to \cite{2017MNRAS.471.1428V}, and z0MGS to \cite{2019ApJS..244...24L}. M$_{K\mathrm{s}}$ refers to masses estimated from the galaxies $K_{\rm s}$-band magnitude using Equation 2 of \cite{2013ApJ...778L...2C}. Column 14 contains the reference where the ALMA data for a given WISDOM galaxy was previously (or soon to be) published, if applicable.}
\end{center}

\end{table*}

\subsection{Simulations}
 \label{simdetails}

To help distinguish the physical mechanisms shaping the ISM in the centres of our sample objects, we compare with simulated galaxies from \cite{2020MNRAS.495..199G}. These authors conducted high-resolution hydrodynamic simulations of a set of galaxies using the \textsc{Arepo} moving-mesh code \citep{2010MNRAS.401..791S}. 
 All simulated galaxies are initialised with a stellar mass of 4.71$\times$10$^{10}$\,\msun, a \cite{1990ApJ...356..359H} dark matter halo of 2$\times$10$^{12}$\,\msun\ and a gas fraction of 5~per~cent. The discs of gas and stars are initially setup with an exponential surface density profile, with a scale radius of 4.6 kpc. The only differences between the simulations are the bulge-to-disc mass ratios and the bulge scale radii. Therefore, the effective stellar surface density varies between the simulated galaxies, despite their identical total stellar mass.

 As discussed in detail in Gensior et al. in prep, the gas discs in the centres of these simulated galaxies appear visually similar to those in the WISDOM systems, becoming more smooth and symmetric in the bulge-dominated galaxies. 
  Here we analyse face-on projected gas mass surface density maps which were output from these simulations. 
Formally the simulations do not track molecular gas, so we create mock moment zero maps by imposing a gas mass surface density cut of 10 M$_{\odot}$\,pc$^{-2}$, where the gas is expected to become primarily molecular \citep[e.g.][]{2009ApJ...693..216K}. The exact value we take for this cut does not change our conclusions. We then analyse these simulated gas maps in the same way as the observations.  Table \ref{simtable} lists the galaxy properties that were varied for each simulation, and the resulting galaxy properties we compare with the observations. 

The simulations performed by \cite{2020MNRAS.495..199G} are not perfectly matched to our sample galaxies. By construction the simulated galaxies have somewhat lower stellar surface densities within their effective radii (see Column 10 of Table \ref{datatable} and Column 4 of Table \ref{simtable}), and the gas surface densities within a radius of one kiloparsec are significantly lower (as the ISM is by construction more radially extended; see Column 5 of Tables \ref{datatable} and \ref{simtable}). We will comment further on these differences and how they affect our analyses where appropriate throughout the paper.

\begin{table*}

\caption{Simulated galaxy properties.}
\begin{center}
\begin{tabular*}{0.7\textwidth}{@{\extracolsep{\fill}} l r r r r r r r}
\hline
Name & $B$/$T$ & R$_{\rm bulge}$ & log $\mu_*$  & ${\Sigma_{\rm H_2}}$ & SFR & log sSFR & $\sigma_*$\\
 & & (kpc) & (\msun\ kpc$^{-2}$) & (\msun\ pc$^{-2}$) & (\msun\ yr$^{-1}$)  & (yr$^{-1}$) & (km s$^{-1}$)\\
 (1) & (2) &(3) &(4) & (5) & (6) & (7) & (8) \\
\hline
noB & 0 & 0.0 & 8.09 & 45.09 & 0.93 & -10.70 & 75.9 \\
B\_M30\_R1 & 30 & 1.0 & 8.33 & 25.21 & 0.59 & -10.90 & 98.3 \\
B\_M30\_R2 & 30 & 2.0 & 8.24 & 37.28 & 0.66 & -10.86 & 93.1\\
B\_M30\_R3 & 30 & 3.0 & 8.17 & 20.31 & 0.20 & -11.37 & 93.4 \\
B\_M60\_R1 & 60 & 1.0 & 8.74 & 21.90 & 0.33 & -11.16 & 122.4 \\
B\_M60\_R2 & 60 & 2.0 & 8.46 & 23.30 & 0.28 & -11.23 & 105.5\\
B\_M60\_R3 & 60 & 3.0 & 8.30 & 38.07 & 0.70 & -10.83 & 101.9 \\
B\_M90\_R1 & 90 & 1.0 & 9.28 & 24.48 & 0.17 & -11.45 & 142.9\\
B\_M90\_R2 & 90 & 2.0 & 8.76 & 23.26 & 0.42 & -11.04 & 116.9\\
B\_M90\_R3 & 90 & 3.0 & 8.49 & 30.38 & 0.49 & -10.98 & 108.8\\
\hline
\end{tabular*}
\label{simtable}\vspace{0.01cm}
\parbox[t]{0.7\textwidth}{ \textit{Notes:} Column 1 contains the name of the simulation. Column 2 lists the bulge-to-total mass ratio of the system. The bulge scale radius and the resulting effective stellar mass surface density of the system are shown in Columns 3 and 4. Column 5 lists the median gas surface density in the inner kiloparsec of each system, while columns 6 and 7 contain the total star formation rate and specific star formation rate of each system at the time the snapshot was output. }
\end{center}
\end{table*}

\section{Non-parametric gas morphology measurements}
\label{nonparamorpho}
To quantitatively investigate the morphology of the ISM in the WISDOM galaxies (and compare to the simulated galaxies), we make use of modified versions of the Asymmetry statistic, Smoothness statistic and Gini coefficient used extensively by optical astronomers \citep[e.g][]{2003ApJS..147....1C,2019MNRAS.483.4140R} and also in \hi\ studies  \citep[e.g.][]{2011MNRAS.416.2415H}. 
The concentration and M20 statistics (which both measure the central concentration of the gas; \citealt{2000AJ....119.2645B,Conselice2003}) typically used alongside Asymmetry, Smoothness and Gini do not show strong correlations with any of the physical variables probed here, and hence we do not consider them further. 

In each case these measurements are performed on the moment-zero map, which covers the inner $3\times3$\,kpc$^{2}$ of each system (see Table \ref{castable}). The moment maps are further masked using elliptical apertures (defined using the position angle and inclination of each system) that contain 90\% of the CO flux. This ensures our measurements are not biased by large areas in the outer regions of some galaxies where no gas is detected. The choice of 90\% for this flux threshold is arbitrary, but our results remain robust to any reasonable choice of this parameter (75-95\%). We calculate uncertainties that correspond to the change in each statistic when altering this flux threshold by $\pm5$\%, and list these in Table \ref{castable}. For the simulations the uncertainties are defined differently, and reflect the variation when performing each measurement on different snapshots of the simulation. 

\subsection{Asymmetry}

The Asymmetry index ($A$; \citealt{1995ApJ...451L...1S,1996MNRAS.279L..47A}) quantifies the degree of rotational (a)symmetry present in a distribution, and hence can be used to assess how asymmetric the gas discs in our sources are. {It also has the advantage that it is insensitive to all but the most extreme inclination changes.} It is obtained by subtracting the moment map rotated by $180^{\circ}$ from the original map:
\begin{flalign}
A \equiv \frac{\sum_{i,j}|I_{ij} - I_{ij}^{180}|}{\sum_{i,j}|I_{ij}|}, &&
\label{eq:Asymmetry}
\end{flalign}
where $I_{ij}$ and $I_{ij}^{180}$ are the surface brightness of pixel $i,j$ in the original and rotated moment map, respectively, and the sum is taken over all pixels in the map $i,j$. We note that a background Asymmetry term is typically included when calculating this parameter in optical images, but this is neglected here as the correction is minimal (we apply the statistics on our highly resolved, high signal-to-noise masked moment-zero maps where the background has already been removed).

\subsection{Smoothness}

The Smoothness index ($S$; \citealt{2003ApJS..147....1C}) quantifies how smooth a two-dimensional distribution is on a given spatial scale, and hence can be used to investigate the Smoothness of the ISM of our galaxies. It is obtained by subtracting the moment-zero map smoothed with a boxcar filter of width $\sigma$ from the original map:
\begin{flalign}
S \equiv \frac{\sum_{i,j}I_{ij} - I_{ij}^{\rm S}}{\sum_{i,j}I_{ij}}, &&
\label{eq:Smoothness}
\end{flalign}
where $I_{ij}$ and $I_{ij}^{\rm S}$ are the surface brightness of pixel $i,j$ in the original and smoothed images, respectively, and the sum is taken over all pixels in the map $i,j$. The boxcar filter width $\sigma$ is set to 500\,pc here, and we explore the impact of this choice in Section \ref{nonpara_morphs}. Once again thebackground term typically used when working with optical images can be neglected. 
We also note that as defined here, larger $S$ corresponds to galaxies that are {less} smooth (i.e. more clumpy).

\subsection{Gini coefficient}

The Gini coefficient ($G$) quantifies the (in)equality in a distribution, and hence can provide an alternative measurement for the Smoothness of the ISM. This statistic is primarily used in economics to quantify the wealth inequality in a population, but it has also been used widely in astronomy \citep[e.g.][]{1996MNRAS.279L..47A}. For a set of $n$ pixel fluxes $X_i$, where $i = 1, 2, ..., n$, the Gini coefficient can be computed as in \cite{2019MNRAS.483.4140R}
\begin{flalign}
G \equiv \frac{1}{\bar{X} n (n-1)} \sum_{i=1}^{n} (2i - n - 1) X_i,&&
\label{eq:gini1}
\end{flalign}
where $\bar{X}$ is the mean flux measured over all pixels. A value of $G=1$ is obtained when all of the flux is concentrated in a single pixel, while a homogeneous brightness distribution yields $G=0$. Small values of $G$ thus again imply smoother gas distributions.

\subsection{The impact of resolution}
\label{effect_of_res}
In this work we are combining data for galaxies at different distances, using datasets from two different surveys. The physical scales probed by our datasets therefore vary. In principle this could affect the derived non-parametric morphology measurements. For instance, one would expect that as the spatial resolution of the data used becomes coarser galaxies will appear smoother, and less asymmetric as fine detail is not resolved. 

To avoid this issue, in this paper we present measurements of the above statistics made on moment zero maps convolved to a fixed physical resolution of 120\,pc. Measurements made on the maps at their original resolution show that the retrieved non-parametric morphologies are only marginally affected by this change. The derived asymmetries measured {at 120\,pc are $\approx$0.1 smaller on average than those measured at the original resolution of the data.} Smoothness and Gini were less affected (reducing by 0.05 and 0.03 on average, respectively). None of the correlations found in this work would change if we adopted the values derived at the highest possible spatial resolution.

We note that, in principle, the correct way to obtain lower resolution moment-zero maps is by tapering the observed visibilities in the $uv$-plane within the individual interferometric datasets, and then re-imaging the data. As we do not have access to the calibrated interferometric data for all our sources, however, we here simply convolved each map to a fixed physical resolution of 120\,pc. Tests on several of the WISDOM objects suggests that the non-parametric morphology measurements obtained from tapered data differ very little {(by $<10$\%)} from those obtained on convolved datasets, and hence we do not expect this to bias our results.

Twelve of the galaxies observed in this work have native spatial resolutions of 130-220\,pc, significantly larger than our 120\,pc limit. We chose to include these galaxies here to maximise our statistics, but we have verified that removing them would not change any of our conclusions.

\section{Results}
\label{results}

In Figure \ref{mom0s} we show six examples of the integrated intensity maps of CO(2-1) or CO(3-2) from the WISDOM survey at their original spatial resolution, which in these cases is $\approx$30\,pc. The top row includes galaxies classified using \textit{optical imaging} as early-type, while the second row contains spiral galaxies. It is clear even by eye that the typical molecular gas morphology of ETGs is different from that of spiral galaxies. In ETGs the gas appears generally fairly smoothly distributed on 100\,pc scales. Small clumps are visible, but these are embedded in the larger-scale smooth discs, which can also show some low-level ring structures. In the spirals, on the other hand, at the same physical scale the gas is concentrated into strong spiral structures, that are often asymmetric and clumpy. The remaining moment-zero maps are shown in Appendix \ref{allmaps}, which is available online. Similar trends are obvious in these maps as well (see Section \ref{discuss} for a visual representation of this trend). 

\begin{figure*}
\includegraphics[width=0.75\textwidth,trim=0cm 0cm 0cm 0cm,clip]{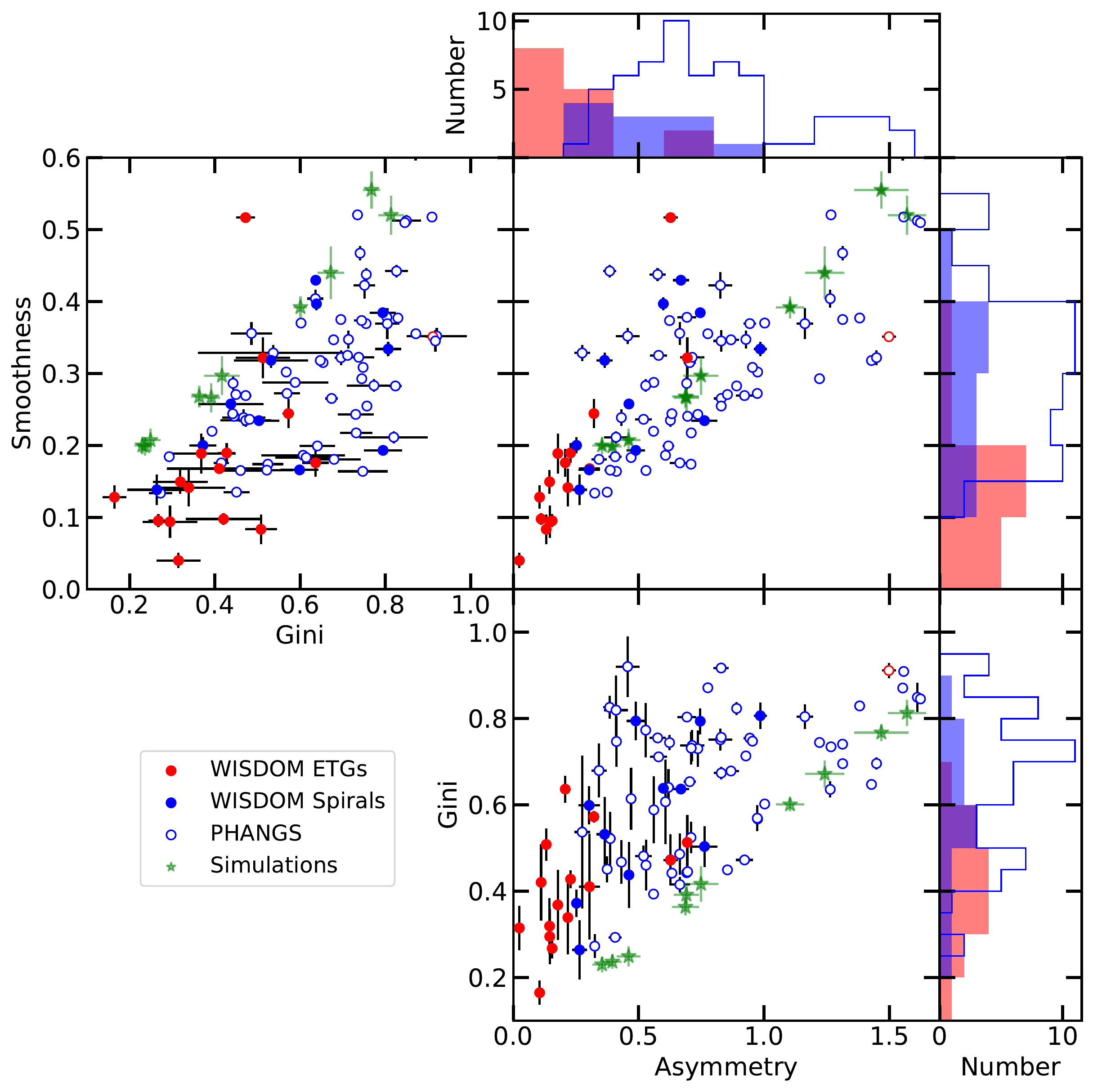}
\caption{Asymmetry, Smoothness and Gini coefficient  measurements for the molecular medium in our observed and simulated galaxies at 120\,pc spatial resolution, plotted against each other (three main panels). WISDOM early-type galaxies are indicated as red circles, while spiral galaxies are shown in blue. PHANGS late-type galaxies are shown as open blue symbols. Simulated galaxies are indicated by green stars. We find that all three non-parametric morphology parameters are strongly correlated with each other.  The marginal distribution for each class of system is shown in the histograms at the top and right of each panel. WISDOM early and late-type galaxies are shown as red and blue shaded histograms, respectively, while PHANGS spiral galaxies are shown as open blue histograms.  These marginal distributions clearly show that each galaxy type (spiral, ETG)  has a distinct distribution in each non-parametric morphology measure.} 
\label{cas}
\end{figure*}
\begin{figure*}
\includegraphics[height=12cm,trim=0.5cm 0cm 0.5cm 0cm,clip]{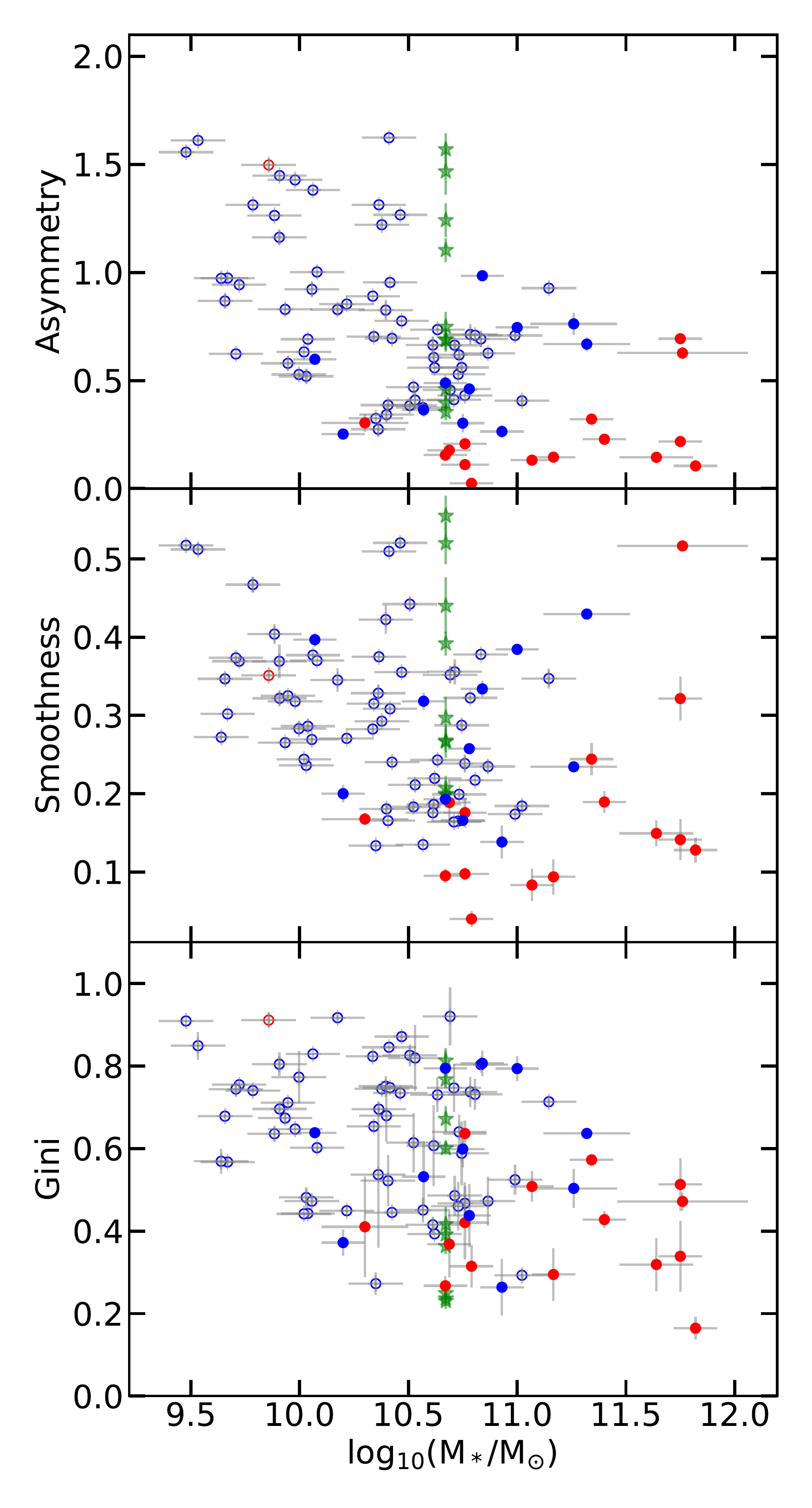}
\includegraphics[height=12cm,trim=2.7cm 0cm 0.5cm 0cm,clip]{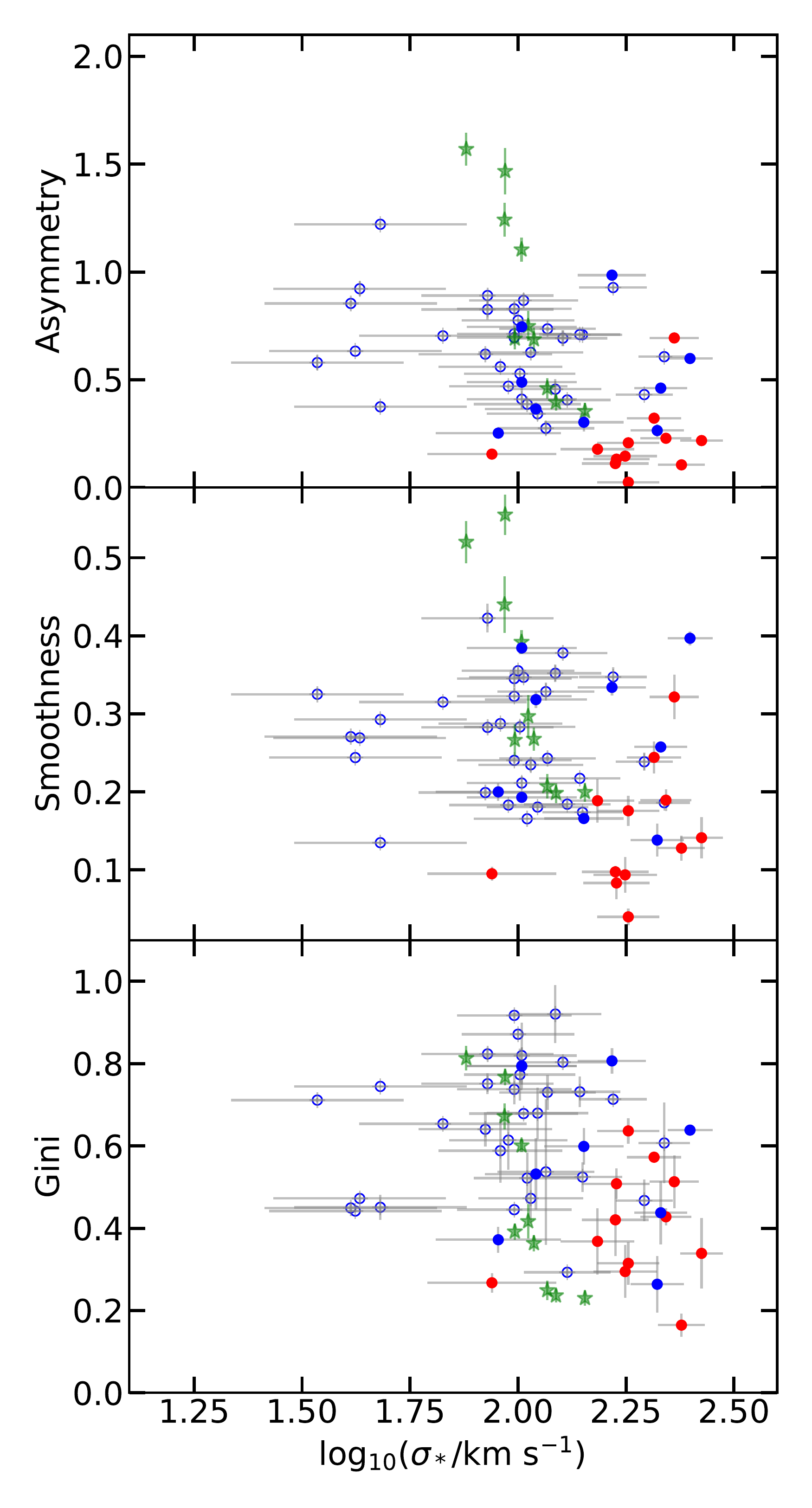}
\includegraphics[height=12cm,trim=2.7cm 0cm 0.5cm 0cm,clip]{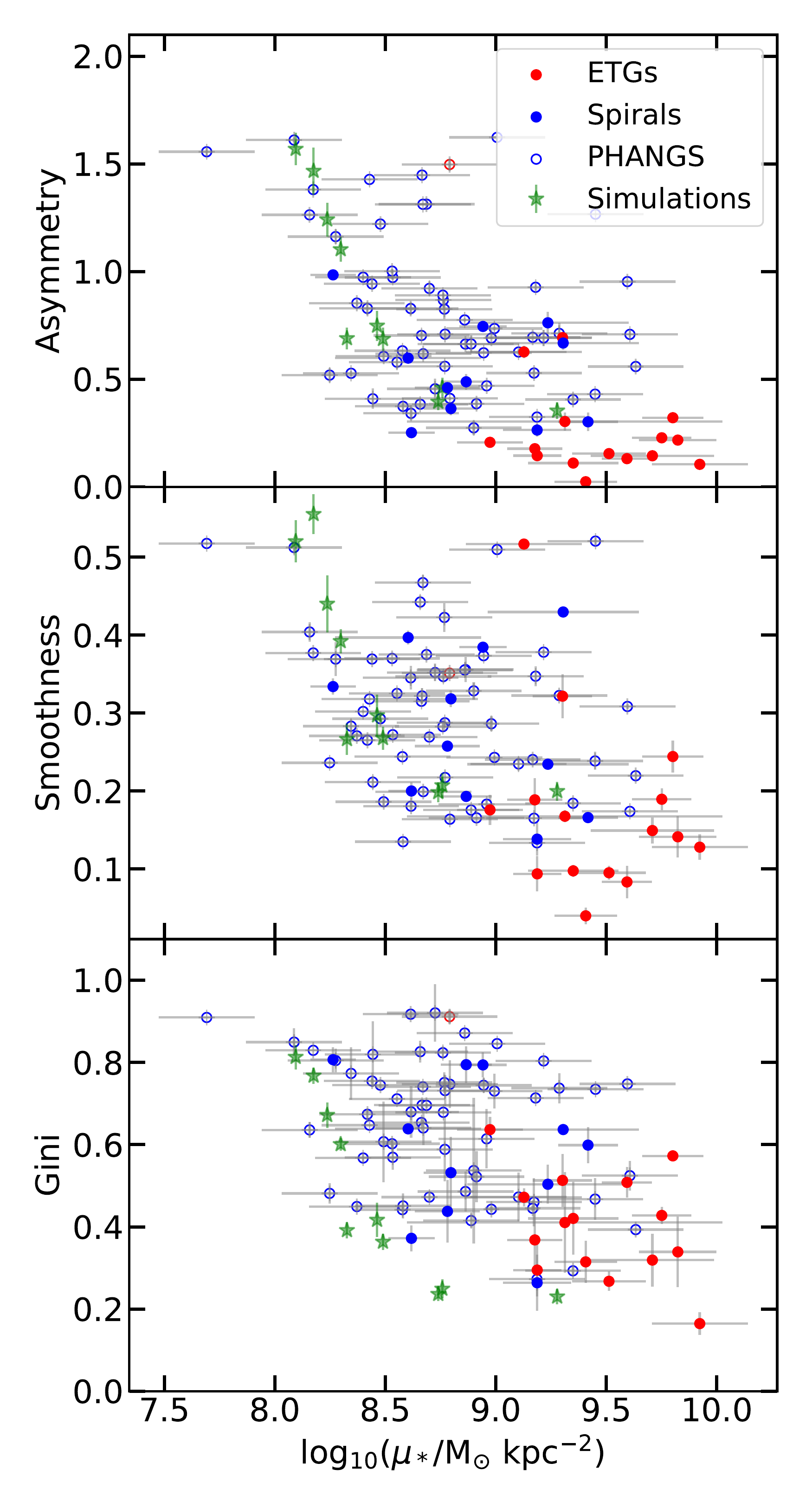}
\caption{Non-parametric morphology measurements (Asymmetry; top panel, Smoothness; middle panel, and Gini; bottom panel) for the molecular ISM at 120\,pc scales in the sample galaxies, plotted against the total stellar mass (left column), stellar velocity dispersion (central column) and the stellar mass surface density of the galaxy (right column). Symbols are as in Figure \ref{cas}. All three morphology measures (and the optical morphology) correlate with the stellar mass, velocity dispersion and the stellar mass surface density. The morphology of the ISM in the simulated systems also show a strong correlation with stellar mass surface density with a similar slope, albeit offset in surface density. The observed correlations are highly significant (see Table \ref{corr_table}), suggesting that the depth of the potential well of these galaxies influences the structure of the ISM.} 
\label{mu_a_s}
\end{figure*}

At first glance it may seem unsurprising that the molecular medium is dominated by spiral features in spiral galaxies (and not in ETGs). However, there is no obvious physical reason the ISM in galaxy centres should follow the large-scale structure of the galaxy. The molecular ISM is typically dense in these regions, and molecular clouds with high surface densities may become decoupled from their surroundings. This is especially true at small scales close to the galaxy centre, where spiral arms are unimportant, bars may have rearranged the gas, and the gravitational potential is expected to be significantly more spherical due to the impact of the bulge. A comparison of these maps with those of the simulated galaxies shows that the latter follow similar trends, with the gas fragmenting significantly more in low bulge fraction galaxies (see Figures 7 and 13 in \citealt{2020MNRAS.495..199G}).

\subsection{Non-parametric morphologies}
\label{nonpara_morphs}
To quantify the visual impression arising from Figure \ref{mom0s} that ETGs have a smoother molecular ISM morphology we utilise our non-parametric morphology measurements. These measurements were made from our CO integrated-intensity maps with a fixed spatial resolution of 120\,pc. 
In Figure \ref{cas} we show the Asymmetry, Smoothness and Gini coefficients for the WISDOM galaxies (solid circles), PHANGS systems (open circles) and the simulations (green stars), here plotted against each other. Early-type systems are shown in red, while spiral galaxies are shown in blue.  
All three of these parameters are strongly correlated with one another (see Table \ref{corr_table} for statistical test results on the observational samples). Some correlations are expected (e.g. between Gini and Smoothness, as these are different statistics that attempt to estimate the same thing), but it seems that galaxies with smooth ISMs also tend to be symmetric, while galaxies with clumpy ISMs can display a wider range of asymmetries. These measures also seem to correlate with galaxy type, as is shown by the marginal distribution for each class of system in the histograms at the top and right of each panel.
It seems that early type systems have a smooth, symmetric ISM distribution (low Asymmetry, Smoothness and Gini), while the ISM in spiral galaxy centres is clumpy and asymmetric (high Asymmetry, Smoothness and Gini). This confirms the trend  identified by eye and described in Section \ref{results}.

{Two galaxies classified as ETGs are outliers from the main trends defined above. These are PHANGS galaxy NGC4694, and WISDOM galaxy NGC0612. NGC4694 is host to an ongoing gas rich merger (which is forming a tidal dwarf galaxy; \citealt{2007A&A...475..187D,2013MNRAS.432.1796A}), explaining its disturbed state. NGC0612 is a gas rich lenticular galaxy, hosting an exceedingly large gas disc (with a radius of $\approx$10\,kpc) which is warped in the outer parts, likely due to interaction with its environment \citep{2008MNRAS.387..197E,2019MNRAS.484.4239R}.}

We note that the Gini and Asymmetry indices are robust, in that their values do not depend on any free parameters. The Smoothness index, however, depends both on the smoothing kernel chosen, and (in extreme cases) how this compares to the total image size. In this work we have adopted a smoothing scale 500\,pc. If we had instead chosen a fixed multiple of the interferometric beam or of the total molecular disc size the absolute values derived for the Smoothness index would change, but the overall trends remain. 

In the remainder of this section we will correlate these non-parametric morphology measurements with physical properties of the galaxies to try and identify the drivers of these observed trends between galaxy morphology and ISM morphology.

\subsection{Correlations with physical properties}

\subsubsection{Stellar mass and the galaxy potential}

One possible driver of the correlation between ISM and galaxy morphology is the (shapes and depths) of the gravitational potential wells of these galaxies. Indeed, the `external' (i.e. galactic) gravitational forces on the gas can help to stabilise it against gravitational collapse (see Section \ref{mu_explain}). In principle, one would wish to correlate measures of ISM morphology directly with the properties of each galaxy's circular velocity curve. However, suitable rotation curves for these galaxies are not available uniformly\footnote{Rotation curves for the PHANGS sample galaxies were extracted by modelling moment-one maps in \cite{2020ApJ...897..122L}, however the two-dimensional fitting procedure used was optimised for the disc regions, and is likely to be less robust in the galaxy centres we probe here. Full three-dimensional modelling of the CO data-cube is typically required to derive accurate rotation curves for bulge regions.}, and obtaining them is beyond the scope of this work. However, in galaxy centres the gravitational potential is likely dominated by the potential of the stars. In this section, we thus compare our non-parametric morphology measurements with stellar properties that should act as proxies for the potential well shapes and depths.

In Figure \ref{mu_a_s} we plot the Asymmetry (top panels), Smoothness (middle panels) and Gini (bottom panels) parameters as a function of the total stellar mass of each galaxy (left column), the stellar velocity dispersion ($\sigma_*$; central column), and of the effective stellar mass surface density ($\mu_*$; right column) of each galaxy. Symbols are as in Figure \ref{cas}.
Negative correlations are seen between our non-parametric morphology measures and all of these quantities. The Spearman's rank correlation coefficients (listed in Table \ref{corr_table}) show that these correlations are all significant, with the {stellar mass and} effective stellar mass surface density correlations being the strongest. 

It is not possible to probe correlations with stellar mass for the simulated galaxies (shown as green stars in Fig. \ref{mu_a_s} ), as they all have the same total stellar mass, but the simulated galaxies do show very similar strongly decreasing trends with $\sigma_*$ and $\mu_*$ (albeit these systems are somewhat offset  in central surface density compared to the observed galaxies due to the simulation setup; see Section \ref{simdetails}). This suggests that the shape of the gravitational potential may matter, as objects with higher stellar velocity dispersions and stellar mass surface densities have smoother, and more symmetric molecular ISM morphologies.  
While all these parameters are closely linked (e.g. galaxies with dominant bulges generally have large M$_*$, high $\sigma_*$, high $\mu_*$ and are typically classified as ETGs), these results suggest that the mass distribution of each galaxy is a factor that influences the structure of the ISM.

\begin{figure*}
\includegraphics[height=12cm,trim=0.5cm 0cm 0.5cm 0cm,clip]{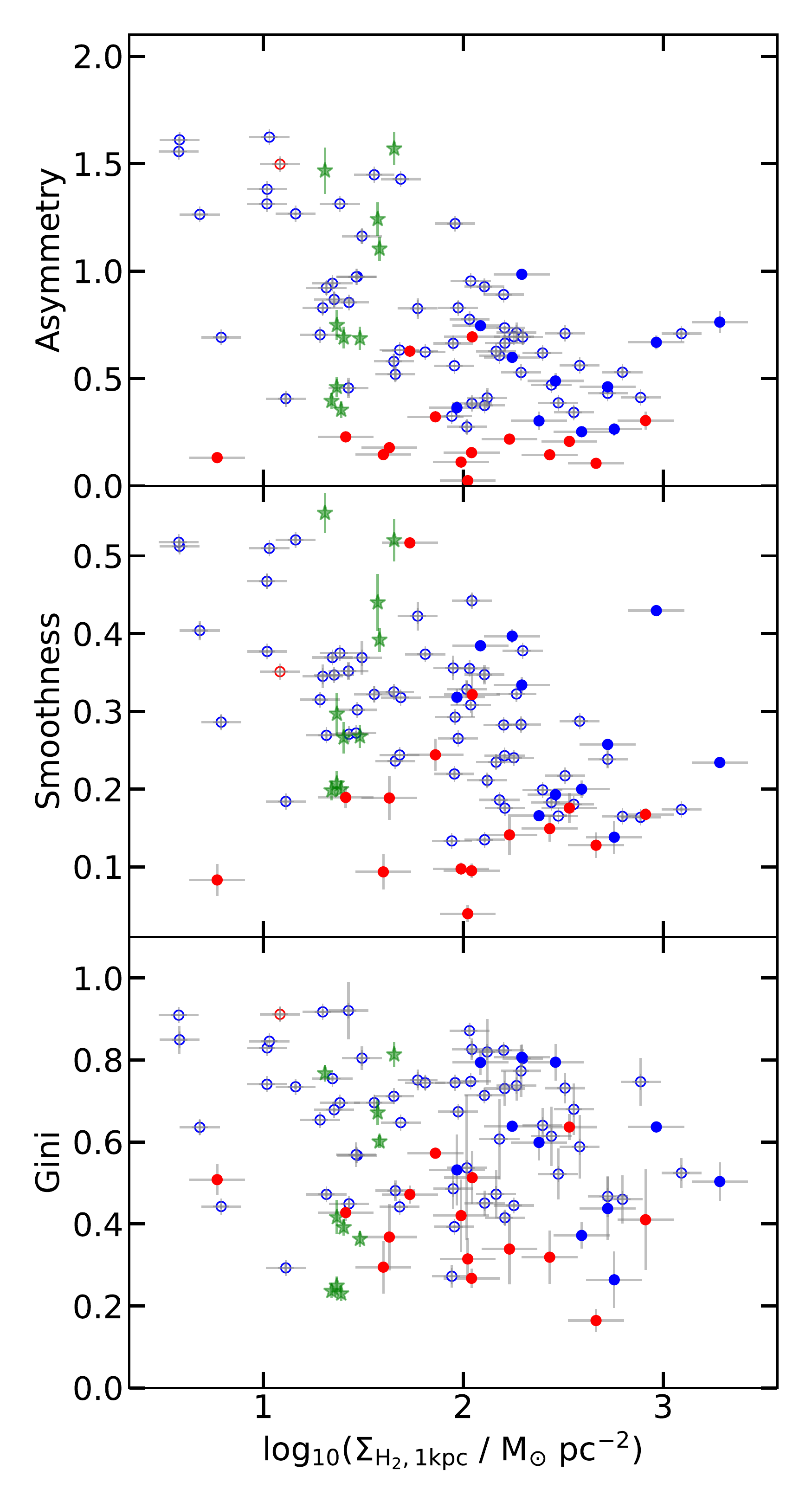}
\includegraphics[height=12cm,trim=2.7cm 0cm 0.5cm 0cm,clip]{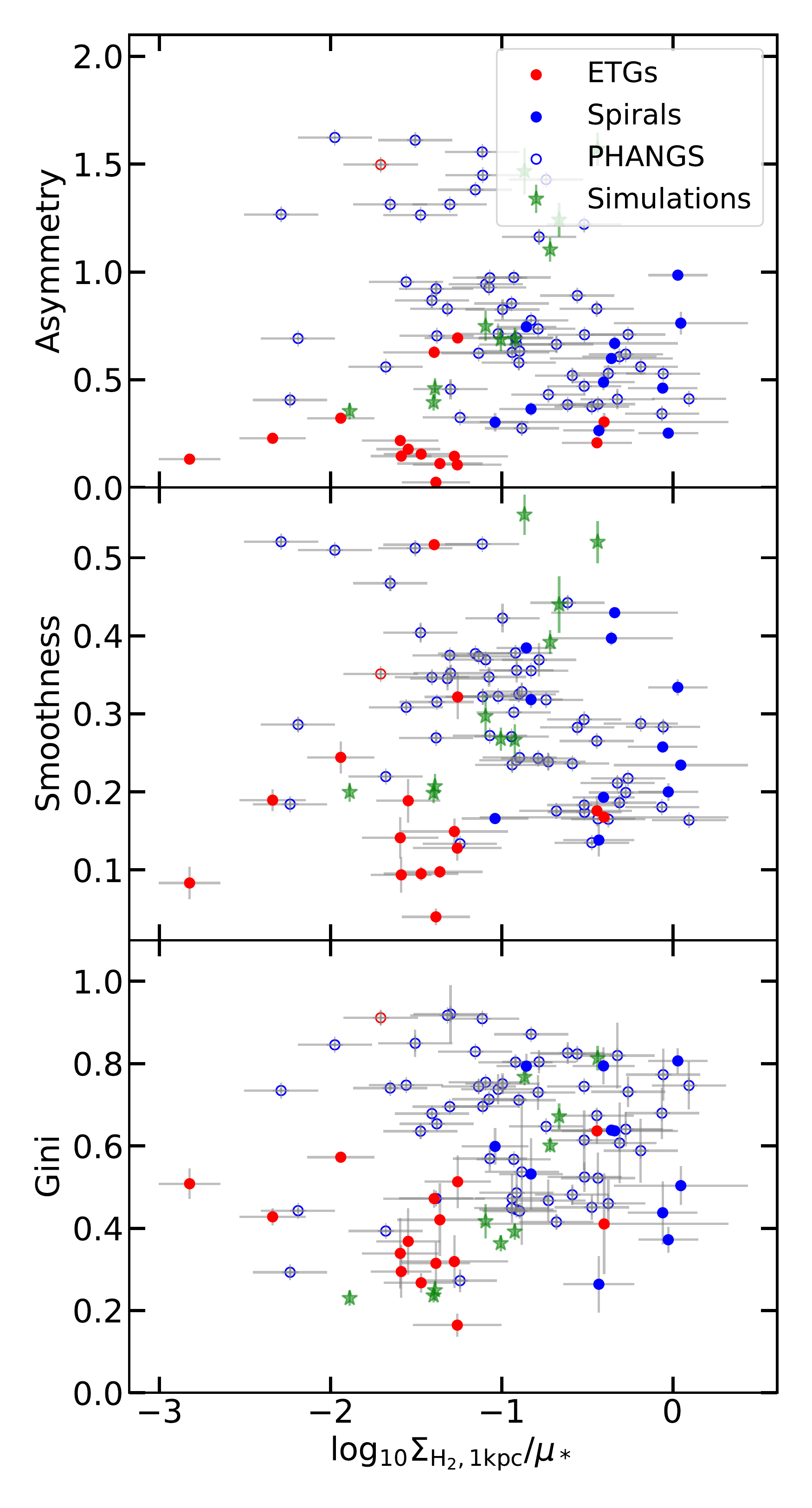}
\includegraphics[height=12cm,trim=2.7cm 0cm 0.5cm 0cm,clip]{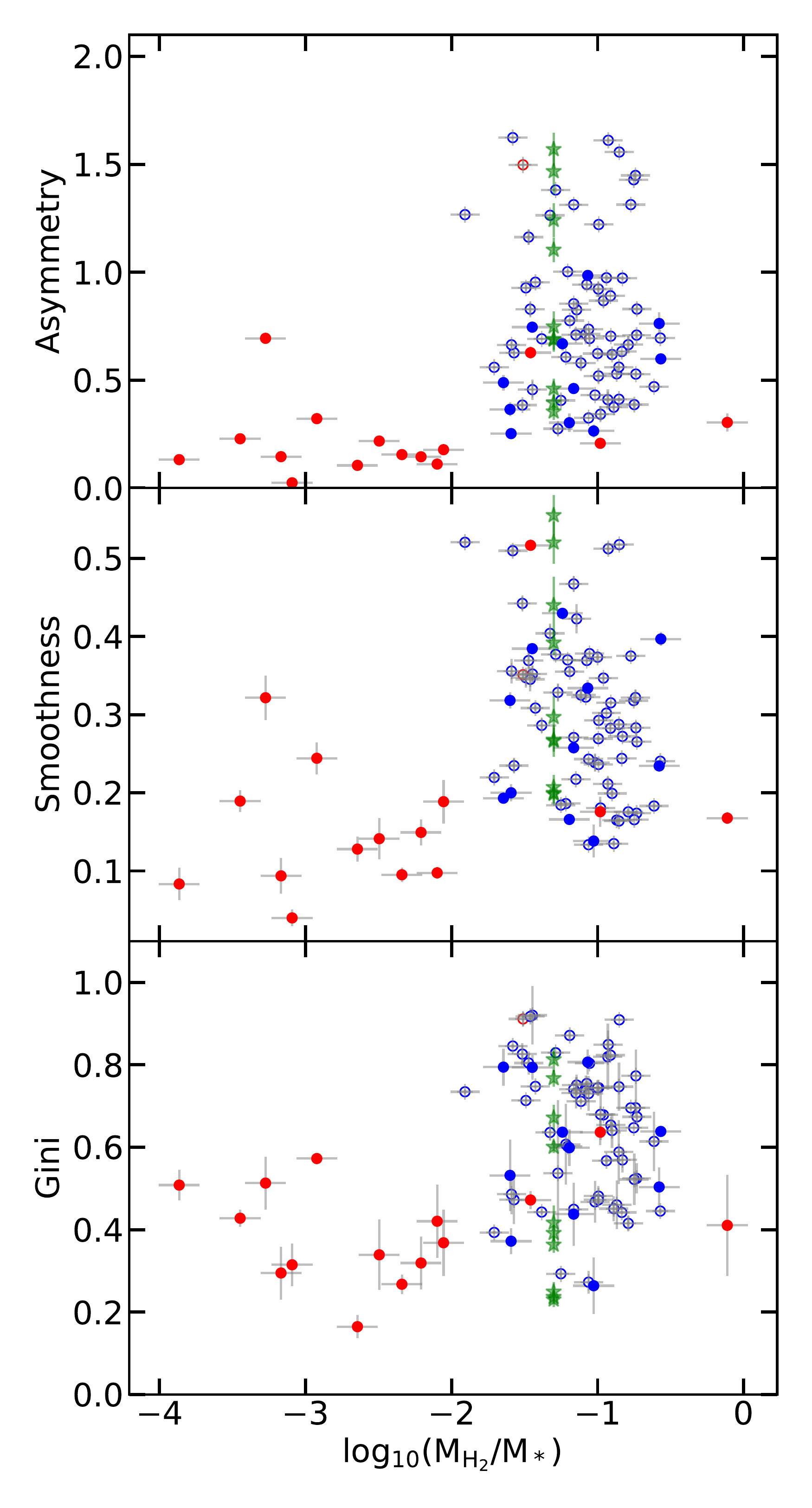}
\caption{As Figure \ref{mu_a_s}, but plotting the non-parametric morphology measurements against the mean molecular gas surface density within the inner kiloparsec of each galaxy (${\Sigma_{\rm H_2,1kpc}}$; left column), the resolved gas fraction (${\Sigma_{\rm H_2,1kpc}}$ divided by $\mu_*$; central column), and the total gas fraction (right column). A negative correlation is seen between molecular gas surface density and gas morphology for the observed galaxies. No strong correlations are found between our non-parametric morphology measurements and the resolved or total gas fractions (once the obvious correlation between total gas fraction and morphological type is taken into account). } 
\label{surfdens_a_s}
\label{resolvedgasfrac_a_s}
\end{figure*}

\subsubsection{Central molecular gas mass surface density}
\label{surfdens}
 Another explanation of our results would be if the molecular ISM of ETGs was lower surface density, reducing the impact of self-gravity, and thus suppressing fragmentation. In this case, all else being equal, we would expect a positive correlation between our ISM morphology measures and the molecular gas mass surface density. We plot the Asymmetry, Smoothness and Gini parameters as a function of the mean molecular gas mass surface density of our galaxies in the left column of Figure \ref{surfdens_a_s}. This molecular gas surface density is calculated within an elliptical aperture 1\,kpc in radius. We choose this radius as molecular gas is not detected beyond this in some of the WISDOM ETGs, and it ensures the measurements are comparable between galaxies. We assume that the surface density of the gas in non-detected regions is zero, and include this in our averages. Calculating the mean surface density of only the detected material instead makes very little difference (increasing the calculated surface densities by less than a factor of two on average).  
 
  Contrary to the naive expectation above, we find a weak \textit{negative} correlation between the non-parametric morphology measures and the central molecular gas surface density, which would suggest as the molecular medium becomes denser it fragments \textit{less}. This implies that the gas mass surface density is not the dominant factor determining the ability of the molecular ISM to fragment in these galaxy centres.  
  This result will be discussed further in Section \ref{discuss_driver}.
 
We note that the mass surface densities measured here are for molecular gas only, and rely intimately on our assumptions for the CO-to-H$_2$ conversion factor. While CO-to-H$_2$ conversion factor variations are almost certainly present within (and between) our sample objects, it seems unlikely they can drive the trend seen in Figure \ref{surfdens_a_s}, which extends over four orders of magnitude in mean molecular gas surface density.  Atomic gas is likely present in these regions too, meaning our gas mass surface density estimates are formally lower limits. In the centres of massive galaxies molecular gas typically dominates over atomic gas, as the later saturates at mass surface densities of $\approx$10 \msun\ pc$^{-2}$ \citep{2008AJ....136.2846B}.  For the majority of sample galaxies, which have mean central molecular gas mass surface densities at least an order of magnitude above this, we thus do not expect the inclusion of \hi\ to change our results.

{Due to the initial conditions of the simulations (see Section \ref{simdetails}), the simulated galaxies (again shown as green stars) have significantly lower central molecular gas mass surface densities than the bulk of our observed systems. No obvious trend is present between the ISM surface density of the simulated galaxies and all of our quantitative morphology measures (Asymmetry, Smoothness and Gini), although further simulations probing a larger range of surface density would be required to test if this result is fully general.}

\subsubsection{Gas-to-stellar mass fraction}
\label{gas-to-stars_frac}
Given that we have shown above that the effective stellar mass surface density is important in setting the structure of the molecular ISM in galaxy centres, it is also natural to consider the impact of the gas mass fraction. Simulations which track the impact of large bulges on star formation have revealed that the shape of the galaxy potential matters mainly when the global gas fraction is low (see e.g. \citealt{2009ApJ...707..250M,2013MNRAS.432.1914M,2021MNRAS.500.2000G}). In objects with high gas fractions, self gravity overcomes any stabilising effect of the potential and star formation proceeds normally. Only in low gas fraction objects can the more subtle effects caused by the galaxy's potential be measured. 

We therefore construct two different measurements that allow us to estimate the central and global gas fraction within our galaxies. In the central panel of Figure \ref{resolvedgasfrac_a_s} we plot the Asymmetry, Smoothness and Gini parameters versus the ratio of the mean gas surface density (measured within a radius of 1\,kpc, as above) to the mean effective stellar mass surface density.  While this measurement is not a true reflection of the resolved gas fraction at any specific location in our galaxy (which would require resolved stellar mass measurements which are beyond the scope of this work), it should provide some indication of the relative importance of the gas and stellar components. 
 In addition, in the right panel of Figure \ref{resolvedgasfrac_a_s} we show the same coefficients versus the total molecular gas mass divided by the total stellar mass (M$_{\rm H_2}$/M$_*$).

The global gas mass fraction measure (M$_{\rm H2}$/M$_*$) is weakly correlated {with the Asymmetry morphology measurement}, but this seems entirely driven by changes in the gas fraction due to morphology, rather than reflecting the change in the ISM properties. When considering spiral systems alone no trend is seen. Given this, we conclude that the gas-to-stellar mass fraction is unlikely to be an important parameter driving the morphology of the ISM in these galaxy centres. This conclusion is supported by the simulations, where each simulated galaxy has the same global gas fraction but can show a vastly different ISM morphology. This suggests gas mass fraction alone cannot be the physical mechanism driving the correlations we observe.

\begin{figure*}
\includegraphics[height=12cm,trim=0.5cm 0cm 0.5cm 0cm,clip]{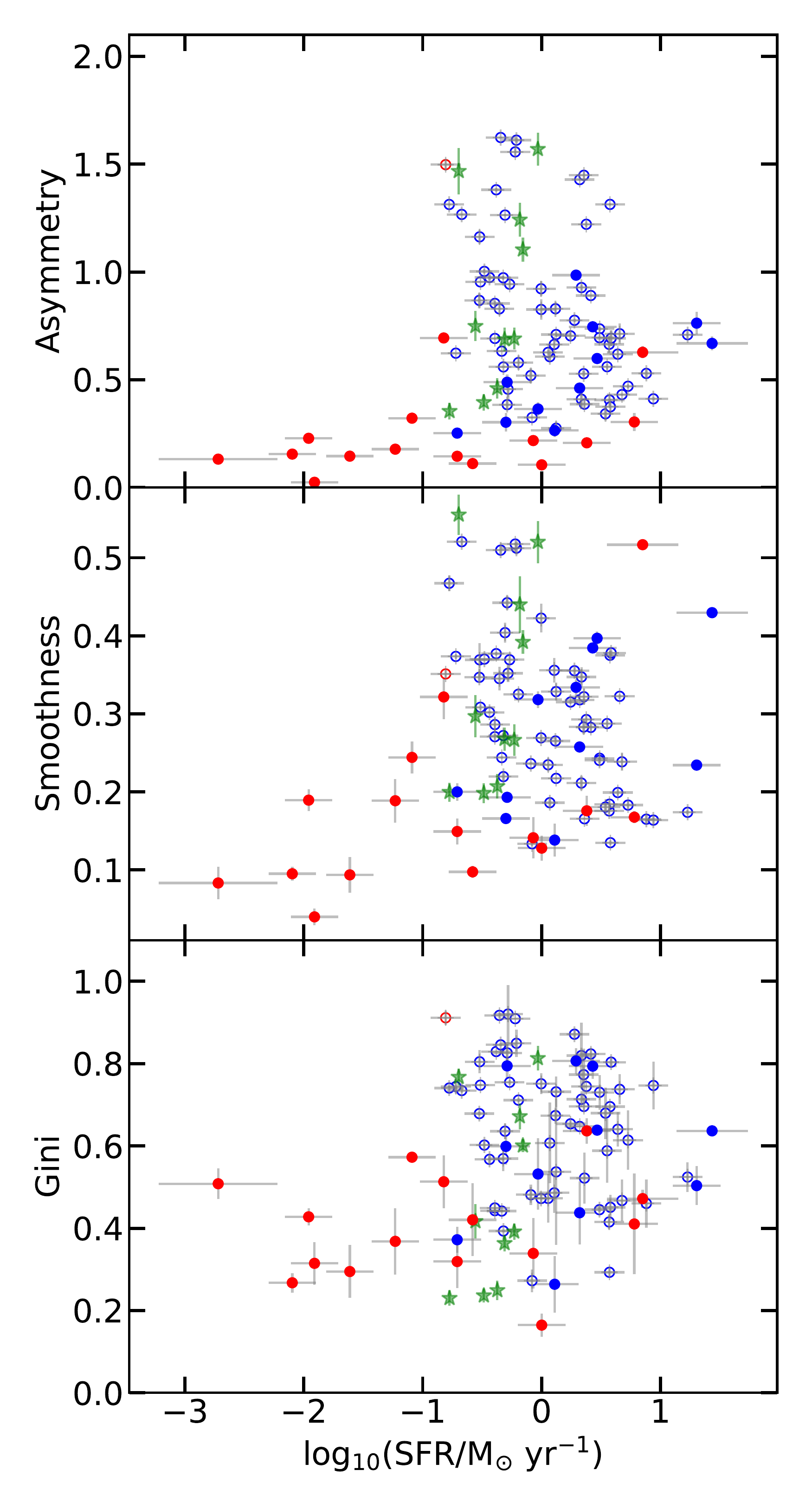}
\includegraphics[height=12cm,trim=2.7cm 0cm 0.5cm 0cm,clip]{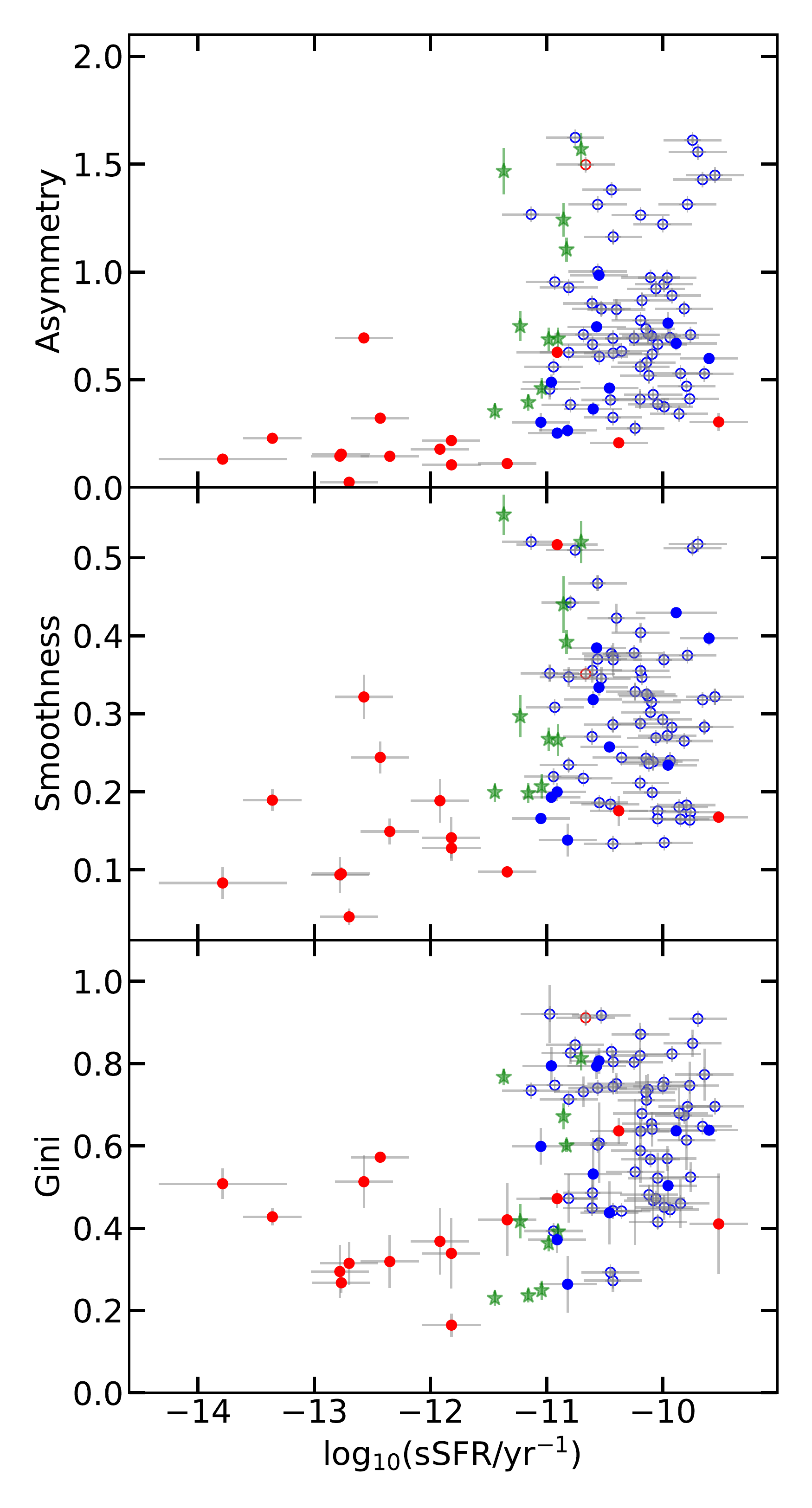}
\includegraphics[height=12cm,trim=2.7cm 0cm 0.5cm 0cm,clip]{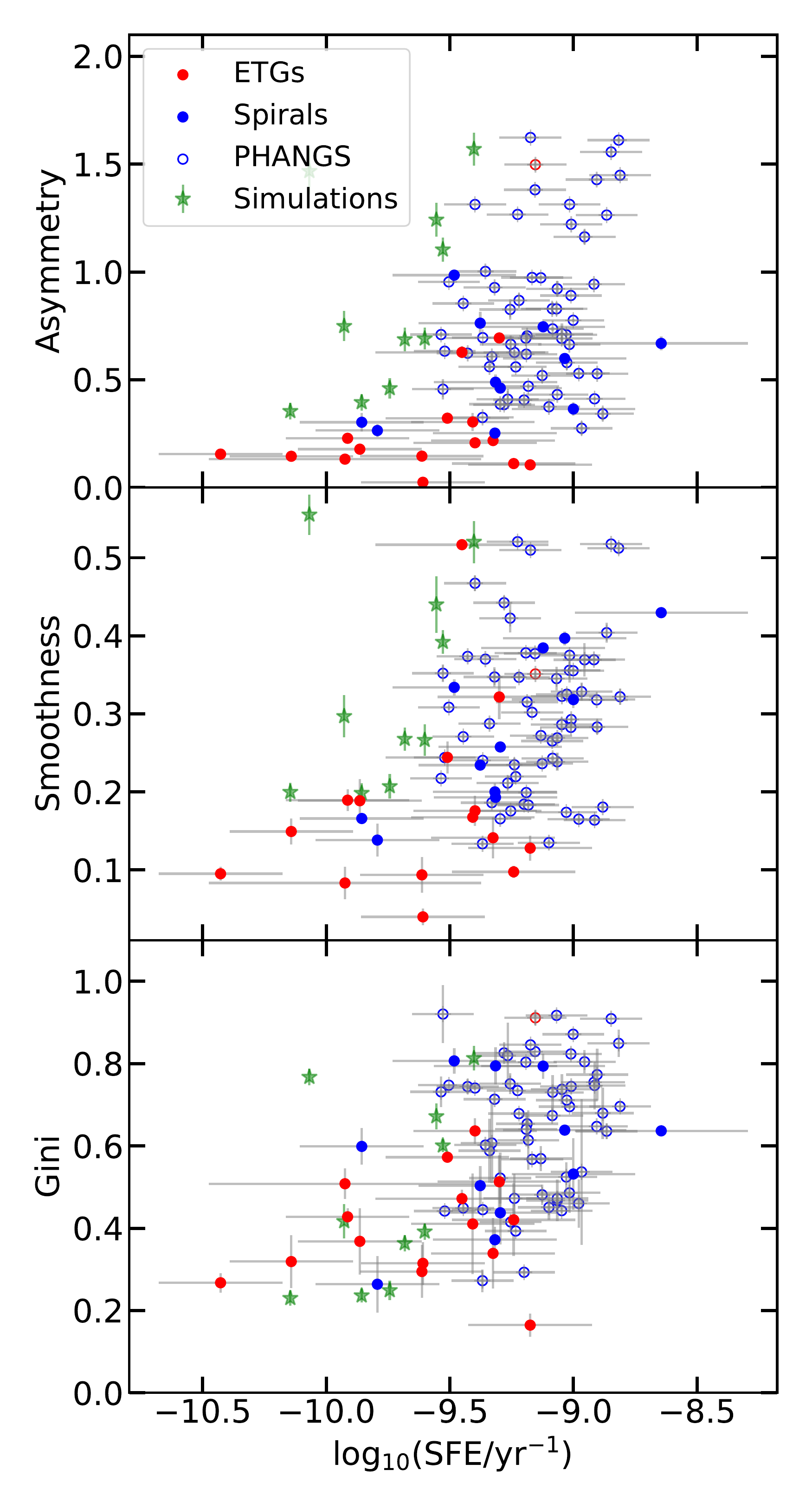}
\caption{As Figure \ref{mu_a_s}, but plotting the non-parametric morphology measurements against the total star formation rate (left panel), the specific star formation rate (sSFR; SFR/M$_*$; central column), and the star formation efficiency (SFE: SFR/M$_{H_2}$; right column) of each galaxy. The observational data and the simulations show correlations between all the morphology measurements and these star formation related parameters. While the SFR and sSFR trends may be driven by morphology, the SFE trend is robust to the removal of ETGs. This suggests that star formation feedback and/or supernovae may play a role in shaping the morphology of the ISM. } 
\label{ssfr_a_s}
\end{figure*}

\subsubsection{Star formation}
\label{sf_relations}
Another process that can change the morphology of the ISM is turbulence. While the sources of turbulence in the molecular ISM are still debated, it is clear that feedback from recently-formed stars is an important contributor \citep[e.g.][]{2004RvMP...76..125M,2012A&ARv..20...55H}. Direct, homogeneous measurements of the molecular gas turbulence of our sample galaxies unfortunately are not available, and obtaining them is beyond the scope of this work. However, radiation, winds and supernova explosions can act to heat, expel and/or accumulate the ISM in certain regions. As such, we expect that as star formation inputs more feedback energy in a given galaxy this will impact the turbulence, and perhaps thus the morphology of the cold ISM.

In Figure \ref{ssfr_a_s} we plot the Asymmetry, Smoothness and Gini parameters versus the global star formation rate (left column), specific star formation rate (sSFR$\equiv$SFR/M$_{\rm *}$; central column) and star formation efficiency (SFE$\equiv$SFR/M$_{\rm H_2}$; right column) of each galaxy. All panels of Figure \ref{ssfr_a_s} reveal positive correlations, and those with SFE are significant for all indicators (see Table \ref{corr_table} for Spearman's rank correlation coefficients). 
However, once again if one removes the ETGs from the analysis, the trend in sSFR is removed completely (as is the weak correlation with the SFR itself). It thus seems unlikely that the absolute amount of star formation, or the ratio of the energy input from feedback to the stellar mass is crucial in setting the morphology of the ISM. The trend with SFE does remain when considering only the spirals, however. This suggests that the amount of feedback energy per unit gas mass does play a role in shaping the morphology of the ISM.

However, it should be noted that the SFE is expected to be correlated with the other parameters in our analysis. 
{A variety of observational and simulation works have suggested that the presence of a deep potential well could directly reduce star formation efficiencies \citep[due to the increased Toomre- or shear-stability of gas in such a situation e.g.][]{2009ApJ...707..250M,2011MNRAS.415...61S,2014MNRAS.444.3427D,2020MNRAS.495..199G,2021MNRAS.500.2000G}.}
Further investigation is thus required to determine if this correlation reflects causation.

\begin{table}
\caption{Correlation measurements for the observed galaxies}
\begin{center}
\begin{tabular*}{0.48\textwidth}{@{\extracolsep{\fill}} l r l c}
\hline
Correlation & $\rho$ & $p$ & $p$$<$0.05 \\
 (1) & (2) &(3)&(4) \\
\hline 
Asymmetry vs. Smoothness & 0.72 & \num{2e-15} & \checkmark \\
Gini vs. Smoothness & 0.68 & \num{3e-13}& \checkmark\\
Gini vs. Asymmetry & 0.61 & \num{1e-10}& \checkmark\\
\hline
log$_{10}$(M$_*$/M$_{\odot}$) vs Asymmetry &  -0.54  & \num{1e-07} & \checkmark\\
log$_{10}$(M$_*$/M$_{\odot}$) vs Smoothness &  -0.42  & \num{7e-05} & \checkmark\\
log$_{10}$(M$_*$/M$_{\odot}$) vs Gini &  -0.38  & \num{0.0003} & \checkmark\\
\hline
log$_{10}$($\sigma_*$/km s$^{-1}$) vs Asymmetry &  -0.49  & \num{2e-06} & \checkmark\\
log$_{10}$($\sigma_*$/km s$^{-1}$) vs Smoothness &  -0.38  & \num{0.0003} & \checkmark\\
log$_{10}$($\sigma_*$/km s$^{-1}$) vs. Gini &  -0.25  & \num{0.04} & \checkmark\\
\hline
log$_{10}$($\mu_*$/M$_{\odot}$ kpc$^{-2}$) vs Asymmetry &  -0.52  & \num{4e-07} & \checkmark\\
log$_{10}$($\mu_*$/M$_{\odot}$ kpc$^{-2}$) vs Smoothness &  -0.43  & \num{3e-05} & \checkmark\\
log$_{10}$($\mu_*$/M$_{\odot}$ kpc$^{-2}$) vs Gini &  -0.49  & \num{2e-06} & \checkmark\\
\hline
log$_{10}$(${\Sigma_{\rm H_2,1kpc}}$ / M$_{\odot}$ pc$^{-2}$) vs Asymmetry &  -0.41  & \num{0.0001} & \checkmark\\
log$_{10}$(${\Sigma_{\rm H_2,1kpc}}$ / M$_{\odot}$ pc$^{-2}$) vs Smoothness &  -0.43  & \num{4e-05} & \checkmark\\
log$_{10}$(${\Sigma_{\rm H_2,1kpc}}$ / M$_{\odot}$ pc$^{-2}$) vs Gini &  -0.22  & \num{0.04} & \checkmark\\
\hline
log$_{10}$${\Sigma_{\rm H_2,1kpc}}$/$\mu_*$ vs Asymmetry &  -0.07  & \num{0.5} & \text{\sffamily X}\\
log$_{10}$${\Sigma_{\rm H_2,1kpc}}$/$\mu_*$ vs Smoothness &  -0.13  & \num{0.2} & \text{\sffamily X}\\
log$_{10}$${\Sigma_{\rm H_2,1kpc}}$/$\mu_*$ vs Gini &  0.11  & \num{0.3} & \text{\sffamily X}\\
\hline
log$_{10}$(M$_{\rm H_2}$/M$_*$) vs Asymmetry &  0.31  & \num{0.004} & \checkmark\\
log$_{10}$(M$_{\rm H_2}$/M$_*$) vs Smoothness &  0.03  & \num{0.8} & \text{\sffamily X}\\
log$_{10}$(M$_{\rm H_2}$/M$_*$) vs Gini &  0.17  & \num{0.1} & \text{\sffamily X}\\
\hline
log$_{10}$(SFR/M$_{\odot}$ yr$^{-1}$) vs Asymmetry &  0.05  & \num{0.6} & \text{\sffamily X}\\
log$_{10}$(SFR/M$_{\odot}$ yr$^{-1}$) vs Smoothness &  -0.03  & \num{0.8} & \text{\sffamily X}\\
log$_{10}$(SFR/M$_{\odot}$ yr$^{-1}$) vs Gini &  0.08  & \num{0.5} & \text{\sffamily X}\\
\hline
log$_{10}$(sSFR/yr$^{-1}$) vs Asymmetry &  0.40  & \num{0.0002} & \checkmark\\
log$_{10}$(sSFR/yr$^{-1}$) vs Smoothness &  0.18  & \num{0.1} & \text{\sffamily X}\\
log$_{10}$(sSFR/yr$^{-1}$) vs Gini &  0.27  & \num{0.01} & \checkmark\\
\hline
log$_{10}$(SFE/yr$^{-1}$) vs Asymmetry &  0.44  & \num{2e-05} & \checkmark\\
log$_{10}$(SFE/yr$^{-1}$) vs Smoothness &  0.39  & \num{0.0002} & \checkmark\\
log$_{10}$(SFE/yr$^{-1}$) vs Gini &  0.40  & \num{0.0001} & \checkmark\\
\hline\end{tabular*}\vspace{0.01cm}
\parbox[t]{0.48\textwidth}{ \textit{Notes:} Column 1 lists the correlation variables, while column 2 and 3 contain Spearman's-rank correlation coefficients ($\rho$) and their associated $p$-values. Column 4 acts as a guide, highlighting significant correlations ($p<0.05$).}
\end{center}
\label{corr_table}
\end{table}

\subsubsection{Bars}
\label{bars}
\begin{figure*}
\includegraphics[width=0.9\textwidth,trim=0.0cm 0cm 0.0cm 0cm,clip]{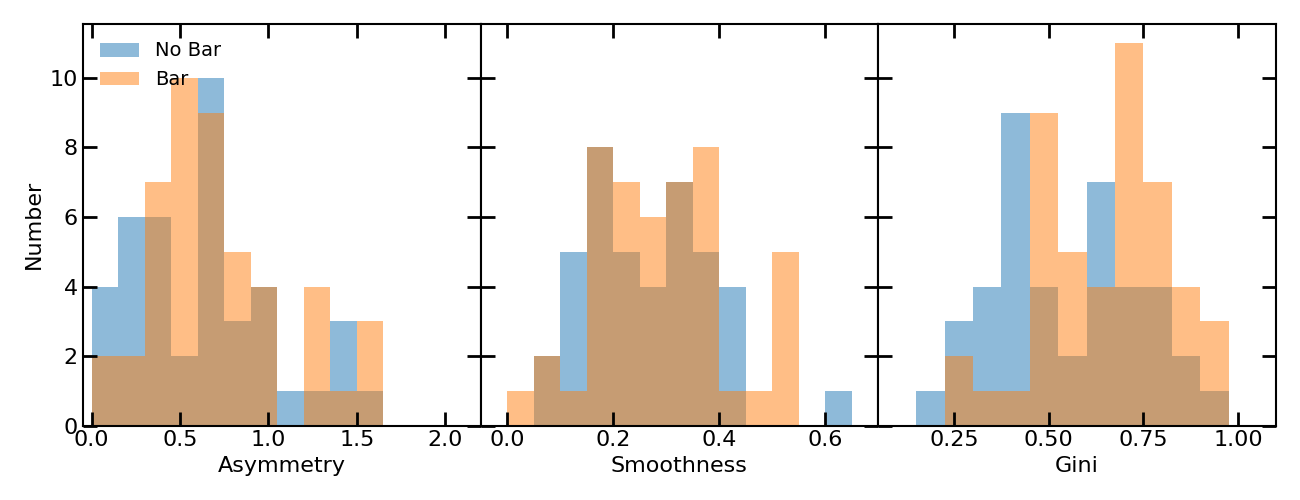}
\caption{Histograms showing the distribution of Asymmetry (left panel), Smoothness (middle panel) and Gini (right panel) statistics for galaxies classified by eye as barred (orange histogram) and non-barred (blue histogram). The molecular ISM in barred galaxies has similar Asymmetry and Smoothness, but the Gini parameter is significantly different than that found for non-barred systems.} 
\label{cas_bars}
\end{figure*}

Bars (and other non-axisymmetric perturbations to the potentials of galaxies) are known to affect the molecular gas, causing resonance rings, gas flows, etc. Bars are thus expected to be important in shaping the morphology of the ISM. Given that bars are present in a significant number of the spiral and ETGs in our sample, it is important to consider the effects they may have.  In Figure \ref{cas_bars} we show histograms of the Asymmetry (left panel), Smoothness (middle panel) and Gini (right panel) coefficients for galaxies classified by eye as barred (orange histogram) or non-barred (blue histogram) using HST (mostly near-infrared) images.

Kolmogorov-Smirnov (KS) tests doe not reveal any evidence of a difference in the Asymmetry and Smoothness values of barred and non-barred galaxies (KS statistic distance ($D$) of 0.25 and 0.14, and probability ($p$) of 0.09 and 0.75 respectively). The distribution of the Gini estimator does show significant difference between barred and unbarred galaxies: suggesting that the central ISM in barred galaxies is less smooth ($D=0.32$, $p=$\num{0.01}). 
While at first glance the lack of a strong correlation with Asymmetry seems unexpected, bars do have bifold symmetry, and as our Asymmetry measure only rotates 180 degrees around the galaxy centre it would not be sensitive to the higher order asymmetries bars may induce.

Within our sample bars are more prevalent in lower effective stellar surface density galaxies ($D$=0.36, $p$=0.01; a reflection of the fact that bars are more common in dynamically cold disks), but they are not more common in any specific galaxy type, stellar mass, gas surface density or star formation (SFR/sSFR/SFE) parameter range (all show $p>$0.05). Given this, and the fact that all the correlations reported in this work are present in Asymmetry/Smoothness as well as Gini, it is likely that bars do not drive the observed trends, but rather add scatter to them.

\subsubsection{AGN}

AGN are able to output significant amounts of energy into their surroundings, and are thought to be important in regulating star formation in galaxies over cosmic time \citep[e.g.][]{2006ApJS..163....1H}. The PHANGS galaxies have AGN classifications from \cite{2010A&A...518A..10V,2021A&A...653A.172S}, and we checked if these known AGN show different ISM morphologies. 
{Fourteen (24\%) of the PHANGS systems included here are are identified as having AGN in \cite{2010A&A...518A..10V}, four are classed as Seyfert 1, 6 are Seyfert 2, and 4 are LINERS (two of which also show broad Balmer lines). No AGN are identified in the lowest mass galaxies (M$_*$<10$^{10}$ \msun), but they are present in systems covering the full range of $\mu_*$ and SFR.} 

No significant differences in the non-parametric gas morphology coefficients were found between galaxies with and without AGN. {We also checked if any differences were seen between AGN of different classes. No differences were found, however,  due to the low number of systems with AGN these tests should be treated with caution.
We conclude that, at least in the spiral galaxies probed by PHANGS, optically classified AGN are not strongly perturbing the morphology of the molecular ISM.}
 Given that these are local galaxies, not e.g. strong quasars, this is perhaps unsurprising. {Extending this comparison to ETGs, and to objects with more powerful AGN would be a productive line of future enquiry.}

\subsection{Selection effects and uncertanties}

\subsubsection{Sample selection}
\label{select_effect}
As discussed in Section \ref{wisdomdata} it is possible that some of the results of this paper are affected by selection effects. 
The WISDOM spiral and ETG samples were both selected to look as regular as possible in optical imaging. It is possible that this may bias their molecular gas morphologies. A relaxed optical morphology is no guarantee of a relaxed molecular gas morphology (see e.g. NGC3607, NGC4826 in Figures \ref{mom0s_etgs} and \ref{mom0s_ltgs}), but we still must consider this carefully.

Comparing the WISDOM spirals with those from the PHANGS survey shows that on average they do tend to have more regular ISM morphologies. However the WISDOM spiral galaxies do not stand out in our various property plots, lying at the same locations as some of the PHANGS objects, and following the same trends with galaxy parameters. The only exception to this is in mean central molecular gas surface density (${\Sigma_{\rm H_2,1kpc}}$) and resolved central gas fraction (${\Sigma_{\rm H_2,1kpc}}$/$\mu_*$). In these cases the WISDOM spirals tend to have larger molecular gas surface densities (and local molecular gas fractions) than the PHANGS systems (see Figure \ref{surfdens_a_s}), likely because this made them good targets for the survey (see Section \ref{selection_wisdom}). Taking WISDOM alone would suggest that these surface density parameters positively correlate with our non-parametric morphology indicators, while an anti-correlation is seen in the PHANGS sample. This could be due to selection effects, and should be treated with caution. 

While the above analysis gives us confidence in our results for spiral galaxies, the PHANGS selection criteria lead to only a single early-type galaxy being included in their first data release primary sample. WISDOM dominates the ETG sample used here, thus the impact of selection effects may be larger in this population. 
{The fraction of disturbed molecular gas at kiloparsec scales in complete surveys of ETGs is low ($<$20 per cent; \citealt{2013MNRAS.432.1796A}), and initial analysis of a complete volume limited sample of high-mass ETGs from the MASSIVE survey \citep{2014ApJ...795..158M} suggests that this fraction remains low when observing ETGs with higher spatial resolution (Davis et al. in prep). 
Given that the type of regular ETGs selected by WISDOM dominate the ETG population, we thus do not expect the addition of a small number of disturbed systems to significantly affect our results. 
Despite this, it is clear that revisiting this issue with larger, homogeneously selected samples of ETGs would be beneficial to confirm this. }

\subsubsection{Galaxy inclination}
The molecular discs of the galaxies observed by WISDOM (and to a lesser extent PHANGS) have a large range of inclinations. This could, in principle, affect the non-parametric morphology indicators we measure. We tested how much this could affect our results both by checking for residual correlations with inclination (see Section \ref{discuss_driver}), and by repeating our entire analysis with all systems at $i>60^{\circ}$ removed. {No residual correlations were identified in Section \ref{discuss_driver}, suggesting the inclination of our galaxies is not strongly affecting our measurements of non-parametric morphology indicators.}

All of the {strong} correlations between galaxy properties and non-parametric morphology indicators remain when only considering the more face-on systems. {The weak correlations between sSFR and Asymmetry/Gini, and the global gas fraction and Asymmetry are not significant once the inclined systems are removed. However, as we already suspected these correlations were not reflecting a physical driver this does not change our results.}  

{Even imposing an extreme inclination cut by considering galaxies with $i<45^{\circ}$ does not change our key results (it only reduces the significance of the correlations with SFE above $p=0.05$).} 
As such we conclude that inclination uncertainties are not unduly affecting our results.

\subsubsection{Faint sightlines}

In this work we have measured the morphology of all the material robustly detected by ALMA in each datacube. However varying sensitivity limits could potential affect the non-parametric morphology indicators, and thus our conclusions. To test this we repeated the analysis described above after removing the faintest 30\% of sightlines for each object. This did not change any of our conclusions.

\begin{figure*}
\includegraphics[width=\textwidth,trim=0.0cm 0cm 0.0cm 0cm,clip]{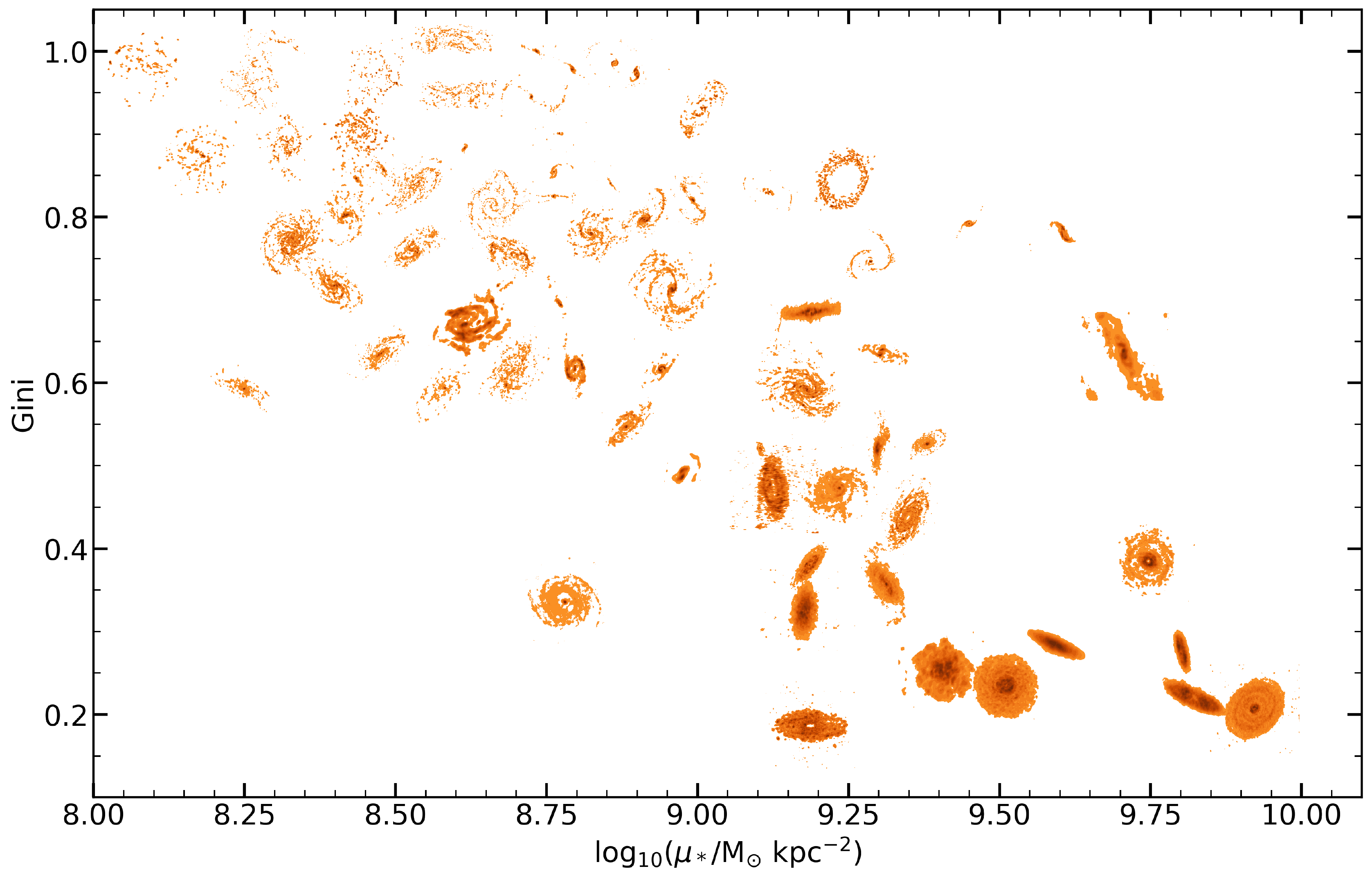}
\caption{Integrated intensity maps of 62 of the sample galaxies, plotted at their approximate location in the $\mu_*$-Gini diagram (see Figure \ref{mu_a_s}). The exact position of each galaxy has been allowed to vary by $\approx$5\% for display purposes. Each galaxy has been scaled to the same size, and the gas distribution is traced by 10 contours starting at 5\% of the peak integrated intensity of each map. This figure demonstrates visually the changes we see in the gas distributions of our galaxies across this space, from flocculent systems in the top left to smooth discs in the lower-right. }
\label{mu_gini_imageplot}
\end{figure*}

\section{Discussion}
\label{discuss}

We have shown above that the morphology of the ISM in the centres of nearby galaxies changes systematically as a function of various large scale galaxy parameters, including galaxy morphological type, and the depth of the potential well (as traced by the stellar mass, stellar velocity dispersion and effective stellar mass surface density). 
 The morphology of the molecular ISM also shows weak correlations with its surface density, although the sense of these correlations suggests that self-gravity itself is not the important driver. 
There are also significant correlations between the star formation efficiency and ISM morphology. Bars have an effect, likely because they concentrate material into rings etc, but do not seem to drive the major trends observed. 

A visual representation of these results can be seen in Figure \ref{mu_gini_imageplot}, where we show integrated intensity images of the gas discs in 62 of our sample galaxies, 
plotted here at their approximate location in the $\mu_*$-Gini diagram. The changes in the gas distributions of our galaxies across this space are clearly visible. 
In this section we discuss these results, attempting to determine what drives the observed correlations, and compare to the literature.

\subsection{What sets the morphology of the ISM?}
\label{discuss_driver}
As discussed above, the morphology of the ISM is expected to be set through a competition between attractive (e.g. gravity) and destructive forces (e.g. internal kinetic energy, the external gravitational field, feedback). Given the correlations described above it seems that, at least in galaxy centres, these destructive processes are dominant in setting the structure of the ISM.

However, determining the relative importance of these processes in setting the ISM structure is a difficult task. Effective stellar surface density and galaxy morphological type are closely linked (e.g. galaxies with large bulges will have high $\mu_*$), and galaxies with larger bulges typically have lower star formation rates \citep[e.g.][]{2013MNRAS.432.1862C}. In addition, deep potential wells have been suggested to directly reduce star formation efficiencies \citep[e.g.][]{2009ApJ...707..250M,2011MNRAS.415...61S,2014MNRAS.444.3427D,2020MNRAS.495..199G,2021MNRAS.500.2000G}. Thus these processes could be directly connected. 
Indeed, in the idealised simulations of \cite{2020MNRAS.495..199G} we find qualitatively similar trends in ISM morphology when only the depth of the potential well of each galaxy is varied. 
Below we outline three broad scenarios which could be at play in shaping the morphology of the ISM in our galaxy centres. 
\begin{enumerate}
\item The gravitational potential dominates: The deep potential well affects the ISM morphology directly, suppressing fragmentation. This smoother, less fragmented ISM is less efficient at forming stars, creating the correlations with SFE.
\item The gravitational potential and turbulence driven by star formation are important: As (i), but the feedback from the stars that do form drives turbulence which also plays an important role in regulating the state of the ISM.
\item Star formation feedback dominates: Galaxies with high star formation efficiencies experience more feedback per unit gas mass, driving turbulence and disrupting/blowing holes in the ISM, and effects caused by the gravitational potential are unimportant. 
\end{enumerate}

In the next section we attempt to determine which, if any, of these scenarios is driving the observed correlations between the ISM morphology and galaxy properties.

\subsubsection{Optimal predictors}

In order to attempt to distinguish the primary drivers of the observed correlations, and their relative importance, we can make use of statistical tools to determine which combination of parameters provides an optimal predictor of non-parametric morphology. 

We begin by using the linear regressor in \textsc{scikit-learn} \citep{scikit-learn} to fit linear relations between our non-parametric morphology measures and observations quantities. 
We include the following observed parameters, which we believe could potentially drive the observed correlations (either physically, or via observational effects): stellar mass, effective radius, inclination, SFR, sSFR, SFE, ${\Sigma_{\rm H_2,1kpc}}$, M$_{H2}$, $\mu_*$, ${\Sigma_{\rm H_2,1kpc}}$/$\mu_*$, and the beam size in parsecs and arcseconds. We then quantified the root-mean squared (RMS) scatter around each relation, to determine which quantities can be used to predict the morphology of the ISM most accurately. For the asymmetry and smoothness we found that $\mu_*$ and ${\Sigma_{\rm H_2,1kpc}}$ are the quantities that best predict the ISM morphology, resulting in almost identical RMS scatters. For the Gini coefficient $\mu_*$ alone was the best predictor.

We then attempt to extend this by using the multivariate linear regressor (again from \textsc{scikit-learn}) which finds the $n$-dimensional hyperplane that best fits our observed data using an ordinary least-squares regression. We describe here regressing the parameters described above against the Asymmetry parameter (as it shows the strongest correlations in many of the figures above and is less affected by bars). The best-fitting relation using all 12 of these parameters can predict the Asymmetry with an RMS scatter of 0.24. 

In order to determine which parameters provide the most diagnostic power we make use of `Sequential Feature Selection' \citep{FERRI1994403} within \textsc{scikit-learn}. This algorithm attempts to find the optimum combination of features to include in a feature subset using cross-validation. We find that $\mu_*$ is the single feature that can predict the Asymmetry parameter most strongly, with an RMS scatter of 0.29. If allowed to include a second parameter the algorithm chooses ${\Sigma_{\rm H_2,1kpc}}$ (resulting in predictions with an RMS scatter of 0.26). The third important parameter is the SFE, resulting in predictions with an RMS scatter of 0.25. Adding additional parameters beyond this point does not substantially improve the prediction. We obtain similar results when regressing on the other parameters. The optimal predictor for Asymmetry is

\begin{multline}
A=-0.18\log\left(\frac{\mu_*}{\mathrm{M_{\odot} kpc^{-2}}}\right) -0.27\log\left(\frac{{\Sigma_{\rm H_2,1kpc}}}{\mathrm{M_{\odot} pc^{-2}}}\right)  + 0.44\log\left(\frac{\rm SFE}{\mathrm{yr^{-1}}}\right)\\ + 6.85,
\end{multline}

\noindent the optimal predictor for Smoothness is
\begin{multline}
S=0.03\log\left(\frac{\mu_*}{\mathrm{M_{\odot} kpc^{-2}}}\right) -0.07\log\left(\frac{{\Sigma_{\rm H_2,1kpc}}}{\mathrm{M_{\odot} pc^{-2}}}\right)  + 0.12\log\left(\frac{\rm SFE}{\mathrm{yr^{-1}}}\right)\\ + 1.81,
\end{multline}

\noindent and the optimal predictor for Gini is
\begin{multline}
Gini=-0.16\log\left(\frac{\mu_*}{\mathrm{M_{\odot} kpc^{-2}}}\right) -0.03\log\left(\frac{{\Sigma_{\rm H_2,1kpc}}}{\mathrm{M_{\odot} pc^{-2}}}\right)  + 0.12\log\left(\frac{\rm SFE}{\rm yr^{-1}}\right)\\ + 3.18.
\end{multline}

Given the above, it seems that $\mu_*$, the gas surface density and the SFE provide information which can help predict the morphology of the ISM, but $\mu_*$ appears {on average} to be the most informative.

The surface density of the gas is found to be the second most important parameter in both our simple linear fitting and `Sequential Feature Selection' analysis. It seems that surface density variations at fixed $\mu_*$ are important in driving scatter in the observed relations. However, as before, this correlation goes in the sense that at higher surface densities the ISM is smoother and less asymmetric. As previously discussed, this is hard to understand if self-gravity is the driving force behind these relations. We thus suspect that self-gravity is not important in determining the morphology of the ISM itself in these galaxy centres, and the observed correlation is created thanks to secondary correlations with other variables (see section \ref{mu_explain}). Indeed, in the simulations of \cite{2020MNRAS.495..199G} the simulated galaxies have very different morphologies, despite initially having very similar surface densities by construction. 

The star formation efficiency was selected as the other parameter which was most predictive of the ISM morphology in all cases. 
As above, it seems that star formation efficiency variations at fixed $\mu_*$ are important in driving scatter in the observed relations. 
The sense of this correlation is physically intuitive, in that objects with higher SFE have a more disturbed ISM morphology, as more stars are formed that can drive feedback and hence turbulence in the ISM. 
This is in agreement with e.g. the work of \cite{2020MNRAS.493.2872C,2022MNRAS.509..272C}, who recently investigated the lifecycle of molecular clouds in the discs of the PHANGS galaxies. They showed the importance of early stellar feedback in setting the lifetime of molecular clouds. Here we find that feedback may also be important in galaxy centres (along with other mechanisms linked to the galaxy potential) although we cannot rule out that the correlation seen in Figure \ref{ssfr_a_s} may be driven by an underlying relation between $\mu_*$ and the SFE \citep[e.g.][]{2009ApJ...707..250M,2011MNRAS.415...61S,2014MNRAS.444.3427D,2020MNRAS.495..199G}. 

Overall our results seem to favour scenario ii) discussed above- the morphology of the gas in galaxy centres seems to be set not by self-gravity, but by both the depth of the potential well, and turbulence driven by star formation.

\subsubsection{Understand the correlations with $\mu_*$}
\label{mu_explain}

Above we showed that the effective stellar surface density is a strong predictor for the ISM morphology in galaxy centres. It is interesting to consider what physical mechanisms could be causing this correlation. 

One potential effect that should be considered is the 3D structure of the stellar component. For instance, galaxies with high $\mu_*$ are typically bulge dominated, and as such have rounder potentials. A rounder potential reduces the forces felt by material in the mid-plane of the disc, suppressing both the formation of stellar features (e.g. spiral arms and bars; which themselves can act to disturb the ISM) and direct fragmentation of the gas disc. Spherical or nearly spherical stellar components (such as those of ETGs or classical bulges in spirals) could thus act to make ISM morphology more smooth either directly, or by reducing the non-axisymmetric stellar features that could disturb them. 

Stellar (spiral) density waves and stellar bars in galaxies act to collect molecular gas which can then fragment, collapse, and form stars. The increased stability of the stars in galaxies with large bulges could thus be driving our observed correlations with ISM morphology. While the early-type galaxies studied here do not have spiral features (by selection), many of them do have stellar bars, suggesting (along with the evidence in Section \ref{bars}) that these may not play a dominant role in setting the morphology of the ISM. The importance of stellar spiral arms is harder to quantify, and will be probed further in a future work. Here we conclude it is possible that the stability of the stellar structure of galaxies at small radii could be an important driver of the results we obtain. 

\begin{figure}
\includegraphics[width=0.5\textwidth,trim=0cm 0cm 0cm 0cm,clip]{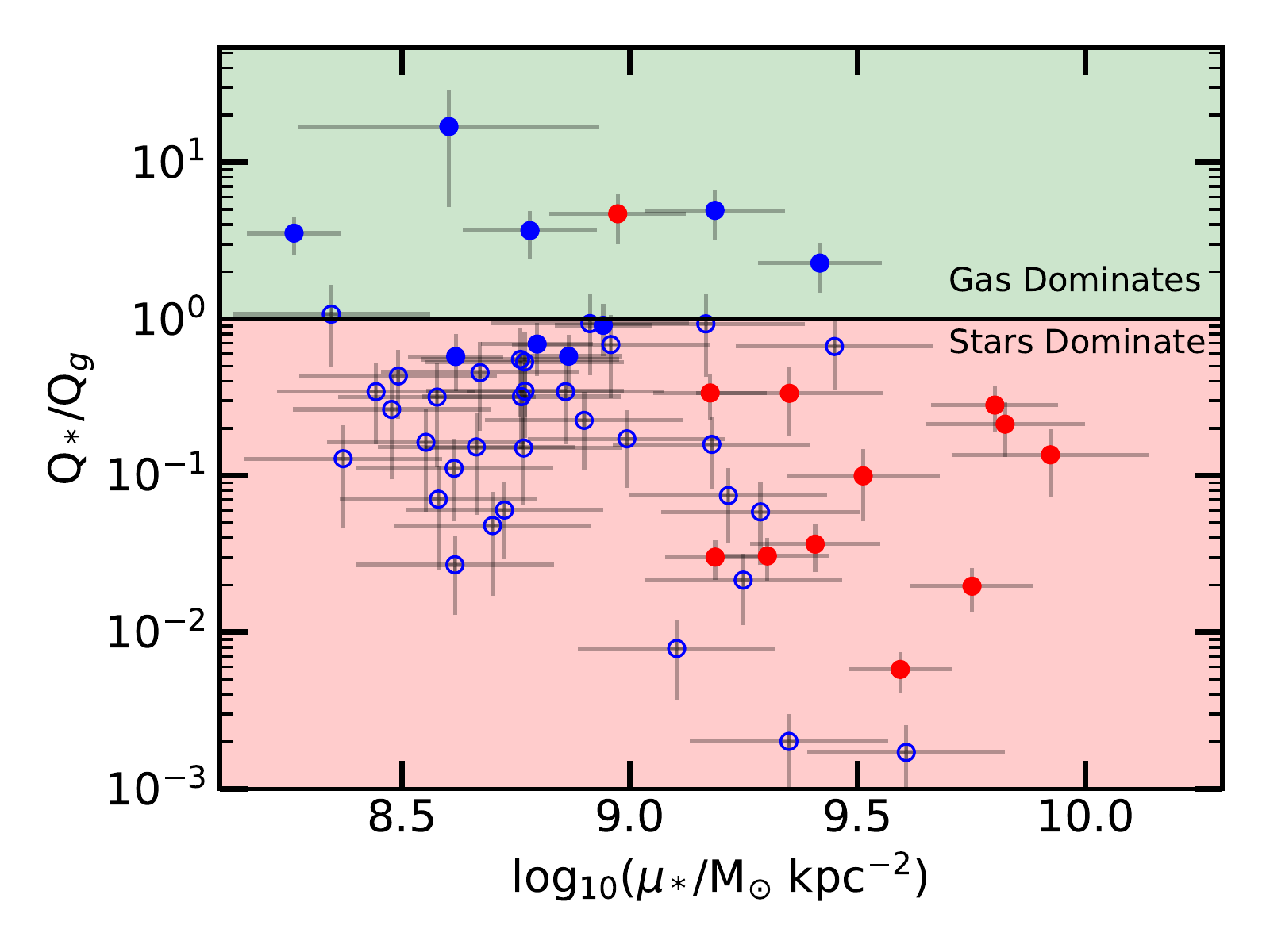}
\caption{Ratio of the Toomre stability parameter for the stellar and gaseous discs in our sample galaxies, plotted as a function of their stellar mass surface densities. WISDOM early-type galaxies are shown as red circles, WISDOM spiral galaxies as blue circles and PHANGS spirals as blue open circles. The green shaded area shows where the stellar disc is more stable than the gaseous disc (and thus the gas dominates the total stability of the system), while the red shaded area indicates the region where the opposite is true. The majority of our objects with dense bulge regions have the stability of the system dominated by that of the stars. } 
\label{qratio_mu}
\end{figure}

In a similar way to the stability of an isolated gas disc, whose stability can be classified using the $Q$ parameter \citep{1964ApJ...139.1217T}, the stability of a gas disc rotating inside a stellar potential depends on the rotation curve of the galaxy, as well as the velocity dispersion and the surface density of both the stars and the gas \citep[e.g.][]{2011MNRAS.416.1191R}. Although (as discussed above) we don't have robust rotation curves for all our galaxies we can still determine if the stellar or gas component dominates the overall stability of the system, as  
\begin{equation}
\frac{Q_*}{Q_g}=\frac{\sigma_* \Sigma_g}{\sigma_g \Sigma_*},
\end{equation}
where $Q_*$ and $Q_g$ are the classic Toomre Q parameters for the gas and stars, respectively, $\Sigma_*$ and $\Sigma_g$ are the surface densities of the stars and gas, and $\sigma_*$ and $\sigma_g$ are their velocity dispersions. 
In order to calculate this quantity we need to estimate $\Sigma_g$ within the same effective radius aperture as used for $\mu_*$. The PHANGS sources typically have been mapped to $R_e$, and so we can estimate this quantity directly. For the WISDOM sources we only have single pointings towards each galaxies centre. However, given that all the ETGs in the WISDOM sample have centrally concentrated gas that is all contained within the effective radius we can estimate the mean surface density using the total masses and radii presented in Table \ref{datatable}. We are forced to calculate the mean surface density for the spiral galaxies in the WISDOM sample in the same way, however we caution that this could lead us to overestimate $\frac{Q_*}{Q_g}$ if their gas were to extend significantly beyond one effective radius. Assuming $\sigma_g$ is $\approx$8\,\kms (e.g. \citealt{2013AJ....146..150C}), in Figure \ref{qratio_mu} we show that $\approx$75\% of our sample galaxies have $\frac{Q_*}{Q_g}<1$, and thus the stability of their gas disc is dominated by that of the stellar component. The cases where the gas disc dominates are almost all spirals, with the exception of gas-rich edge on lenticular galaxy NGC7172, and almost all have low stellar surface densities.  This may naturally explain why the gas surface density does not correlate in the expected manner with our non-parametric morphology indicators, but does correlate with  $\sigma_*$ and  $\mu_*$.

However, Toomre-type stability is not the only way in which the potential of the galaxy can affect the gas disc. Shear is another such process that may be important. Objects with large bulge-to-disk ratios have concentrated mass profiles, and steeply rising rotation curves. In the rising part of these steep rotation curves shear is high, which can suppress fragmentation, star formation \citep[e.g.][]{2014MNRAS.444.3427D,2019MNRAS.484.5734K,2020MNRAS.495..199G}, and pull existing molecular clouds apart (see e.g. \citealt{2021arXiv210604327L}). It should be noted that high shear environments would also be more Toomre-stable, due to the mutual dependence of these quantities on the shape of the rotation curve of the system. As such so these two mechanisms discussed here are not entirely independent.

The amount of shear present depends on the rotation curve of the system, and its derivatives. These depend not on $\mu_*$,  but on the stellar volume density
\begin{equation}
 \rho_* \equiv \frac{M_*}{2 \pi q R_e^3} = \frac{\mu_*}{qR_e},
 \end{equation}
  where $q$ is the ratio of the length of the short axis of the galaxy to the major axis and the other symbols are defined above. Unfortunately we do not know the 3D geometry of our galaxies, and thus $q$, and cannot estimate $\rho_*$ with certainty. 

\begin{figure*}
\includegraphics[height=14cm,trim=0cm 0cm 0cm 0cm,clip]{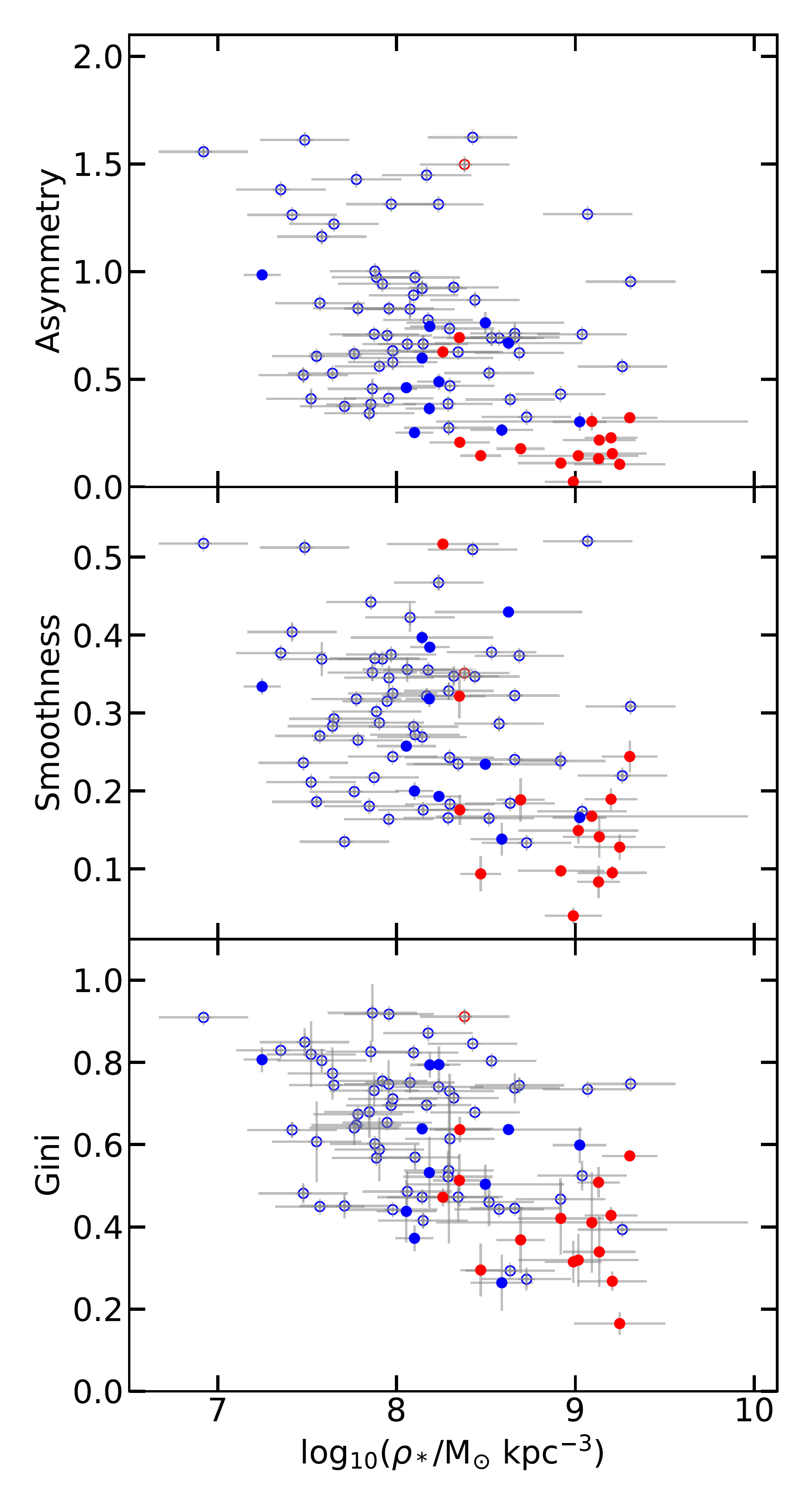}
\includegraphics[height=14cm,trim=2.7cm 0cm 0.5cm 0cm,clip]{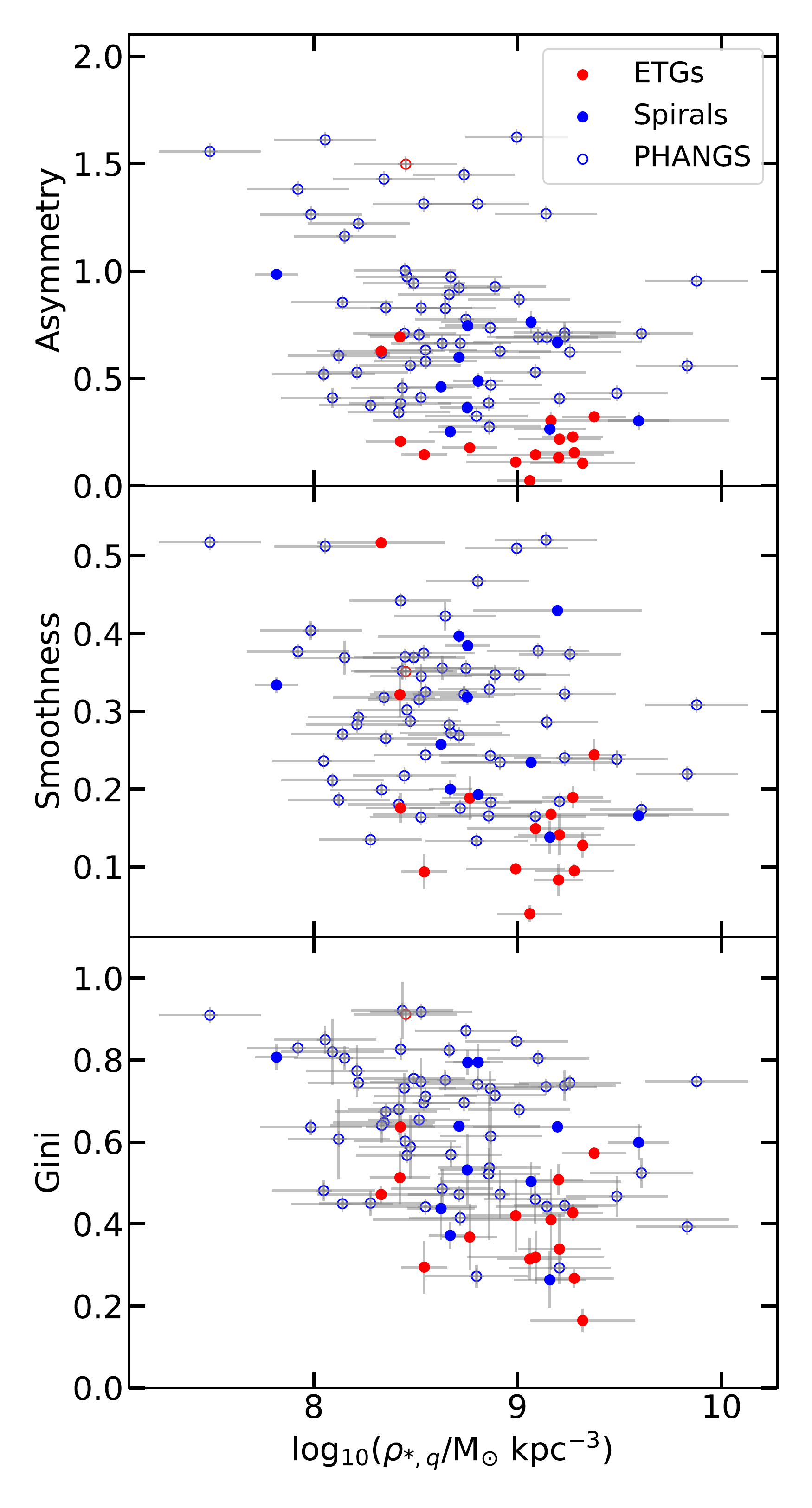}
\caption{As Figure \ref{mu_a_s}, but plotting the non-parametric morphology measurements against the stellar volume density assuming the galaxy is spherical (left column), and that spirals and ETGs have intrinsic axial ratios ($q$) of 0.27 and 0.8, respectively (right column). The morphology measurements for the observational data correlate strongly with these stellar volume densities, again highlighting that the depth of the potential well itself may be important in shaping the morphology of the ISM. } 
\label{rhostar_a_s}
\end{figure*}

Figure \ref{rhostar_a_s} shows the correlation between ISM morphology and the stellar volume density estimated making two extreme assumptions. In the left column we estimate the volume density assuming the galaxies are spherical ($q=1$), and denote this measurement $\rho_*$. Given that almost all these systems have central bulges (even the spiral galaxies), this may not be totally unphysical. Here the resulting correlations with ISM morphology are very strong (see Table \ref{corr_table_rho} for their statistical descriptions). 
In the right panel of Figure \ref{rhostar_a_s} we assume that our ETGs are as spherical as the roundest slow-rotators in \cite{2017MNRAS.472..966F}, while all spirals are as flat as the flattest fast-rotating galaxies from the same work ($q=$0.8 and 0.27, respectively). We denote this measurement $\rho_{*,q}$.
This weakens the correlations, but does not erase them entirely. In reality the true intrinsic shapes of these systems will not be this extreme, suggesting that the depth of the potential well, as traced by $\rho_*$, may be a driver of the observed ISM morphology correlations, and hence shear may also be important in setting the morphology of the ISM. This is in agreement with the simulations of \cite{2020MNRAS.495..199G}, who show that shear is the dominant contributor to the low SFE of their simulated galaxies with large bulges. 

It is also possible that gas flows within galaxies could play a role in setting the morphology of the ISM. In some of the spiral galaxies studied here a large fraction of the total molecular gas mass is located outside of the 3$\times$3\,kpc region we are studying. The average PHANGS spiral galaxy has only 51\% of its gas present in this region, with a wide scatter present from galaxy-to-galaxy. In contrast this region contains essentially all of the molecular gas in our early-type systems. If some of this gas is inflowing in spiral galaxies then this may perturb the inner gas morphology, driving asymmetries and promoting fragmentation. One might even expect inflow rates to (anti-)correlate with the presence of a bulge component, as the rounder potential suppresses the formation of bars and spiral arms (as discussed above). We do not find any correlation between our non-parametric morphology indicators and the fraction of molecular gas inside 3\,kpc, but without knowing the inflow rates of individual systems we cannot rule out the possibility that gas flows are important in shaping the morphology of the ISM.

\begin{table}
\caption{Correlation measurements with $\rho_*$.}
\begin{center}
\begin{tabular*}{0.48\textwidth}{@{\extracolsep{\fill}} l r l c}
\hline
Correlation & $\rho$ & $p$ & $p$$<$0.05 \\
 (1) & (2) &(3)&(4)  \\
\hline
log$_{10}$($\rho_*$/M$_{\odot}$ kpc$^{-3}$) vs Asymmetry &  -0.44  & \num{2e-05} & \checkmark\\
log$_{10}$($\rho_*$/M$_{\odot}$ kpc$^{-3}$) vs Smoothness &  -0.38  & \num{0.0003} & \checkmark\\
log$_{10}$($\rho_*$/M$_{\odot}$ kpc$^{-3}$) vs Gini &  -0.49  & \num{2e-06} & \checkmark\\
\hline
log$_{10}$($\rho_{*,q}$/M$_{\odot}$ kpc$^{-3}$) vs Asymmetry &  -0.36  & \num{0.0007} & \checkmark\\
log$_{10}$($\rho_{*,q}$/M$_{\odot}$ kpc$^{-3}$) vs Smoothness &  -0.32  & \num{0.003} & \checkmark\\
log$_{10}$($\rho_{*,q}$/M$_{\odot}$ kpc$^{-3}$) vs Gini &  -0.41  & \num{9e-05} & \checkmark\\
\hline
\end{tabular*}\vspace{0.01cm}
\parbox[t]{0.48\textwidth}{ \textit{Notes:} Column 1 lists the correlation variables, while Columns 2 and 3 contain Spearman's rank correlation coefficient ($\rho$) and its associated $p$-value. Column 4 acts as a guide for the eye, highlighting significant ($p<0.05$) correlations.}
\end{center}
\label{corr_table_rho}
\end{table}

\subsection{Comparison with other studies}

Here we compare our results with the conclusions of other studies, who have approached these questions with different techniques or using different ISM tracers.

\subsubsection{Other molecular gas studies}

Our conclusion that the stellar potential of galaxies is important in setting the structure of the ISM is similar to that found by \cite{2021ApJ...913..113M}. They used cross-correlations between molecular gas observations and 3.6$\mu$m maps of the stellar mass distribution in the discs of PHANGS galaxies to show that stellar dynamical features appear to play an important role in setting the cloud-scale gas density, with gas self-gravity playing a weaker role.  Earlier studies by the same group \cite[e.g.][]{2018ApJ...854..100M,2020ApJ...892...73M} showed that the larger scale potential can affect molecular cloud properties, and even suppress star formation. 
Similarly \cite{2021arXiv210904491Q} showed that stellar structures in galaxies strongly affect the organisation of molecular gas and star formation. Overall these results agree well with those derived here.

On the other hand, \cite{2020NatAs...4.1064H} conducted an analysis of the velocity fluctuations in molecular gas in the Milky Way and a nearby spiral galaxy NGC 4321. While they conclude that a variety of scales and processes matter for the assembly of the ISM, they find characteristic velocity fluctuations on scales suggestive of fragmentation due to self-gravity. This seems somewhat counter to our findings here. Exploring the velocity power-spectra of our sample galaxies, and thus determining if this difference is due to the small sample size of \cite{2020NatAs...4.1064H} or some other physical process, will be the focus of a future work.

\subsubsection{Atomic gas}

The atomic gas in galaxies cannot typically be observed at high angular resolution, but has a much larger radial extent than molecular gas, meaning it can often be resolved well enough to study the processes that drive its large-scale morphology. 
For instance, \cite{2011MNRAS.416.2401H} and \cite{2011MNRAS.416.2415H} demonstrated {(using the same non-parametric indicators used in this work)} that Asymmetry in \hi\ is a useful tracer of galaxy interactions. Larger scale environment is an interesting parameter which may impact the morphology of the molecular gas that we do not explore here. However, the dynamical times at the centres of our galaxies are very short, so it is likely that interactions play a smaller role here.

\cite{2013MNRAS.435.1020H} showed that in dwarf galaxies non-parametric indicators of  \hi\ morphology correlate weakly with star formation rate, as we found here for the molecular component. They conclude that local physics dominates when setting the appearance of the atomic ISM, in concordance with our findings for more massive galaxies.

\cite{2020MNRAS.492.4697N} showed the morphology of the HI in galaxies {(classifed by eye)} is affected strongly by the presence of bars, apart from at high gas fractions where they seem to play little role. In molecular gas we find that bars have {some} effect on the smoothness of the ISM {(as measured by the Gini parameter)}, in agreement with that work, but do not observe a molecular gas fraction dependence. 

\subsubsection{Dust}

Dust is another tracer of the cold ISM. It can be observed both in emission directly (in the mid- to far-infrared) and also in absorption at optical wavelengths, when it is silhouetted against the stellar component of a galaxy. With current technology dust emission maps are only able to be used to reveal the structure of the ISM on parsec scales in the very nearest galaxies. Dust absorption maps created by e.g. HST can, however, be used to image the structure of the ISM in a wider range of systems. 
For instance, \cite{1999AJ....117.2676R} used HST to study the morphology of the dusty ISM in late-type Seyfert galaxies, finding that spiral features are common in their nuclear ISM. As these systems are typically lower stellar mass this would match the picture painted here, where the ISM breaks up into spiral structures more easily in low mass galaxies. 

The structure of the ISM has also been studied using dust as a tracer in edge-on galaxies, where its vertical scale height can be resolved. \cite{2004ApJ...608..189D} found that smooth dust lanes were more common in more massive galaxies, and that this effect was present even at fixed dust surface density.  They attributed this to the balance between turbulence driving mechanisms changing in these systems, with star formation feedback becoming more dominant at lower stellar masses. \cite{2019AJ....158..103H} found further evidence for this change in the structure of the ISM with stellar mass, and also found that regular dust lanes were present more often in galaxies with larger bulges, and those with lower SFRs. 

{In this work we also see a significant change in the ISM structure as a function of stellar mass and effective stellar surface density (a proxy for a more dominant bulge). We have shown this could arise because the stellar potential dominates over the gas potential at higher stellar surface densities in our galaxy centres, which seems to lead to a smoother ISM. If this same effect could be observed in the dust component of galaxies when they are seen edge-on is not clear. This could, However, be tested by dust radiative transfer simulations.}

\section{Conclusions}
\label{conclude}

In this paper we used maps of the molecular ISM in the centres of eighty-six galaxies from the WISDOM and PHANGS surveys to investigate the physical mechanisms setting the morphology of the ISM at molecular cloud scales. We compared these observations with idealised simulations from \cite{2020MNRAS.495..199G}.

Visual classifications showed that early-type galaxies have smooth, regular molecular gas morphologies in their centres (even when observed with resolutions of 10's of parsecs), while the ISM in spiral galaxies is much more asymmetric and clumpy on the same scales. We quantified this using non-parametric morphology measures (as often used in optical morphology studies). We showed that the morphology of the ISM in the centres of nearby galaxies changes systematically as a function of various large scale galaxy parameters. {Negative correlations were seen between ISM morphology measures and galaxy properties that correlate with the stellar potential well depth (galaxy morphology, stellar mass, stellar velocity dispersion, effective stellar mass surface density). This confirmed that massive galaxies with large bulges have smoother, more symmetric ISM morphologies than lower mass systems. Correlations are also present between ISM morphology measures and surface density of the gas, the specific star formation rate, star formation efficiency and (at least for one measure) the presence of a bar.}

We attempted to disentangle which of these correlated parameters are truly important in setting the morphology of the ISM. While this task is fraught with difficulty due to internal correlations, a statistical analysis suggests the stellar surface (or volume) density is the strongest predictor of the morphology of the gas, while the efficiency of star formation and the presence of a bar may be important drivers of scatter. It would seem that self-gravity is not the dominant processes shaping the morphology of the gas in galaxy centres, as expected from previous works. In these regions the stellar potential typically dominates, and in denser bulges this seems to keep the gas more stable. 
We posit that the molecular gas in galaxies with large bulges could be in a smooth disk, because of (i) high shear and/or (ii) absence of stellar spiral density waves, and/or (iii) absence of inflowing gas.
These conclusions are supported by work at other wavelengths, although significant uncertainties remain. 

In order to fully understand the results presented here, additional observational and theoretical work is required. More realistic simulations which span a greater range of galaxy properties could help shed light on which of the correlations we observe are causal, while including the cosmological context could also be important. Observationally, obtaining high resolution observations of the ISM of larger samples of galaxies is key, especially complete volume-limited samples of early-type galaxies. Studies with similar resolution of other gas components (e.g. atomic gas, ionised gas, metallicity) would also be useful. The results of this paper suggest that the structure of the molecular ISM in galaxy centres is set by the properties of the galaxy. By combining these theoretical and observational advances we can better understand the importance of the various physical processes causing this result. 

\section*{Acknowledgements}

TAD and IR acknowledge support from the UK Science and Technology Facilities Council through grants ST/S00033X/1 and ST/W000830/1.
JG gratefully acknowledges financial support from the Swiss National Science Foundation (grant no. CRSII5\_193826)".
JG and JMDK gratefully acknowledge funding from the DFG through an Emmy Noether Grant (grant number KR4801/1-1), as well as from the European Research Council (ERC) under the European Union's Horizon 2020 research and innovation programme via the ERC Starting Grant MUSTANG (grant agreement number 714907).
TGW acknowledges funding from the European Research Council (ERC) under the European Union’s Horizon 2020 research and innovation programme (grant agreement No. 694343).

This paper makes use of ALMA data from projects 2013.1.00493.S, 2012.1.00650.S, 2013.1.00803.S, 2013.1.01161.S, 2015.1.00121.S, 2015.1.00466.S, 2015.1.00782.S, 2015.1.00925.S, 2015.1.00956.S, 2016.2.00053.S, 2016.1.00386.S, 2016.1.00419.S, 2016.1.00437.S, 2017.1.00277, 2017.1.00391.S, 2017.1.00392.S, 2017.1.00766.S, 2017.1.00886.L, 2017.1.00904.S, 2018.1.00484.S, 2018.1.00517.S, 2018.1.01321.S, 2018.1.01651.S, 2018.A.00062.S, 2019.1.00363.S, 2019.1.01235.S, and 2019.2.00129.S. ALMA is a partnership of ESO (representing its member states), NSF (USA) and NINS (Japan), together with NRC (Canada), MOST and ASIAA (Taiwan), and KASI (Republic of Korea), in cooperation with the Republic of Chile. The Joint ALMA Observatory is operated by ESO, AUI/NRAO and NAOJ. 

We acknowledge the usage of the HyperLeda database (http://leda.univ-lyon1.fr).
This research made use of \textsc{numpy} \citep{harris2020array}, \textsc{scipy} \citep{2020SciPy-NMeth}, \textsc{astropy} \citep{astropy:2013, astropy:2018} and \textsc{matplotlib} \cite{Hunter:2007}. 

\section*{Data availability}
The raw data underlying this article is available from the ALMA archive at \url{https://almascience.eso.org/aq} using the project codes listed above. The WISDOM data products will be made available from \url{https://www.wisdom-project.org} and the PHANGS data products are available from \url{https://www.phangs.org}. 

\bsp	
\bibliographystyle{mnras}
\bibliography{bibMASSIVE_smbh.bib}

\begin{thebibliography}{}
\makeatletter
\relax
\def\mn@urlcharsother{\let\do\@makeother \do\$\do\&\do\#\do\^\do\_\do\%\do\~}
\def\mn@doi{\begingroup\mn@urlcharsother \@ifnextchar [ {\mn@doi@}
  {\mn@doi@[]}}
\def\mn@doi@[#1]#2{\def\@tempa{#1}\ifx\@tempa\@empty \href
  {http://dx.doi.org/#2} {doi:#2}\else \href {http://dx.doi.org/#2} {#1}\fi
  \endgroup}
\def\mn@eprint#1#2{\mn@eprint@#1:#2::\@nil}
\def\mn@eprint@arXiv#1{\href {http://arxiv.org/abs/#1} {{\tt arXiv:#1}}}
\def\mn@eprint@dblp#1{\href {http://dblp.uni-trier.de/rec/bibtex/#1.xml}
  {dblp:#1}}
\def\mn@eprint@#1:#2:#3:#4\@nil{\def\@tempa {#1}\def\@tempb {#2}\def\@tempc
  {#3}\ifx \@tempc \@empty \let \@tempc \@tempb \let \@tempb \@tempa \fi \ifx
  \@tempb \@empty \def\@tempb {arXiv}\fi \@ifundefined
  {mn@eprint@\@tempb}{\@tempb:\@tempc}{\expandafter \expandafter \csname
  mn@eprint@\@tempb\endcsname \expandafter{\@tempc}}}

\bibitem[\protect\citeauthoryear{{Abraham}, {Tanvir}, {Santiago}, {Ellis},
  {Glazebrook}  \& {van den Bergh}}{{Abraham}
  et~al.}{1996}]{1996MNRAS.279L..47A}
{Abraham} R.~G.,  {Tanvir} N.~R.,  {Santiago} B.~X.,  {Ellis} R.~S.,
  {Glazebrook} K.,   {van den Bergh} S.,  1996, \mn@doi [\mnras]
  {10.1093/mnras/279.3.L47}, \href
  {https://ui.adsabs.harvard.edu/abs/1996MNRAS.279L..47A} {279, L47}

\bibitem[\protect\citeauthoryear{{Alatalo} et~al.,}{{Alatalo}
  et~al.}{2013}]{2013MNRAS.432.1796A}
{Alatalo} K.,  et~al., 2013, \mn@doi [\mnras] {10.1093/mnras/sts299}, \href
  {https://ui.adsabs.harvard.edu/abs/2013MNRAS.432.1796A} {432, 1796}

\bibitem[\protect\citeauthoryear{{Astropy Collaboration} et~al.,}{{Astropy
  Collaboration} et~al.}{2013}]{astropy:2013}
{Astropy Collaboration} et~al., 2013, \mn@doi [\aap]
  {10.1051/0004-6361/201322068}, \href
  {http://adsabs.harvard.edu/abs/2013A%26A...558A..33A} {558, A33}

\bibitem[\protect\citeauthoryear{{Astropy Collaboration} et~al.,}{{Astropy
  Collaboration} et~al.}{2018}]{astropy:2018}
{Astropy Collaboration} et~al., 2018, \mn@doi [\aj] {10.3847/1538-3881/aabc4f},
  \href {https://ui.adsabs.harvard.edu/abs/2018AJ....156..123A} {156, 123}

\bibitem[\protect\citeauthoryear{{Audibert} et~al.,}{{Audibert}
  et~al.}{2019}]{2019A&A...632A..33A}
{Audibert} A.,  et~al., 2019, \mn@doi [\aap] {10.1051/0004-6361/201935845},
  \href {https://ui.adsabs.harvard.edu/abs/2019A&A...632A..33A} {632, A33}

\bibitem[\protect\citeauthoryear{{Barth}, {Strigari}, {Bentz}, {Greene}  \&
  {Ho}}{{Barth} et~al.}{2009}]{2009ApJ...690.1031B}
{Barth} A.~J.,  {Strigari} L.~E.,  {Bentz} M.~C.,  {Greene} J.~E.,   {Ho}
  L.~C.,  2009, \mn@doi [\apj] {10.1088/0004-637X/690/1/1031}, \href
  {https://ui.adsabs.harvard.edu/abs/2009ApJ...690.1031B} {690, 1031}

\bibitem[\protect\citeauthoryear{{Bershady}, {Jangren}  \&
  {Conselice}}{{Bershady} et~al.}{2000}]{2000AJ....119.2645B}
{Bershady} M.~A.,  {Jangren} A.,   {Conselice} C.~J.,  2000, \mn@doi [\aj]
  {10.1086/301386}, \href
  {https://ui.adsabs.harvard.edu/abs/2000AJ....119.2645B} {119, 2645}

\bibitem[\protect\citeauthoryear{{Bigiel}, {Leroy}, {Walter}, {Brinks}, {de
  Blok}, {Madore}  \& {Thornley}}{{Bigiel} et~al.}{2008}]{2008AJ....136.2846B}
{Bigiel} F.,  {Leroy} A.,  {Walter} F.,  {Brinks} E.,  {de Blok} W.~J.~G.,
  {Madore} B.,   {Thornley} M.~D.,  2008, \mn@doi [\aj]
  {10.1088/0004-6256/136/6/2846}, \href
  {https://ui.adsabs.harvard.edu/abs/2008AJ....136.2846B} {136, 2846}

\bibitem[\protect\citeauthoryear{{Bolatto}, {Wolfire}  \& {Leroy}}{{Bolatto}
  et~al.}{2013}]{2013ARA&A..51..207B}
{Bolatto} A.~D.,  {Wolfire} M.,   {Leroy} A.~K.,  2013, \mn@doi [\araa]
  {10.1146/annurev-astro-082812-140944}, \href
  {https://ui.adsabs.harvard.edu/abs/2013ARA&A..51..207B} {51, 207}

\bibitem[\protect\citeauthoryear{{Cald{\'u}-Primo}, {Schruba}, {Walter},
  {Leroy}, {Sandstrom}, {de Blok}, {Ianjamasimanana}  \&
  {Mogotsi}}{{Cald{\'u}-Primo} et~al.}{2013}]{2013AJ....146..150C}
{Cald{\'u}-Primo} A.,  {Schruba} A.,  {Walter} F.,  {Leroy} A.,  {Sandstrom}
  K.,  {de Blok} W.~J.~G.,  {Ianjamasimanana} R.,   {Mogotsi} K.~M.,  2013,
  \mn@doi [\aj] {10.1088/0004-6256/146/6/150}, \href
  {https://ui.adsabs.harvard.edu/abs/2013AJ....146..150C} {146, 150}

\bibitem[\protect\citeauthoryear{{Callanan} et~al.,}{{Callanan}
  et~al.}{2021}]{2021MNRAS.505.4310C}
{Callanan} D.,  et~al., 2021, \mn@doi [\mnras] {10.1093/mnras/stab1527}, \href
  {https://ui.adsabs.harvard.edu/abs/2021MNRAS.505.4310C} {505, 4310}

\bibitem[\protect\citeauthoryear{{Cappellari}}{{Cappellari}}{2013}]{2013ApJ...778L...2C}
{Cappellari} M.,  2013, \mn@doi [\apjl] {10.1088/2041-8205/778/1/L2}, \href
  {https://ui.adsabs.harvard.edu/abs/2013ApJ...778L...2C} {778, L2}

\bibitem[\protect\citeauthoryear{{Cappellari} et~al.,}{{Cappellari}
  et~al.}{2013a}]{2013MNRAS.432.1862C}
{Cappellari} M.,  et~al., 2013a, \mn@doi [\mnras] {10.1093/mnras/stt644}, \href
  {http://adsabs.harvard.edu/abs/2013MNRAS.432.1862C} {432, 1862}

\bibitem[\protect\citeauthoryear{Cappellari et~al.,}{Cappellari
  et~al.}{2013b}]{2013MNRAS.432.1709C}
Cappellari M.,  et~al., 2013b, \mnras, 432, 1709

\bibitem[\protect\citeauthoryear{{Chevance} et~al.,}{{Chevance}
  et~al.}{2020}]{2020MNRAS.493.2872C}
{Chevance} M.,  et~al., 2020, \mn@doi [\mnras] {10.1093/mnras/stz3525}, \href
  {https://ui.adsabs.harvard.edu/abs/2020MNRAS.493.2872C} {493, 2872}

\bibitem[\protect\citeauthoryear{{Chevance} et~al.,}{{Chevance}
  et~al.}{2022}]{2022MNRAS.509..272C}
{Chevance} M.,  et~al., 2022, \mn@doi [\mnras] {10.1093/mnras/stab2938}, \href
  {https://ui.adsabs.harvard.edu/abs/2022MNRAS.509..272C} {509, 272}

\bibitem[\protect\citeauthoryear{{Cohen}, {Cong}, {Dame}  \&
  {Thaddeus}}{{Cohen} et~al.}{1980}]{1980ApJ...239L..53C}
{Cohen} R.~S.,  {Cong} H.,  {Dame} T.~M.,   {Thaddeus} P.,  1980, \mn@doi
  [\apjl] {10.1086/183290}, \href
  {https://ui.adsabs.harvard.edu/abs/1980ApJ...239L..53C} {239, L53}

\bibitem[\protect\citeauthoryear{{Conselice}}{{Conselice}}{2003a}]{2003ApJS..147....1C}
{Conselice} C.~J.,  2003a, \mn@doi [\apjs] {10.1086/375001}, \href
  {https://ui.adsabs.harvard.edu/abs/2003ApJS..147....1C} {147, 1}

\bibitem[\protect\citeauthoryear{{Conselice}}{{Conselice}}{2003b}]{Conselice2003}
{Conselice} C.~J.,  2003b, \mn@doi [\apjs] {10.1086/375001}, \href
  {https://ui.adsabs.harvard.edu/abs/2003ApJS..147....1C} {147, 1}

\bibitem[\protect\citeauthoryear{{Cook}, {van Sistine}, {Singer}, {Kasliwal},
  {Kaplan}, {Iptf Collaboration}  \& {Growth Collaboration}}{{Cook}
  et~al.}{2017}]{2017GCN.21707....1C}
{Cook} D.~O.,  {van Sistine} A.,  {Singer} L.,  {Kasliwal} M.~M.,  {Kaplan} D.,
   {Iptf Collaboration}  {Growth Collaboration} 2017, GRB Coordinates Network,
  \href {https://ui.adsabs.harvard.edu/abs/2017GCN.21707....1C} {21707, 1}

\bibitem[\protect\citeauthoryear{{Dalcanton}, {Yoachim}  \&
  {Bernstein}}{{Dalcanton} et~al.}{2004}]{2004ApJ...608..189D}
{Dalcanton} J.~J.,  {Yoachim} P.,   {Bernstein} R.~A.,  2004, \mn@doi [\apj]
  {10.1086/386358}, \href
  {https://ui.adsabs.harvard.edu/abs/2004ApJ...608..189D} {608, 189}

\bibitem[\protect\citeauthoryear{{Dame}}{{Dame}}{2011}]{Dame2011}
{Dame} T.~M.,  2011, arXiv e-prints, \href
  {https://ui.adsabs.harvard.edu/abs/2011arXiv1101.1499D} {p. arXiv:1101.1499}

\bibitem[\protect\citeauthoryear{{Dame}, {Hartmann}  \& {Thaddeus}}{{Dame}
  et~al.}{2001}]{2001ApJ...547..792D}
{Dame} T.~M.,  {Hartmann} D.,   {Thaddeus} P.,  2001, \mn@doi [\apj]
  {10.1086/318388}, \href
  {https://ui.adsabs.harvard.edu/abs/2001ApJ...547..792D} {547, 792}

\bibitem[\protect\citeauthoryear{Davis et~al.,}{Davis
  et~al.}{2013}]{2013MNRAS.429..534D}
Davis T.~A.,  et~al., 2013, \mnras, 429, 534

\bibitem[\protect\citeauthoryear{{Davis} et~al.,}{{Davis}
  et~al.}{2014}]{2014MNRAS.444.3427D}
{Davis} T.~A.,  et~al., 2014, \mn@doi [\mnras] {10.1093/mnras/stu570}, \href
  {https://ui.adsabs.harvard.edu/abs/2014MNRAS.444.3427D} {444, 3427}

\bibitem[\protect\citeauthoryear{{Davis}, {Greene}, {Ma}, {Pandya},
  {Blakeslee}, {McConnell}  \& {Thomas}}{{Davis}
  et~al.}{2016}]{2016MNRAS.455..214D}
{Davis} T.~A.,  {Greene} J.,  {Ma} C.-P.,  {Pandya} V.,  {Blakeslee} J.~P.,
  {McConnell} N.,   {Thomas} J.,  2016, \mn@doi [\mnras]
  {10.1093/mnras/stv2313}, \href
  {http://adsabs.harvard.edu/abs/2016MNRAS.455..214D} {455, 214}

\bibitem[\protect\citeauthoryear{{Davis}, {Bureau}, {Onishi}, {Cappellari},
  {Iguchi}  \& {Sarzi}}{{Davis} et~al.}{2017}]{2017MNRAS.468.4675D}
{Davis} T.~A.,  {Bureau} M.,  {Onishi} K.,  {Cappellari} M.,  {Iguchi} S.,
  {Sarzi} M.,  2017, \mn@doi [\mnras] {10.1093/mnras/stw3217}, \href
  {http://adsabs.harvard.edu/abs/2017MNRAS.468.4675D} {468, 4675}

\bibitem[\protect\citeauthoryear{{Davis} et~al.,}{{Davis}
  et~al.}{2018}]{2018MNRAS.473.3818D}
{Davis} T.~A.,  et~al., 2018, \mn@doi [\mnras] {10.1093/mnras/stx2600}, \href
  {https://ui.adsabs.harvard.edu/abs/2018MNRAS.473.3818D} {473, 3818}

\bibitem[\protect\citeauthoryear{{Davis}, {Greene}, {Ma}, {Blakeslee},
  {Dawson}, {Pandya}, {Veale}  \& {Zabel}}{{Davis}
  et~al.}{2019}]{2019MNRAS.486.1404D}
{Davis} T.~A.,  {Greene} J.~E.,  {Ma} C.-P.,  {Blakeslee} J.~P.,  {Dawson}
  J.~M.,  {Pandya} V.,  {Veale} M.,   {Zabel} N.,  2019, \mn@doi [\mnras]
  {10.1093/mnras/stz871}, \href
  {https://ui.adsabs.harvard.edu/abs/2019MNRAS.486.1404D} {486, 1404}

\bibitem[\protect\citeauthoryear{{Dickman}, {Snell}  \& {Schloerb}}{{Dickman}
  et~al.}{1986}]{Dickman:1986jz}
{Dickman} R.~L.,  {Snell} R.~L.,   {Schloerb} F.~P.,  1986, \mn@doi [\apj]
  {10.1086/164604}, \href {http://adsabs.harvard.edu/abs/1986ApJ...309..326D}
  {309, 326}

\bibitem[\protect\citeauthoryear{{Diniz}, {Riffel}, {Storchi-Bergmann}  \&
  {Winge}}{{Diniz} et~al.}{2015}]{2015MNRAS.453.1727D}
{Diniz} M.~R.,  {Riffel} R.~A.,  {Storchi-Bergmann} T.,   {Winge} C.,  2015,
  \mn@doi [\mnras] {10.1093/mnras/stv1694}, \href
  {https://ui.adsabs.harvard.edu/abs/2015MNRAS.453.1727D} {453, 1727}

\bibitem[\protect\citeauthoryear{{Duc}, {Braine}, {Lisenfeld}, {Brinks}  \&
  {Boquien}}{{Duc} et~al.}{2007}]{2007A&A...475..187D}
{Duc} P.~A.,  {Braine} J.,  {Lisenfeld} U.,  {Brinks} E.,   {Boquien} M.,
  2007, \mn@doi [\aap] {10.1051/0004-6361:20078335}, \href
  {https://ui.adsabs.harvard.edu/abs/2007A&A...475..187D} {475, 187}

\bibitem[\protect\citeauthoryear{{Dumas}, {Mundell}, {Emsellem}  \&
  {Nagar}}{{Dumas} et~al.}{2007}]{2007MNRAS.379.1249D}
{Dumas} G.,  {Mundell} C.~G.,  {Emsellem} E.,   {Nagar} N.~M.,  2007, \mn@doi
  [\mnras] {10.1111/j.1365-2966.2007.12014.x}, \href
  {https://ui.adsabs.harvard.edu/abs/2007MNRAS.379.1249D} {379, 1249}

\bibitem[\protect\citeauthoryear{{Emonts}, {Morganti}, {Oosterloo}, {Holt},
  {Tadhunter}, {van der Hulst}, {Ojha}  \& {Sadler}}{{Emonts}
  et~al.}{2008}]{2008MNRAS.387..197E}
{Emonts} B.~H.~C.,  {Morganti} R.,  {Oosterloo} T.~A.,  {Holt} J.,  {Tadhunter}
  C.~N.,  {van der Hulst} J.~M.,  {Ojha} R.,   {Sadler} E.~M.,  2008, \mn@doi
  [\mnras] {10.1111/j.1365-2966.2008.13142.x}, \href
  {https://ui.adsabs.harvard.edu/abs/2008MNRAS.387..197E} {387, 197}

\bibitem[\protect\citeauthoryear{Ferri, Pudil, Hatef  \& Kittler}{Ferri
  et~al.}{1994}]{FERRI1994403}
Ferri F.,  Pudil P.,  Hatef M.,   Kittler J.,  1994, in Gelsema E.~S.,  Kanal
  L.~S.,  eds, Machine Intelligence and Pattern Recognition, Vol.~16, Pattern
  Recognition in Practice IV.
North-Holland, pp 403--413,
  \mn@doi{https://doi.org/10.1016/B978-0-444-81892-8.50040-7}

\bibitem[\protect\citeauthoryear{{Foster} et~al.,}{{Foster}
  et~al.}{2017}]{2017MNRAS.472..966F}
{Foster} C.,  et~al., 2017, \mn@doi [\mnras] {10.1093/mnras/stx1869}, \href
  {https://ui.adsabs.harvard.edu/abs/2017MNRAS.472..966F} {472, 966}

\bibitem[\protect\citeauthoryear{{Garc{\'\i}a-Burillo}
  et~al.,}{{Garc{\'\i}a-Burillo} et~al.}{2003}]{2003A&A...407..485G}
{Garc{\'\i}a-Burillo} S.,  et~al., 2003, \mn@doi [\aap]
  {10.1051/0004-6361:20030866}, \href
  {https://ui.adsabs.harvard.edu/abs/2003A&A...407..485G} {407, 485}

\bibitem[\protect\citeauthoryear{{Garcia-Burillo} et~al.,}{{Garcia-Burillo}
  et~al.}{2021}]{2021arXiv210410227G}
{Garcia-Burillo} S.,  et~al., 2021, arXiv e-prints, \href
  {https://ui.adsabs.harvard.edu/abs/2021arXiv210410227G} {p. arXiv:2104.10227}

\bibitem[\protect\citeauthoryear{{Gensior} \& {Kruijssen}}{{Gensior} \&
  {Kruijssen}}{2021}]{2021MNRAS.500.2000G}
{Gensior} J.,  {Kruijssen} J.~M.~D.,  2021, \mn@doi [\mnras]
  {10.1093/mnras/staa3453}, \href
  {https://ui.adsabs.harvard.edu/abs/2021MNRAS.500.2000G} {500, 2000}

\bibitem[\protect\citeauthoryear{{Gensior}, {Kruijssen}  \& {Keller}}{{Gensior}
  et~al.}{2020}]{2020MNRAS.495..199G}
{Gensior} J.,  {Kruijssen} J.~M.~D.,   {Keller} B.~W.,  2020, \mn@doi [\mnras]
  {10.1093/mnras/staa1184}, \href
  {https://ui.adsabs.harvard.edu/abs/2020MNRAS.495..199G} {495, 199}

\bibitem[\protect\citeauthoryear{{Gu}, {Melnick}, {Cid Fernandes}, {Kunth},
  {Terlevich}  \& {Terlevich}}{{Gu} et~al.}{2006}]{2006MNRAS.366..480G}
{Gu} Q.,  {Melnick} J.,  {Cid Fernandes} R.,  {Kunth} D.,  {Terlevich} E.,
  {Terlevich} R.,  2006, \mn@doi [\mnras] {10.1111/j.1365-2966.2005.09872.x},
  \href {https://ui.adsabs.harvard.edu/abs/2006MNRAS.366..480G} {366, 480}

\bibitem[\protect\citeauthoryear{{H{\"a}gele}, {D{\'\i}az}, {Cardaci},
  {Terlevich}  \& {Terlevich}}{{H{\"a}gele} et~al.}{2007}]{2007MNRAS.378..163H}
{H{\"a}gele} G.~F.,  {D{\'\i}az} {\'A}.~I.,  {Cardaci} M.~V.,  {Terlevich} E.,
   {Terlevich} R.,  2007, \mn@doi [\mnras] {10.1111/j.1365-2966.2007.11751.x},
  \href {https://ui.adsabs.harvard.edu/abs/2007MNRAS.378..163H} {378, 163}

\bibitem[\protect\citeauthoryear{Harris et~al.,}{Harris
  et~al.}{2020}]{harris2020array}
Harris C.~R.,  et~al., 2020, \mn@doi [Nature] {10.1038/s41586-020-2649-2}, 585,
  357

\bibitem[\protect\citeauthoryear{{Hennebelle} \& {Falgarone}}{{Hennebelle} \&
  {Falgarone}}{2012}]{2012A&ARv..20...55H}
{Hennebelle} P.,  {Falgarone} E.,  2012, \mn@doi [\aapr]
  {10.1007/s00159-012-0055-y}, \href
  {https://ui.adsabs.harvard.edu/abs/2012A&ARv..20...55H} {20, 55}

\bibitem[\protect\citeauthoryear{{Henshaw} et~al.,}{{Henshaw}
  et~al.}{2020}]{2020NatAs...4.1064H}
{Henshaw} J.~D.,  et~al., 2020, \mn@doi [Nature Astronomy]
  {10.1038/s41550-020-1126-z}, \href
  {https://ui.adsabs.harvard.edu/abs/2020NatAs...4.1064H} {4, 1064}

\bibitem[\protect\citeauthoryear{{Hernquist}}{{Hernquist}}{1990}]{1990ApJ...356..359H}
{Hernquist} L.,  1990, \mn@doi [\apj] {10.1086/168845}, \href
  {https://ui.adsabs.harvard.edu/abs/1990ApJ...356..359H} {356, 359}

\bibitem[\protect\citeauthoryear{{Ho}, {Greene}, {Filippenko}  \&
  {Sargent}}{{Ho} et~al.}{2009}]{2009ApJS..183....1H}
{Ho} L.~C.,  {Greene} J.~E.,  {Filippenko} A.~V.,   {Sargent} W. L.~W.,  2009,
  \mn@doi [\apjs] {10.1088/0067-0049/183/1/1}, \href
  {https://ui.adsabs.harvard.edu/abs/2009ApJS..183....1H} {183, 1}

\bibitem[\protect\citeauthoryear{{Holwerda}, {Pirzkal}, {de Blok}, {Bouchard},
  {Blyth}, {van der Heyden}  \& {Elson}}{{Holwerda}
  et~al.}{2011a}]{2011MNRAS.416.2401H}
{Holwerda} B.~W.,  {Pirzkal} N.,  {de Blok} W.~J.~G.,  {Bouchard} A.,  {Blyth}
  S.~L.,  {van der Heyden} K.~J.,   {Elson} E.~C.,  2011a, \mn@doi [\mnras]
  {10.1111/j.1365-2966.2011.18938.x}, \href
  {https://ui.adsabs.harvard.edu/abs/2011MNRAS.416.2401H} {416, 2401}

\bibitem[\protect\citeauthoryear{{Holwerda}, {Pirzkal}, {de Blok}, {Bouchard},
  {Blyth}, {van der Heyden}  \& {Elson}}{{Holwerda}
  et~al.}{2011b}]{2011MNRAS.416.2415H}
{Holwerda} B.~W.,  {Pirzkal} N.,  {de Blok} W.~J.~G.,  {Bouchard} A.,  {Blyth}
  S.~L.,  {van der Heyden} K.~J.,   {Elson} E.~C.,  2011b, \mn@doi [\mnras]
  {10.1111/j.1365-2966.2011.17683.x}, \href
  {https://ui.adsabs.harvard.edu/abs/2011MNRAS.416.2415H} {416, 2415}

\bibitem[\protect\citeauthoryear{{Holwerda}, {Pirzkal}, {de Blok}  \&
  {Blyth}}{{Holwerda} et~al.}{2013}]{2013MNRAS.435.1020H}
{Holwerda} B.~W.,  {Pirzkal} N.,  {de Blok} W.~J.~G.,   {Blyth} S.~L.,  2013,
  \mn@doi [\mnras] {10.1093/mnras/stt1291}, \href
  {https://ui.adsabs.harvard.edu/abs/2013MNRAS.435.1020H} {435, 1020}

\bibitem[\protect\citeauthoryear{{Holwerda} et~al.,}{{Holwerda}
  et~al.}{2019}]{2019AJ....158..103H}
{Holwerda} B.~W.,  et~al., 2019, \mn@doi [\aj] {10.3847/1538-3881/ab2886},
  \href {https://ui.adsabs.harvard.edu/abs/2019AJ....158..103H} {158, 103}

\bibitem[\protect\citeauthoryear{{Hopkins}, {Hernquist}, {Cox}, {Di Matteo},
  {Robertson}  \& {Springel}}{{Hopkins} et~al.}{2006}]{2006ApJS..163....1H}
{Hopkins} P.~F.,  {Hernquist} L.,  {Cox} T.~J.,  {Di Matteo} T.,  {Robertson}
  B.,   {Springel} V.,  2006, \mn@doi [\apjs] {10.1086/499298}, \href
  {http://adsabs.harvard.edu/abs/2006ApJS..163....1H} {163, 1}

\bibitem[\protect\citeauthoryear{Hunter}{Hunter}{2007}]{Hunter:2007}
Hunter J.~D.,  2007, \mn@doi [Computing in Science \& Engineering]
  {10.1109/MCSE.2007.55}, 9, 90

\bibitem[\protect\citeauthoryear{{Jarrett}, {Chester}, {Cutri}, {Schneider}  \&
  {Huchra}}{{Jarrett} et~al.}{2003}]{2003AJ....125..525J}
{Jarrett} T.~H.,  {Chester} T.,  {Cutri} R.,  {Schneider} S.~E.,   {Huchra}
  J.~P.,  2003, \mn@doi [\aj] {10.1086/345794}, \href
  {https://ui.adsabs.harvard.edu/abs/2003AJ....125..525J} {125, 525}

\bibitem[\protect\citeauthoryear{{Krips} et~al.,}{{Krips}
  et~al.}{2005}]{2005A&A...442..479K}
{Krips} M.,  et~al., 2005, \mn@doi [\aap] {10.1051/0004-6361:20041731}, \href
  {https://ui.adsabs.harvard.edu/abs/2005A&A...442..479K} {442, 479}

\bibitem[\protect\citeauthoryear{{Kruijssen} et~al.,}{{Kruijssen}
  et~al.}{2019}]{2019MNRAS.484.5734K}
{Kruijssen} J.~M.~D.,  et~al., 2019, \mn@doi [\mnras] {10.1093/mnras/stz381},
  \href {https://ui.adsabs.harvard.edu/abs/2019MNRAS.484.5734K} {484, 5734}

\bibitem[\protect\citeauthoryear{{Krumholz}, {McKee}  \&
  {Tumlinson}}{{Krumholz} et~al.}{2009}]{2009ApJ...693..216K}
{Krumholz} M.~R.,  {McKee} C.~F.,   {Tumlinson} J.,  2009, \mn@doi [\apj]
  {10.1088/0004-637X/693/1/216}, \href
  {https://ui.adsabs.harvard.edu/abs/2009ApJ...693..216K} {693, 216}

\bibitem[\protect\citeauthoryear{{Lang} et~al.,}{{Lang}
  et~al.}{2020}]{2020ApJ...897..122L}
{Lang} P.,  et~al., 2020, \mn@doi [\apj] {10.3847/1538-4357/ab9953}, \href
  {https://ui.adsabs.harvard.edu/abs/2020ApJ...897..122L} {897, 122}

\bibitem[\protect\citeauthoryear{{Larson}}{{Larson}}{1981}]{1981MNRAS.194..809L}
{Larson} R.~B.,  1981, \mn@doi [\mnras] {10.1093/mnras/194.4.809}, \href
  {https://ui.adsabs.harvard.edu/abs/1981MNRAS.194..809L} {194, 809}

\bibitem[\protect\citeauthoryear{{Leigh}, {Georgiev}, {B{\"o}ker}, {Knigge}  \&
  {den Brok}}{{Leigh} et~al.}{2015}]{2015MNRAS.451..859L}
{Leigh} N. W.~C.,  {Georgiev} I.~Y.,  {B{\"o}ker} T.,  {Knigge} C.,   {den
  Brok} M.,  2015, \mn@doi [\mnras] {10.1093/mnras/stv1012}, \href
  {https://ui.adsabs.harvard.edu/abs/2015MNRAS.451..859L} {451, 859}

\bibitem[\protect\citeauthoryear{{Leroy}, {Walter}, {Brinks}, {Bigiel}, {de
  Blok}, {Madore}  \& {Thornley}}{{Leroy} et~al.}{2008}]{2008AJ....136.2782L}
{Leroy} A.~K.,  {Walter} F.,  {Brinks} E.,  {Bigiel} F.,  {de Blok} W.~J.~G.,
  {Madore} B.,   {Thornley} M.~D.,  2008, \mn@doi [\aj]
  {10.1088/0004-6256/136/6/2782}, \href
  {https://ui.adsabs.harvard.edu/abs/2008AJ....136.2782L} {136, 2782}

\bibitem[\protect\citeauthoryear{{Leroy} et~al.,}{{Leroy}
  et~al.}{2019}]{2019ApJS..244...24L}
{Leroy} A.~K.,  et~al., 2019, \mn@doi [\apjs] {10.3847/1538-4365/ab3925}, \href
  {https://ui.adsabs.harvard.edu/abs/2019ApJS..244...24L} {244, 24}

\bibitem[\protect\citeauthoryear{{Leroy} et~al.,}{{Leroy}
  et~al.}{2021a}]{2021arXiv210911583L}
{Leroy} A.~K.,  et~al., 2021a, arXiv e-prints, \href
  {https://ui.adsabs.harvard.edu/abs/2021arXiv210911583L} {p. arXiv:2109.11583}

\bibitem[\protect\citeauthoryear{{Leroy} et~al.,}{{Leroy}
  et~al.}{2021b}]{2021arXiv210407665L}
{Leroy} A.~K.,  et~al., 2021b, \mn@doi [\apjs] {10.3847/1538-4365/abec80},
  \href {https://ui.adsabs.harvard.edu/abs/2021ApJS..255...19L} {255, 19}

\bibitem[\protect\citeauthoryear{{Leroy} et~al.,}{{Leroy}
  et~al.}{2021c}]{2021arXiv210407739L}
{Leroy} A.~K.,  et~al., 2021c, \mn@doi [\apjs] {10.3847/1538-4365/ac17f3},
  \href {https://ui.adsabs.harvard.edu/abs/2021ApJS..257...43L} {257, 43}

\bibitem[\protect\citeauthoryear{{Levy} et~al.,}{{Levy}
  et~al.}{2018}]{2018ApJ...860...92L}
{Levy} R.~C.,  et~al., 2018, \mn@doi [\apj] {10.3847/1538-4357/aac2e5}, \href
  {http://adsabs.harvard.edu/abs/2018ApJ...860...92L} {860, 92}

\bibitem[\protect\citeauthoryear{{Lewis} \& {Eracleous}}{{Lewis} \&
  {Eracleous}}{2006}]{2006ApJ...642..711L}
{Lewis} K.~T.,  {Eracleous} M.,  2006, \mn@doi [\apj] {10.1086/501419}, \href
  {https://ui.adsabs.harvard.edu/abs/2006ApJ...642..711L} {642, 711}

\bibitem[\protect\citeauthoryear{{Liu}, {Bureau}, {Blitz}, {Davis}, {Onishi},
  {Smith}, {North}  \& {Iguchi}}{{Liu} et~al.}{2021}]{2021arXiv210604327L}
{Liu} L.,  {Bureau} M.,  {Blitz} L.,  {Davis} T.~A.,  {Onishi} K.,  {Smith} M.,
   {North} E.,   {Iguchi} S.,  2021, \mn@doi [\mnras] {10.1093/mnras/stab1537},
  \href {https://ui.adsabs.harvard.edu/abs/2021MNRAS.505.4048L} {505, 4048}

\bibitem[\protect\citeauthoryear{{Longmore} et~al.,}{{Longmore}
  et~al.}{2013}]{2013MNRAS.429..987L}
{Longmore} S.~N.,  et~al., 2013, \mn@doi [\mnras] {10.1093/mnras/sts376}, \href
  {https://ui.adsabs.harvard.edu/abs/2013MNRAS.429..987L} {429, 987}

\bibitem[\protect\citeauthoryear{{Ma}, {Greene}, {McConnell}, {Janish},
  {Blakeslee}, {Thomas}  \& {Murphy}}{{Ma} et~al.}{2014}]{2014ApJ...795..158M}
{Ma} C.-P.,  {Greene} J.~E.,  {McConnell} N.,  {Janish} R.,  {Blakeslee} J.~P.,
   {Thomas} J.,   {Murphy} J.~D.,  2014, \mn@doi [ApJ]
  {10.1088/0004-637X/795/2/158}, \href
  {http://adsabs.harvard.edu/abs/2014ApJ...795..158M} {795, 158}

\bibitem[\protect\citeauthoryear{{Mac Low} \& {Klessen}}{{Mac Low} \&
  {Klessen}}{2004}]{2004RvMP...76..125M}
{Mac Low} M.-M.,  {Klessen} R.~S.,  2004, \mn@doi [Reviews of Modern Physics]
  {10.1103/RevModPhys.76.125}, \href
  {https://ui.adsabs.harvard.edu/abs/2004RvMP...76..125M} {76, 125}

\bibitem[\protect\citeauthoryear{{Makarov}, {Prugniel}, {Terekhova}, {Courtois}
   \& {Vauglin}}{{Makarov} et~al.}{2014}]{2014A&A...570A..13M}
{Makarov} D.,  {Prugniel} P.,  {Terekhova} N.,  {Courtois} H.,   {Vauglin} I.,
  2014, \mn@doi [\aap] {10.1051/0004-6361/201423496}, \href
  {https://ui.adsabs.harvard.edu/abs/2014A&A...570A..13M} {570, A13}

\bibitem[\protect\citeauthoryear{{Martig}, {Bournaud}, {Teyssier}  \&
  {Dekel}}{{Martig} et~al.}{2009}]{2009ApJ...707..250M}
{Martig} M.,  {Bournaud} F.,  {Teyssier} R.,   {Dekel} A.,  2009, \mn@doi
  [\apj] {10.1088/0004-637X/707/1/250}, \href
  {https://ui.adsabs.harvard.edu/abs/2009ApJ...707..250M} {707, 250}

\bibitem[\protect\citeauthoryear{Martig et~al.,}{Martig
  et~al.}{2013}]{2013MNRAS.432.1914M}
Martig M.,  et~al., 2013, \mnras, 432, 1914

\bibitem[\protect\citeauthoryear{{McMullin}, {Waters}, {Schiebel}, {Young}  \&
  {Golap}}{{McMullin} et~al.}{2007}]{2007ASPC..376..127M}
{McMullin} J.~P.,  {Waters} B.,  {Schiebel} D.,  {Young} W.,   {Golap} K.,
  2007, in {Shaw} R.~A.,  {Hill} F.,   {Bell} D.~J.,  eds,  Astronomical
  Society of the Pacific Conference Series Vol. 376, Astronomical Data Analysis
  Software and Systems XVI. p.~127

\bibitem[\protect\citeauthoryear{{Mei} et~al.,}{{Mei}
  et~al.}{2007}]{2007ApJ...655..144M}
{Mei} S.,  et~al., 2007, \mn@doi [\apj] {10.1086/509598}, \href
  {https://ui.adsabs.harvard.edu/abs/2007ApJ...655..144M} {655, 144}

\bibitem[\protect\citeauthoryear{{Meidt} et~al.,}{{Meidt}
  et~al.}{2018}]{2018ApJ...854..100M}
{Meidt} S.~E.,  et~al., 2018, \mn@doi [\apj] {10.3847/1538-4357/aaa290}, \href
  {https://ui.adsabs.harvard.edu/abs/2018ApJ...854..100M} {854, 100}

\bibitem[\protect\citeauthoryear{{Meidt} et~al.,}{{Meidt}
  et~al.}{2020}]{2020ApJ...892...73M}
{Meidt} S.~E.,  et~al., 2020, \mn@doi [\apj] {10.3847/1538-4357/ab7000}, \href
  {https://ui.adsabs.harvard.edu/abs/2020ApJ...892...73M} {892, 73}

\bibitem[\protect\citeauthoryear{{Meidt} et~al.,}{{Meidt}
  et~al.}{2021}]{2021ApJ...913..113M}
{Meidt} S.~E.,  et~al., 2021, \mn@doi [\apj] {10.3847/1538-4357/abf35b}, \href
  {https://ui.adsabs.harvard.edu/abs/2021ApJ...913..113M} {913, 113}

\bibitem[\protect\citeauthoryear{{Newnham}, {Hess}, {Masters}, {Kruk}, {Penny},
  {Lingard}  \& {Smethurst}}{{Newnham} et~al.}{2020}]{2020MNRAS.492.4697N}
{Newnham} L.,  {Hess} K.~M.,  {Masters} K.~L.,  {Kruk} S.,  {Penny} S.~J.,
  {Lingard} T.,   {Smethurst} R.~J.,  2020, \mn@doi [\mnras]
  {10.1093/mnras/staa064}, \href
  {https://ui.adsabs.harvard.edu/abs/2020MNRAS.492.4697N} {492, 4697}

\bibitem[\protect\citeauthoryear{{North} et~al.,}{{North}
  et~al.}{2019}]{2019MNRAS.490..319N}
{North} E.~V.,  et~al., 2019, \mn@doi [\mnras] {10.1093/mnras/stz2598}, \href
  {https://ui.adsabs.harvard.edu/abs/2019MNRAS.490..319N} {490, 319}

\bibitem[\protect\citeauthoryear{{North} et~al.,}{{North}
  et~al.}{2021}]{2021MNRAS.503.5179N}
{North} E.~V.,  et~al., 2021, \mn@doi [\mnras] {10.1093/mnras/stab793}, \href
  {https://ui.adsabs.harvard.edu/abs/2021MNRAS.503.5179N} {503, 5179}

\bibitem[\protect\citeauthoryear{{Onishi}, {Iguchi}, {Davis}, {Bureau},
  {Cappellari}, {Sarzi}  \& {Blitz}}{{Onishi}
  et~al.}{2017}]{2017MNRAS.468.4663O}
{Onishi} K.,  {Iguchi} S.,  {Davis} T.~A.,  {Bureau} M.,  {Cappellari} M.,
  {Sarzi} M.,   {Blitz} L.,  2017, \mn@doi [\mnras] {10.1093/mnras/stx631},
  \href {https://ui.adsabs.harvard.edu/abs/2017MNRAS.468.4663O} {468, 4663}

\bibitem[\protect\citeauthoryear{{Pagotto}, {Corsini}, {Sarzi}, {Pagani},
  {Dalla Bont{\`a}}, {Morelli}  \& {Pizzella}}{{Pagotto}
  et~al.}{2019}]{2019MNRAS.483...57P}
{Pagotto} I.,  {Corsini} E.~M.,  {Sarzi} M.,  {Pagani} B.,  {Dalla Bont{\`a}}
  E.,  {Morelli} L.,   {Pizzella} A.,  2019, \mn@doi [\mnras]
  {10.1093/mnras/sty2918}, \href
  {https://ui.adsabs.harvard.edu/abs/2019MNRAS.483...57P} {483, 57}

\bibitem[\protect\citeauthoryear{{Patra}}{{Patra}}{2019}]{2019MNRAS.484...81P}
{Patra} N.~N.,  2019, \mn@doi [\mnras] {10.1093/mnras/sty3493}, \href
  {https://ui.adsabs.harvard.edu/abs/2019MNRAS.484...81P} {484, 81}

\bibitem[\protect\citeauthoryear{Pedregosa et~al.,}{Pedregosa
  et~al.}{2011}]{scikit-learn}
Pedregosa F.,  et~al., 2011, Journal of Machine Learning Research, 12, 2825

\bibitem[\protect\citeauthoryear{{Perez}, {Casassus}, {Cort{\'e}s}  \&
  {Kenney}}{{Perez} et~al.}{2009}]{2009MNRAS.400.2098P}
{Perez} S.,  {Casassus} S.,  {Cort{\'e}s} J.~R.,   {Kenney} J. D.~P.,  2009,
  \mn@doi [\mnras] {10.1111/j.1365-2966.2009.15603.x}, \href
  {https://ui.adsabs.harvard.edu/abs/2009MNRAS.400.2098P} {400, 2098}

\bibitem[\protect\citeauthoryear{{Popping}, {Somerville}  \&
  {Trager}}{{Popping} et~al.}{2014}]{2014MNRAS.442.2398P}
{Popping} G.,  {Somerville} R.~S.,   {Trager} S.~C.,  2014, \mn@doi [\mnras]
  {10.1093/mnras/stu991}, \href
  {https://ui.adsabs.harvard.edu/abs/2014MNRAS.442.2398P} {442, 2398}

\bibitem[\protect\citeauthoryear{{Querejeta} et~al.,}{{Querejeta}
  et~al.}{2021}]{2021arXiv210904491Q}
{Querejeta} M.,  et~al., 2021, arXiv e-prints, \href
  {https://ui.adsabs.harvard.edu/abs/2021arXiv210904491Q} {p. arXiv:2109.04491}

\bibitem[\protect\citeauthoryear{{Regan} \& {Mulchaey}}{{Regan} \&
  {Mulchaey}}{1999}]{1999AJ....117.2676R}
{Regan} M.~W.,  {Mulchaey} J.~S.,  1999, \mn@doi [\aj] {10.1086/300888}, \href
  {https://ui.adsabs.harvard.edu/abs/1999AJ....117.2676R} {117, 2676}

\bibitem[\protect\citeauthoryear{{Rodriguez-Gomez} et~al.,}{{Rodriguez-Gomez}
  et~al.}{2019}]{2019MNRAS.483.4140R}
{Rodriguez-Gomez} V.,  et~al., 2019, \mn@doi [\mnras] {10.1093/mnras/sty3345},
  \href {https://ui.adsabs.harvard.edu/abs/2019MNRAS.483.4140R} {483, 4140}

\bibitem[\protect\citeauthoryear{{Romeo} \& {Wiegert}}{{Romeo} \&
  {Wiegert}}{2011}]{2011MNRAS.416.1191R}
{Romeo} A.~B.,  {Wiegert} J.,  2011, \mn@doi [\mnras]
  {10.1111/j.1365-2966.2011.19120.x}, \href
  {https://ui.adsabs.harvard.edu/abs/2011MNRAS.416.1191R} {416, 1191}

\bibitem[\protect\citeauthoryear{{Rosolowsky}}{{Rosolowsky}}{2007}]{2007ApJ...654..240R}
{Rosolowsky} E.,  2007, \mn@doi [\apj] {10.1086/509249}, \href
  {https://ui.adsabs.harvard.edu/abs/2007ApJ...654..240R} {654, 240}

\bibitem[\protect\citeauthoryear{{Ruffa} et~al.,}{{Ruffa}
  et~al.}{2019}]{2019MNRAS.484.4239R}
{Ruffa} I.,  et~al., 2019, \mn@doi [\mnras] {10.1093/mnras/stz255}, \href
  {https://ui.adsabs.harvard.edu/abs/2019MNRAS.484.4239R} {484, 4239}

\bibitem[\protect\citeauthoryear{{Saintonge} et~al.,}{{Saintonge}
  et~al.}{2011}]{2011MNRAS.415...61S}
{Saintonge} A.,  et~al., 2011, \mn@doi [\mnras]
  {10.1111/j.1365-2966.2011.18823.x}, \href
  {http://adsabs.harvard.edu/abs/2011MNRAS.415...61S} {415, 61}

\bibitem[\protect\citeauthoryear{{Schade}, {Lilly}, {Crampton}, {Hammer}, {Le
  Fevre}  \& {Tresse}}{{Schade} et~al.}{1995}]{1995ApJ...451L...1S}
{Schade} D.,  {Lilly} S.~J.,  {Crampton} D.,  {Hammer} F.,  {Le Fevre} O.,
  {Tresse} L.,  1995, \mn@doi [\apjl] {10.1086/309677}, \href
  {https://ui.adsabs.harvard.edu/abs/1995ApJ...451L...1S} {451, L1}

\bibitem[\protect\citeauthoryear{{Seidel} et~al.,}{{Seidel}
  et~al.}{2015}]{2015MNRAS.446.2837S}
{Seidel} M.~K.,  et~al., 2015, \mn@doi [\mnras] {10.1093/mnras/stu2295}, \href
  {https://ui.adsabs.harvard.edu/abs/2015MNRAS.446.2837S} {446, 2837}

\bibitem[\protect\citeauthoryear{{Smaji{\'c}}, {Fischer}, {Zuther}  \&
  {Eckart}}{{Smaji{\'c}} et~al.}{2012}]{2012A&A...544A.105S}
{Smaji{\'c}} S.,  {Fischer} S.,  {Zuther} J.,   {Eckart} A.,  2012, \mn@doi
  [\aap] {10.1051/0004-6361/201118256}, \href
  {https://ui.adsabs.harvard.edu/abs/2012A&A...544A.105S} {544, A105}

\bibitem[\protect\citeauthoryear{{Smith} et~al.,}{{Smith}
  et~al.}{2019}]{2019MNRAS.485.4359S}
{Smith} M.~D.,  et~al., 2019, \mn@doi [\mnras] {10.1093/mnras/stz625}, \href
  {https://ui.adsabs.harvard.edu/abs/2019MNRAS.485.4359S} {485, 4359}

\bibitem[\protect\citeauthoryear{{Smith} et~al.,}{{Smith}
  et~al.}{2021}]{2021MNRAS.503.5984S}
{Smith} M.~D.,  et~al., 2021, \mn@doi [\mnras] {10.1093/mnras/stab791}, \href
  {https://ui.adsabs.harvard.edu/abs/2021MNRAS.503.5984S} {503, 5984}

\bibitem[\protect\citeauthoryear{{Springel}}{{Springel}}{2010}]{2010MNRAS.401..791S}
{Springel} V.,  2010, \mn@doi [\mnras] {10.1111/j.1365-2966.2009.15715.x},
  \href {https://ui.adsabs.harvard.edu/abs/2010MNRAS.401..791S} {401, 791}

\bibitem[\protect\citeauthoryear{{Steer} et~al.,}{{Steer}
  et~al.}{2017}]{2017AJ....153...37S}
{Steer} I.,  et~al., 2017, \mn@doi [\aj] {10.3847/1538-3881/153/1/37}, \href
  {https://ui.adsabs.harvard.edu/abs/2017AJ....153...37S} {153, 37}

\bibitem[\protect\citeauthoryear{{Stuber} et~al.,}{{Stuber}
  et~al.}{2021}]{2021A&A...653A.172S}
{Stuber} S.~K.,  et~al., 2021, \mn@doi [\aap] {10.1051/0004-6361/202141093},
  \href {https://ui.adsabs.harvard.edu/abs/2021A&A...653A.172S} {653, A172}

\bibitem[\protect\citeauthoryear{{Sun} et~al.,}{{Sun}
  et~al.}{2020}]{2020ApJ...901L...8S}
{Sun} J.,  et~al., 2020, \mn@doi [\apjl] {10.3847/2041-8213/abb3be}, \href
  {https://ui.adsabs.harvard.edu/abs/2020ApJ...901L...8S} {901, L8}

\bibitem[\protect\citeauthoryear{{Tonry}, {Dressler}, {Blakeslee}, {Ajhar},
  {Fletcher}, {Luppino}, {Metzger}  \& {Moore}}{{Tonry}
  et~al.}{2001}]{2001ApJ...546..681T}
{Tonry} J.~L.,  {Dressler} A.,  {Blakeslee} J.~P.,  {Ajhar} E.~A.,  {Fletcher}
  A.~B.,  {Luppino} G.~A.,  {Metzger} M.~R.,   {Moore} C.~B.,  2001, \mn@doi
  [\apj] {10.1086/318301}, \href
  {https://ui.adsabs.harvard.edu/abs/2001ApJ...546..681T} {546, 681}

\bibitem[\protect\citeauthoryear{{Toomre}}{{Toomre}}{1964}]{1964ApJ...139.1217T}
{Toomre} A.,  1964, \mn@doi [\apj] {10.1086/147861}, \href
  {https://ui.adsabs.harvard.edu/abs/1964ApJ...139.1217T} {139, 1217}

\bibitem[\protect\citeauthoryear{{Utomo}, {Blitz}, {Davis}, {Rosolowsky},
  {Bureau}, {Cappellari}  \& {Sarzi}}{{Utomo}
  et~al.}{2015}]{2015ApJ...803...16U}
{Utomo} D.,  {Blitz} L.,  {Davis} T.,  {Rosolowsky} E.,  {Bureau} M.,
  {Cappellari} M.,   {Sarzi} M.,  2015, \mn@doi [\apj]
  {10.1088/0004-637X/803/1/16}, \href
  {https://ui.adsabs.harvard.edu/abs/2015ApJ...803...16U} {803, 16}

\bibitem[\protect\citeauthoryear{{Veale}, {Ma}, {Greene}, {Thomas},
  {Blakeslee}, {McConnell}, {Walsh}  \& {Ito}}{{Veale}
  et~al.}{2017}]{2017MNRAS.471.1428V}
{Veale} M.,  {Ma} C.-P.,  {Greene} J.~E.,  {Thomas} J.,  {Blakeslee} J.~P.,
  {McConnell} N.,  {Walsh} J.~L.,   {Ito} J.,  2017, \mn@doi [\mnras]
  {10.1093/mnras/stx1639}, \href
  {http://adsabs.harvard.edu/abs/2017MNRAS.471.1428V} {471, 1428}

\bibitem[\protect\citeauthoryear{{Vega Beltr{\'a}n}, {Pizzella}, {Corsini},
  {Funes}, {Zeilinger}, {Beckman}  \& {Bertola}}{{Vega Beltr{\'a}n}
  et~al.}{2001}]{2001A&A...374..394V}
{Vega Beltr{\'a}n} J.~C.,  {Pizzella} A.,  {Corsini} E.~M.,  {Funes} J.~G.,
  {Zeilinger} W.~W.,  {Beckman} J.~E.,   {Bertola} F.,  2001, \mn@doi [\aap]
  {10.1051/0004-6361:20010625}, \href
  {https://ui.adsabs.harvard.edu/abs/2001A&A...374..394V} {374, 394}

\bibitem[\protect\citeauthoryear{{V{\'e}ron-Cetty} \&
  {V{\'e}ron}}{{V{\'e}ron-Cetty} \& {V{\'e}ron}}{2010}]{2010A&A...518A..10V}
{V{\'e}ron-Cetty} M.~P.,  {V{\'e}ron} P.,  2010, \mn@doi [\aap]
  {10.1051/0004-6361/201014188}, \href
  {https://ui.adsabs.harvard.edu/abs/2010A&A...518A..10V} {518, A10}

\bibitem[\protect\citeauthoryear{Virtanen et~al.,}{Virtanen
  et~al.}{2020}]{2020SciPy-NMeth}
Virtanen P.,  et~al., 2020, \mn@doi [Nature Methods]
  {10.1038/s41592-019-0686-2}, \href {https://rdcu.be/b08Wh} {17, 261}

\bibitem[\protect\citeauthoryear{{Yoon}, {Park}, {Chung}  \& {Zhang}}{{Yoon}
  et~al.}{2021}]{2021arXiv211006033Y}
{Yoon} Y.,  {Park} C.,  {Chung} H.,   {Zhang} K.,  2021, \mn@doi [\apj]
  {10.3847/1538-4357/ac2302}, \href
  {https://ui.adsabs.harvard.edu/abs/2021ApJ...922..249Y} {922, 249}

\bibitem[\protect\citeauthoryear{{Young} et~al.,}{{Young}
  et~al.}{2011}]{2011MNRAS.414..940Y}
{Young} L.~M.,  et~al., 2011, \mn@doi [\mnras]
  {10.1111/j.1365-2966.2011.18561.x}, \href
  {http://adsabs.harvard.edu/abs/2011MNRAS.414..940Y} {414, 940}

\bibitem[\protect\citeauthoryear{{van den Bosch}, {Gebhardt}, {G{\"u}ltekin},
  {Y{\i}ld{\i}r{\i}m}  \& {Walsh}}{{van den Bosch}
  et~al.}{2015}]{2015ApJS..218...10V}
{van den Bosch} R. C.~E.,  {Gebhardt} K.,  {G{\"u}ltekin} K.,
  {Y{\i}ld{\i}r{\i}m} A.,   {Walsh} J.~L.,  2015, \mn@doi [\apjs]
  {10.1088/0067-0049/218/1/10}, \href
  {https://ui.adsabs.harvard.edu/abs/2015ApJS..218...10V} {218, 10}

\bibitem[\protect\citeauthoryear{{van der Laan} et~al.,}{{van der Laan}
  et~al.}{2013}]{2013A&A...556A..98V}
{van der Laan} T.~P.~R.,  et~al., 2013, \mn@doi [\aap]
  {10.1051/0004-6361/201321572}, \href
  {https://ui.adsabs.harvard.edu/abs/2013A&A...556A..98V} {556, A98}

\makeatother
\end{thebibliography}
\bibdata{bibMASSIVE_smbh.bib}
\bibstyle{mnras}

\label{lastpage}
% online appendix material
\appendix
\clearpage
\section{Integrated Intensity maps}
\label{allmaps}

\begin{figure*}
\begin{tabular}{ccc}
\subfloat[FRL49]{\includegraphics[height=5cm,trim=0cm 0cm 0cm 0cm,clip]{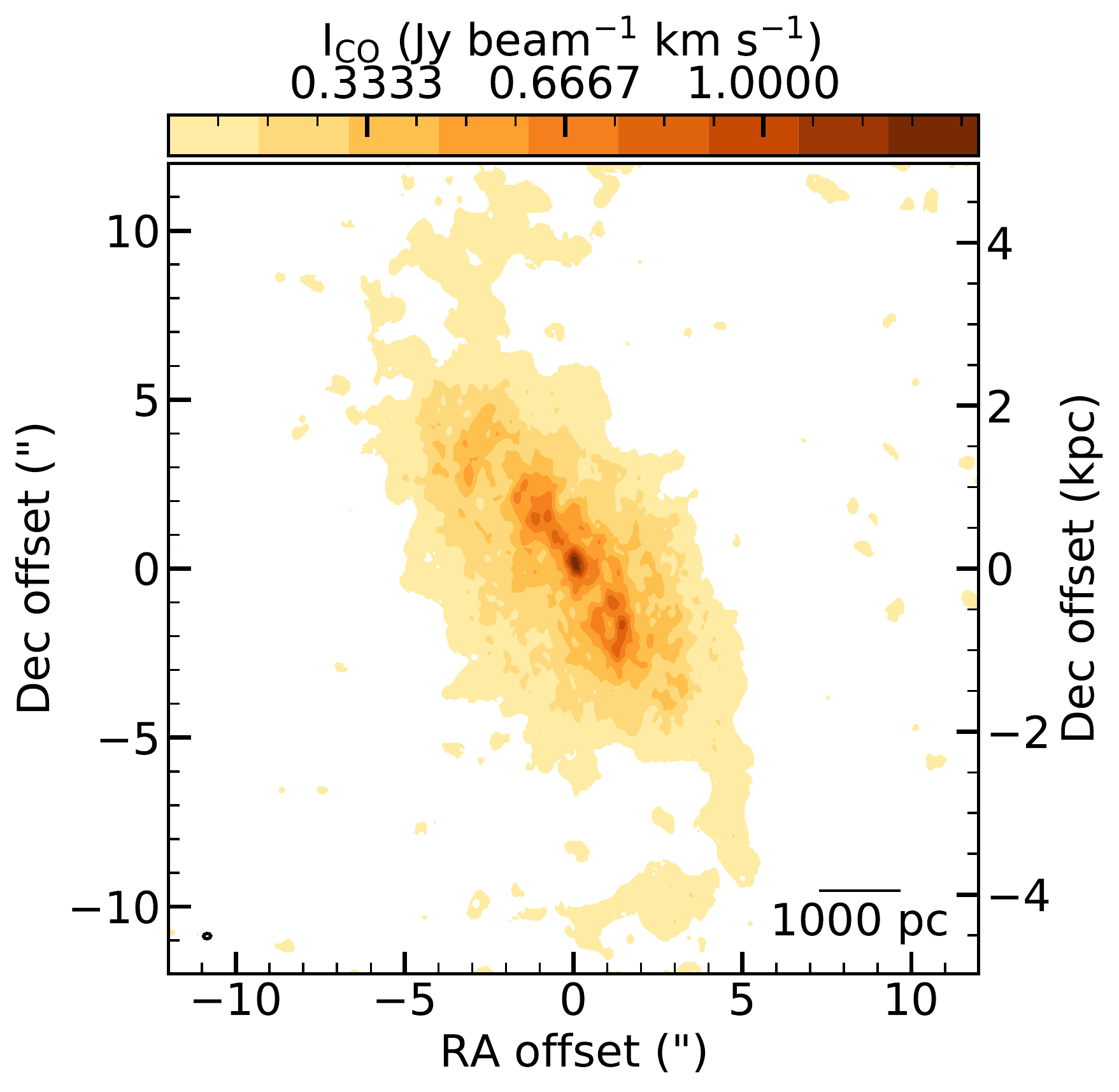}} &
\subfloat[NGC0383]{\includegraphics[height=5cm,trim=0cm 0cm 0cm 0cm,clip]{figs/NGC0383_moment0.pdf}} &
\subfloat[NGC0524]{\includegraphics[height=5cm,trim=0cm 0cm 0cm 0cm,clip]{figs/NGC0524_moment0.pdf}} \\
\subfloat[NGC0708]{\includegraphics[height=5cm,trim=0cm 0cm 0cm 0cm,clip]{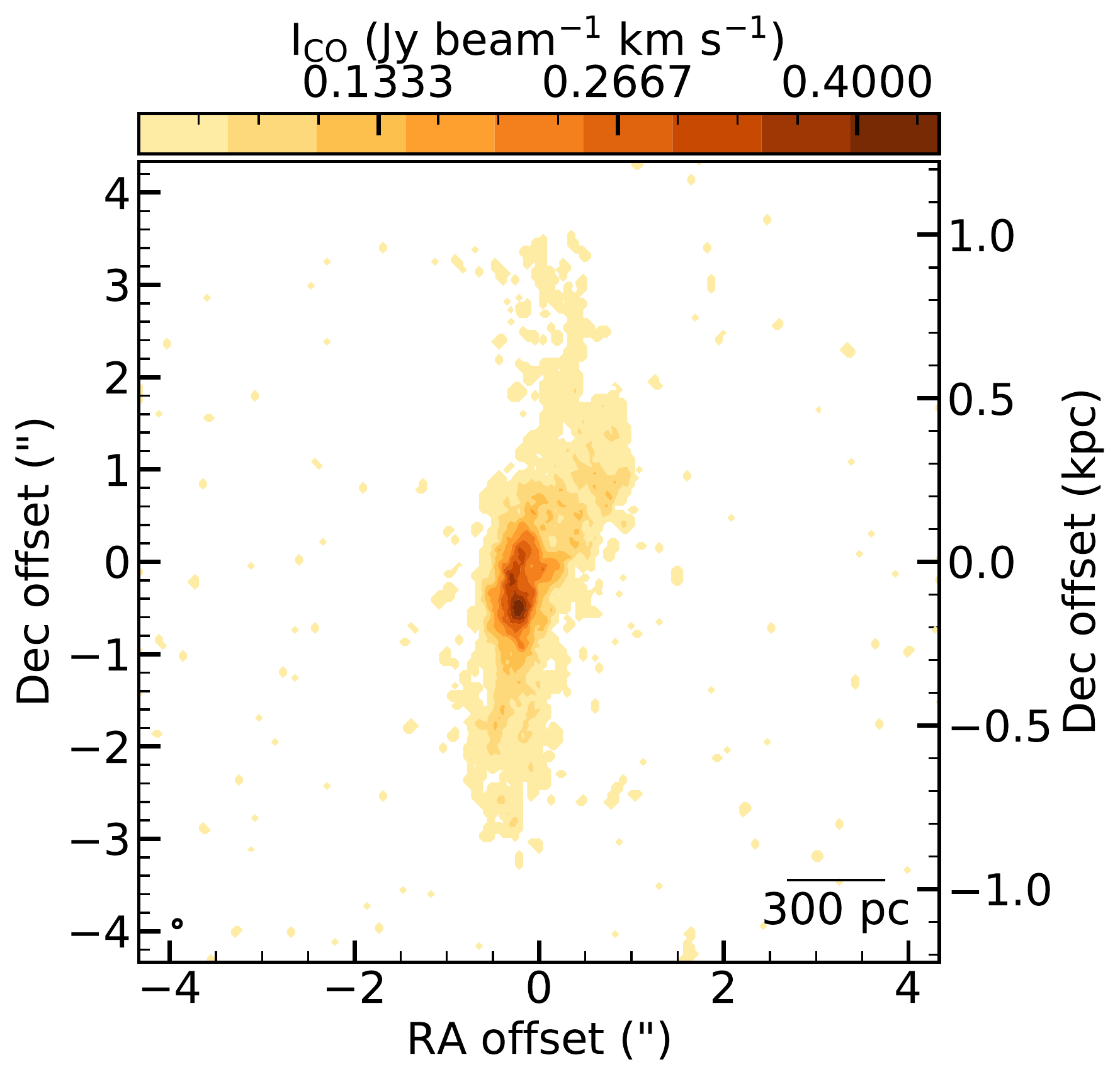}} &
\subfloat[NGC1387]{\includegraphics[height=5cm,trim=0cm 0cm 0cm 0cm,clip]{figs/NGC1387_moment0.pdf}} &
\subfloat[NGC1574]{\includegraphics[height=5cm,trim=0cm 0cm 0cm 0cm,clip]{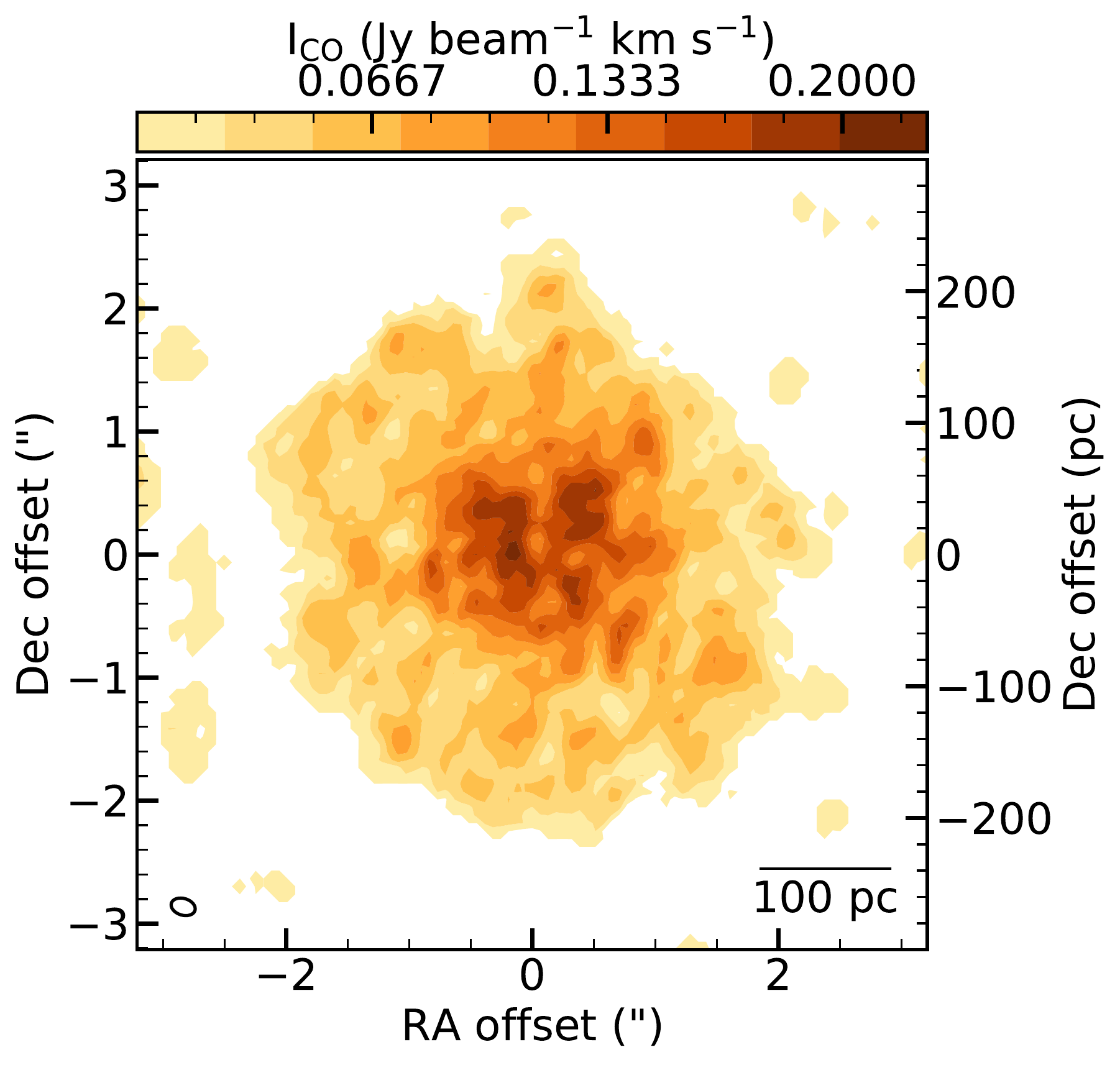}} \\
\subfloat[NGC3607]{\includegraphics[height=5cm,trim=0cm 0cm 0cm 0cm,clip]{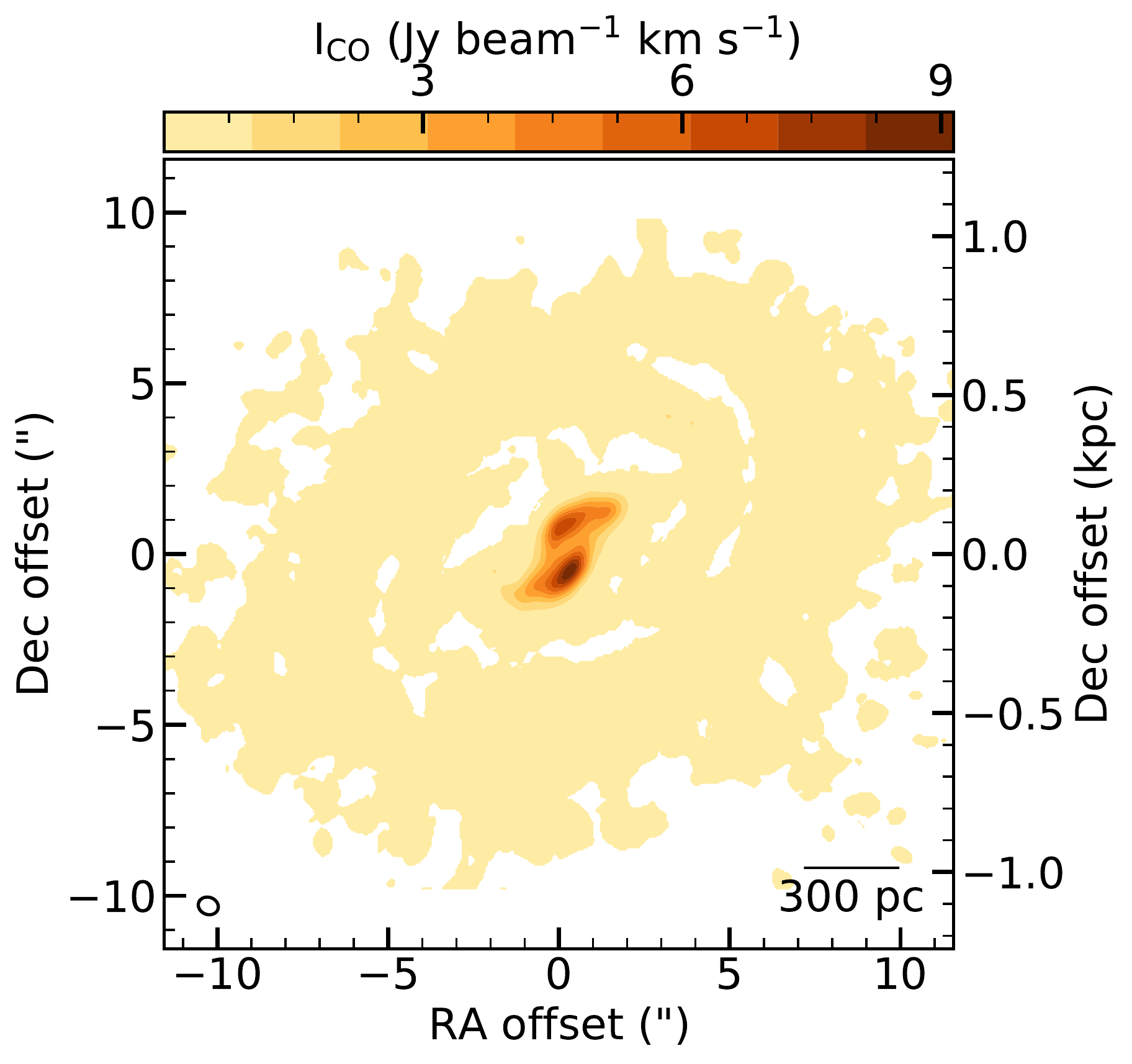}} &
\subfloat[NGC4061]{\includegraphics[height=5cm,trim=0cm 0cm 0cm 0cm,clip]{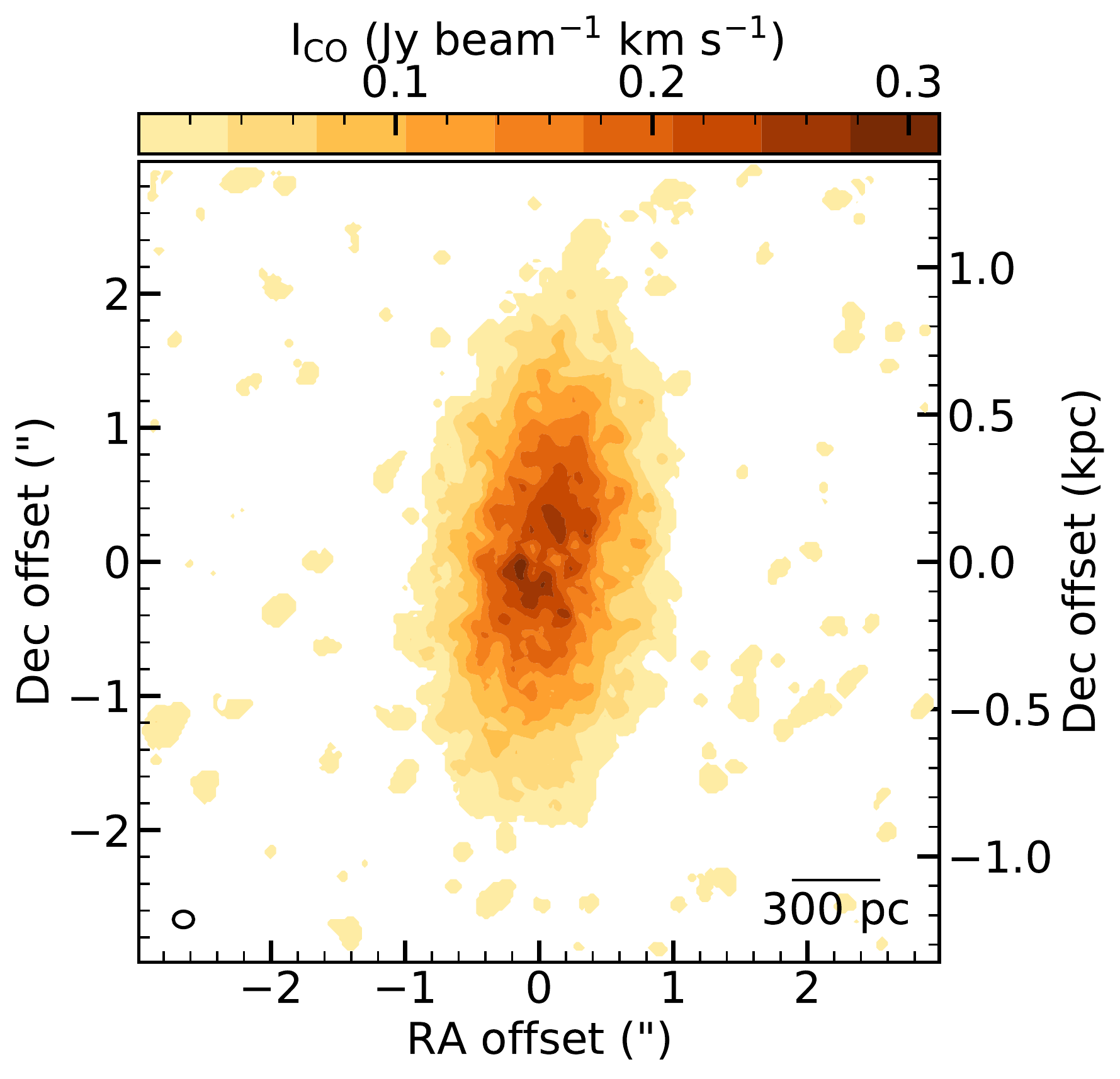}} &
\subfloat[NGC4429]{\includegraphics[height=5cm,trim=0cm 0cm 0cm 0cm,clip]{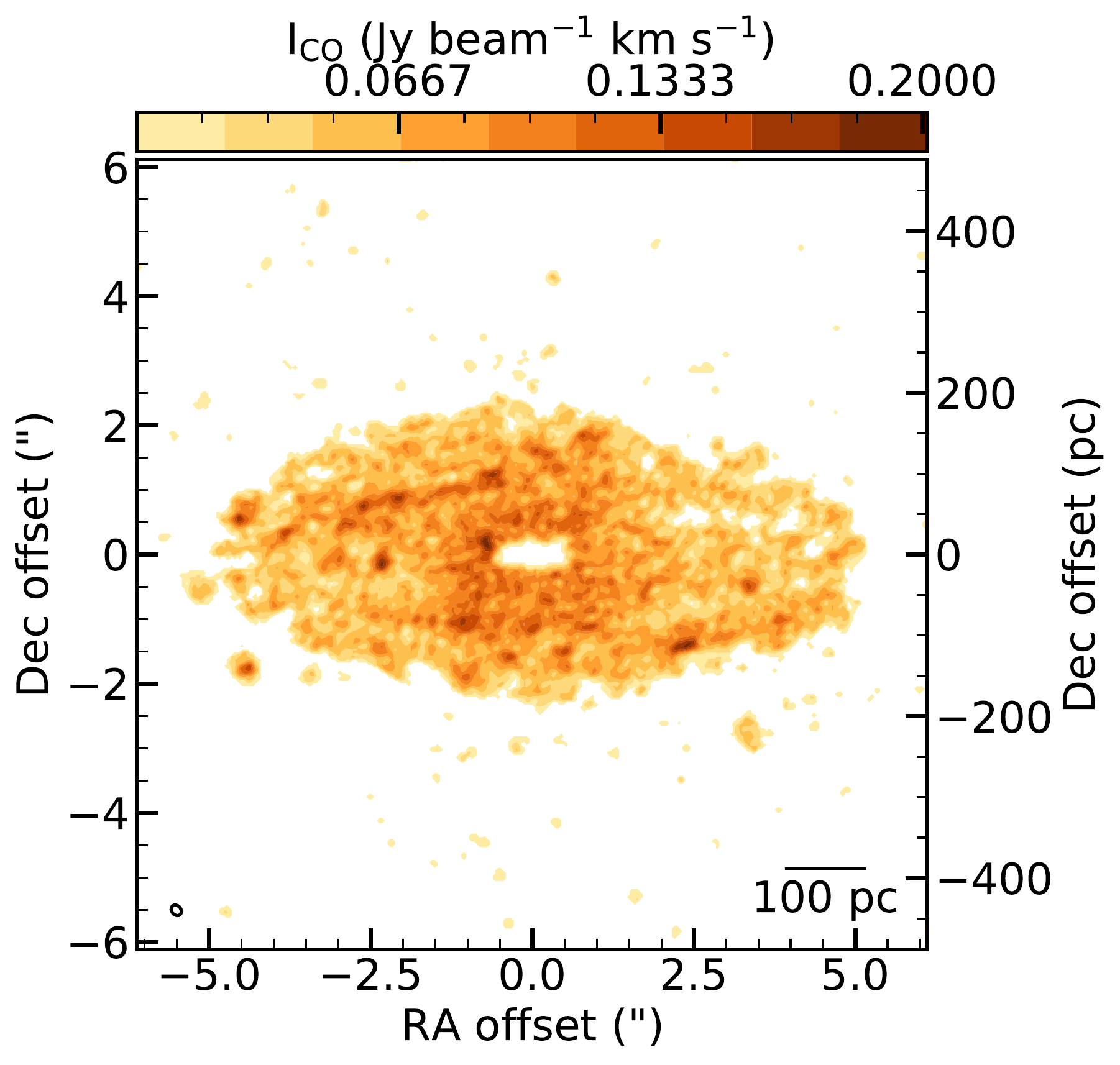}} 
\end{tabular}
\caption{Integrated-intensity maps of the CO(2-1) or CO(3-2) transition for early-type galaxies in the WISDOM survey.} 
\label{mom0s_etgs}
\end{figure*}
\begin{figure*}
\begin{tabular}{ccc}
\subfloat[NGC4435]{\includegraphics[height=5cm,trim=0cm 0cm 0cm 0cm,clip]{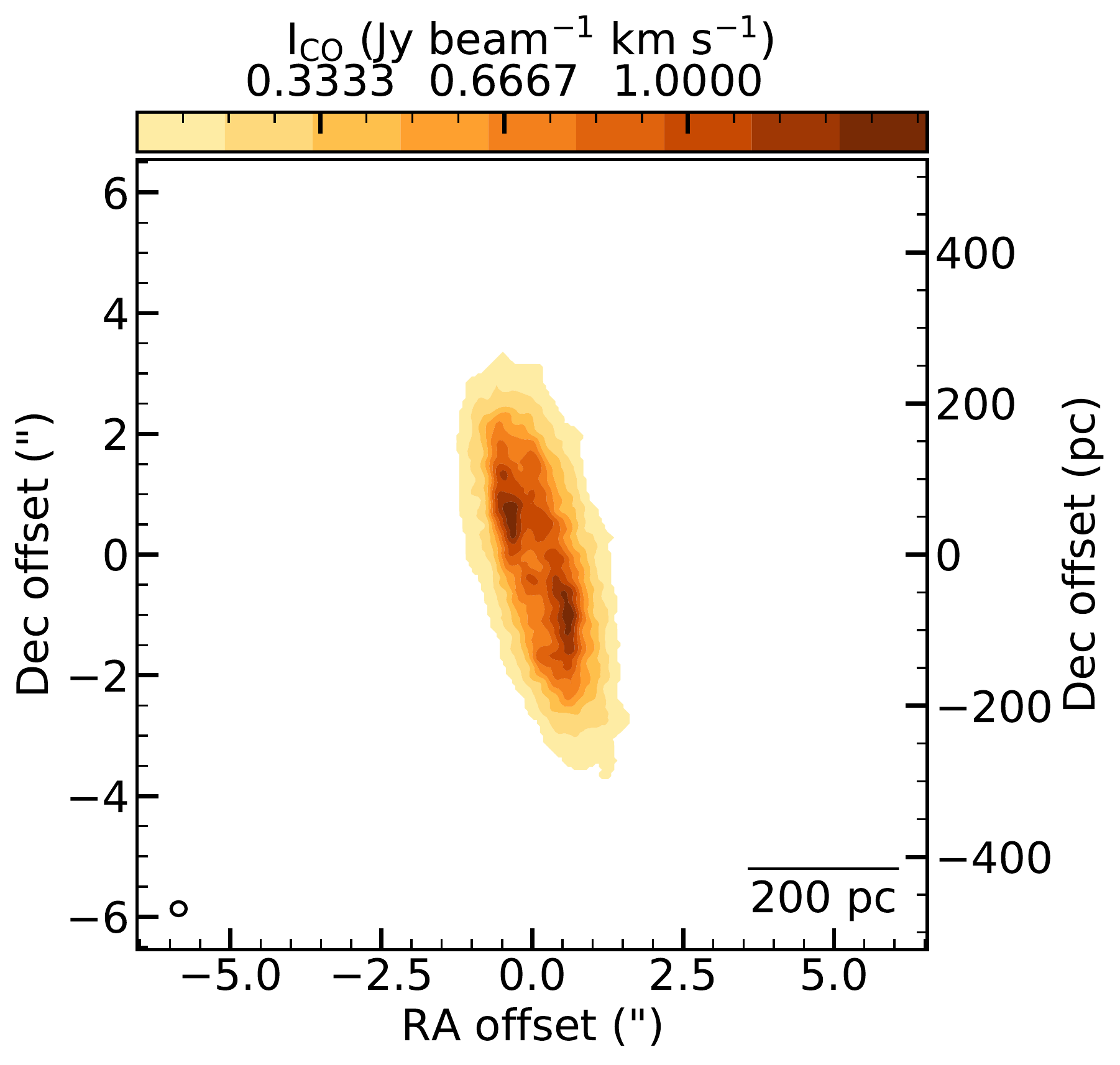}} &
\subfloat[NGC4697]{\includegraphics[height=5cm,trim=0cm 0cm 0cm 0cm,clip]{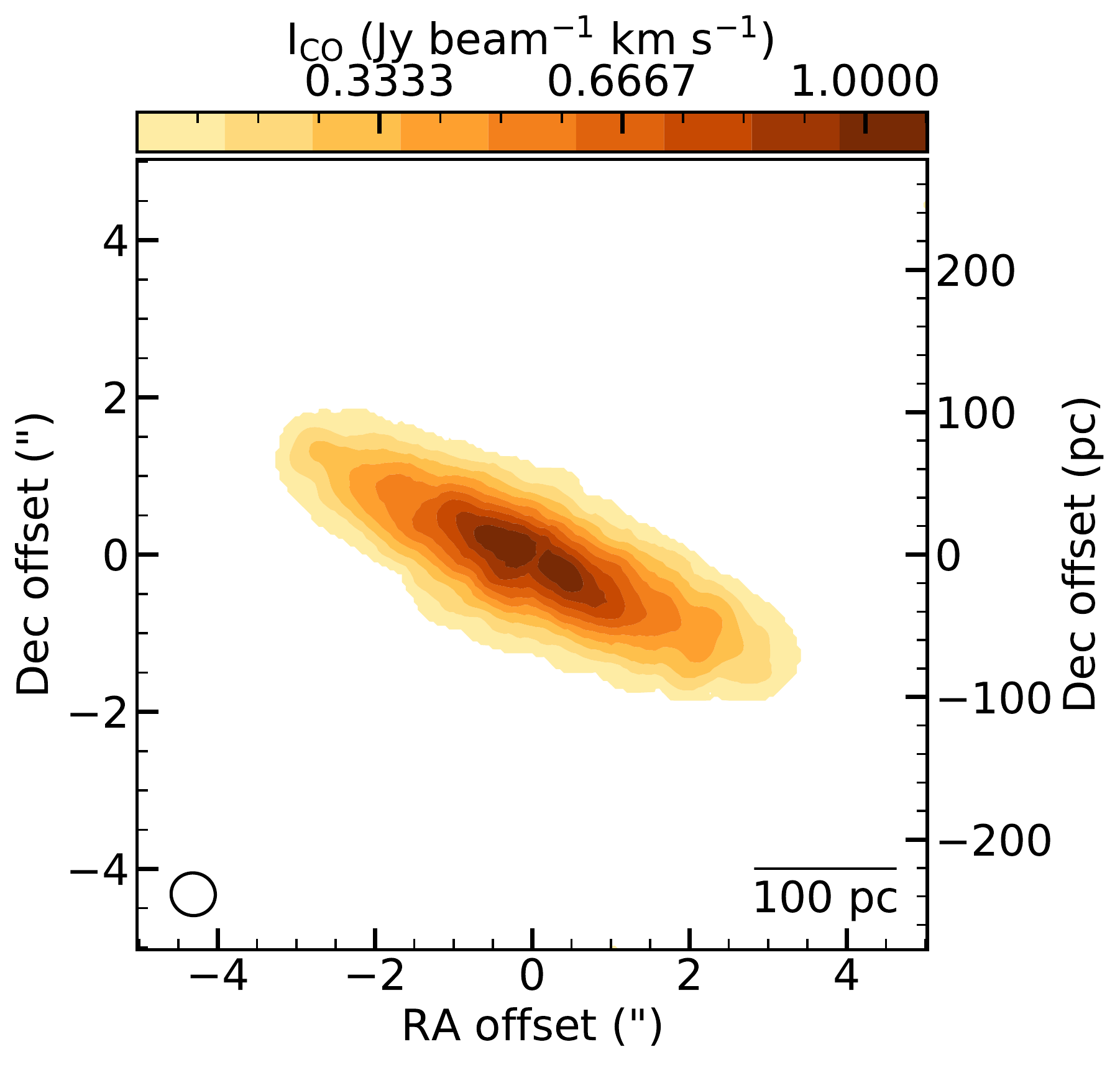}} &
\subfloat[NGC6958]{\includegraphics[height=5cm,trim=0cm 0cm 0cm 0cm,clip]{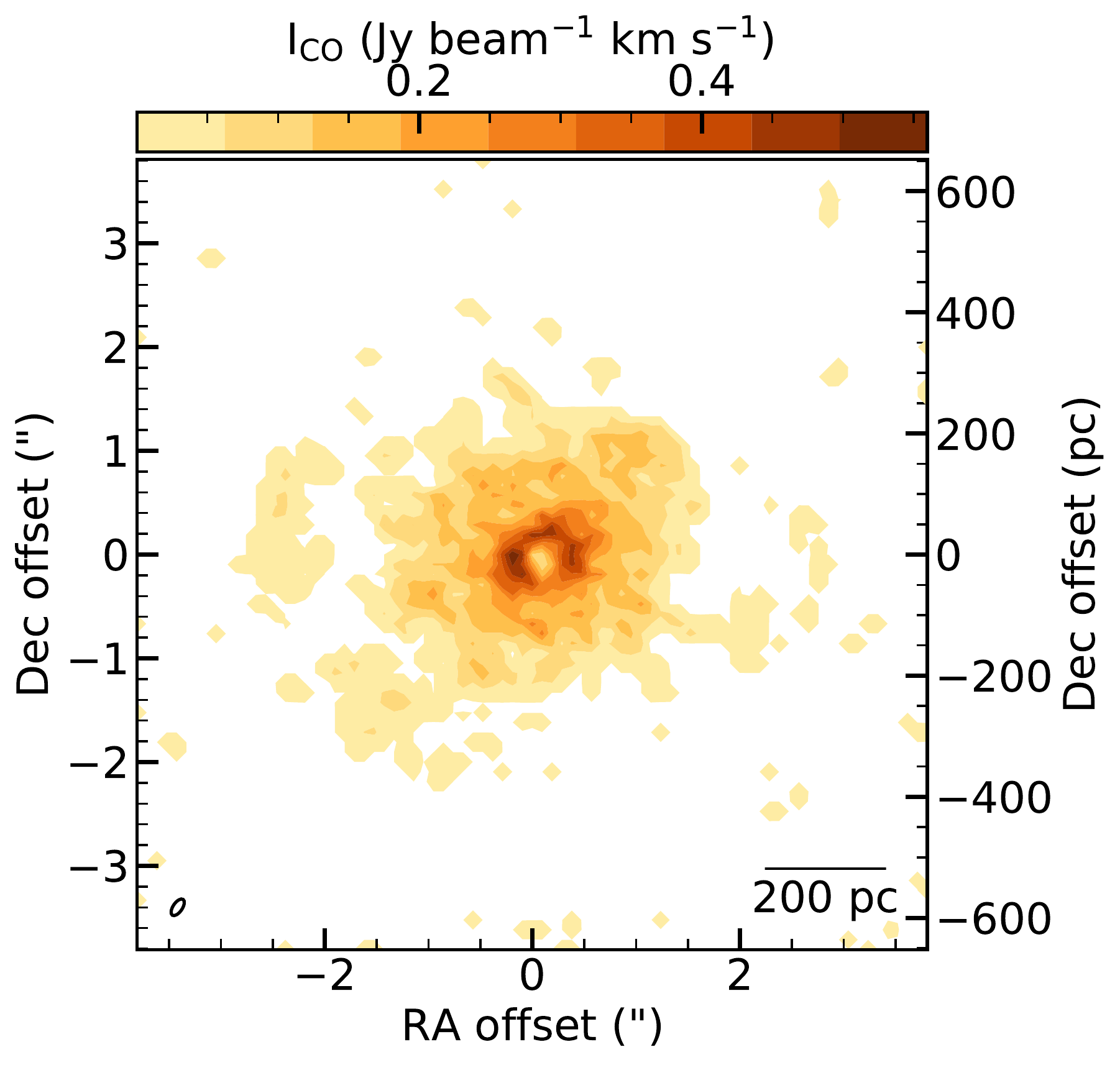}} \\
\subfloat[NGC7052]{\includegraphics[height=5cm,trim=0cm 0cm 0cm 0cm,clip]{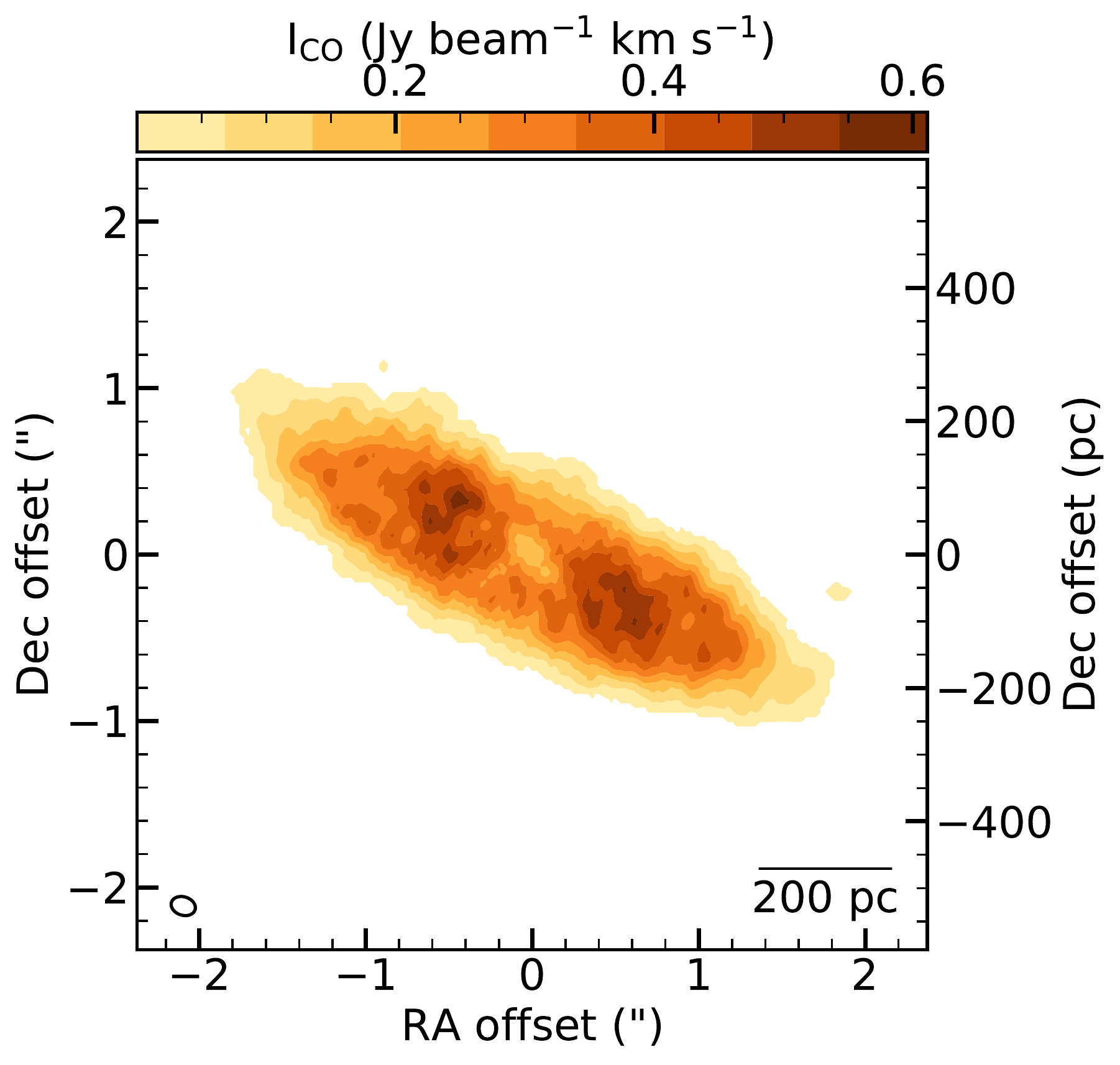}} &
\subfloat[NGC7172]{\includegraphics[height=5cm,trim=0cm 0cm 0cm 0cm,clip]{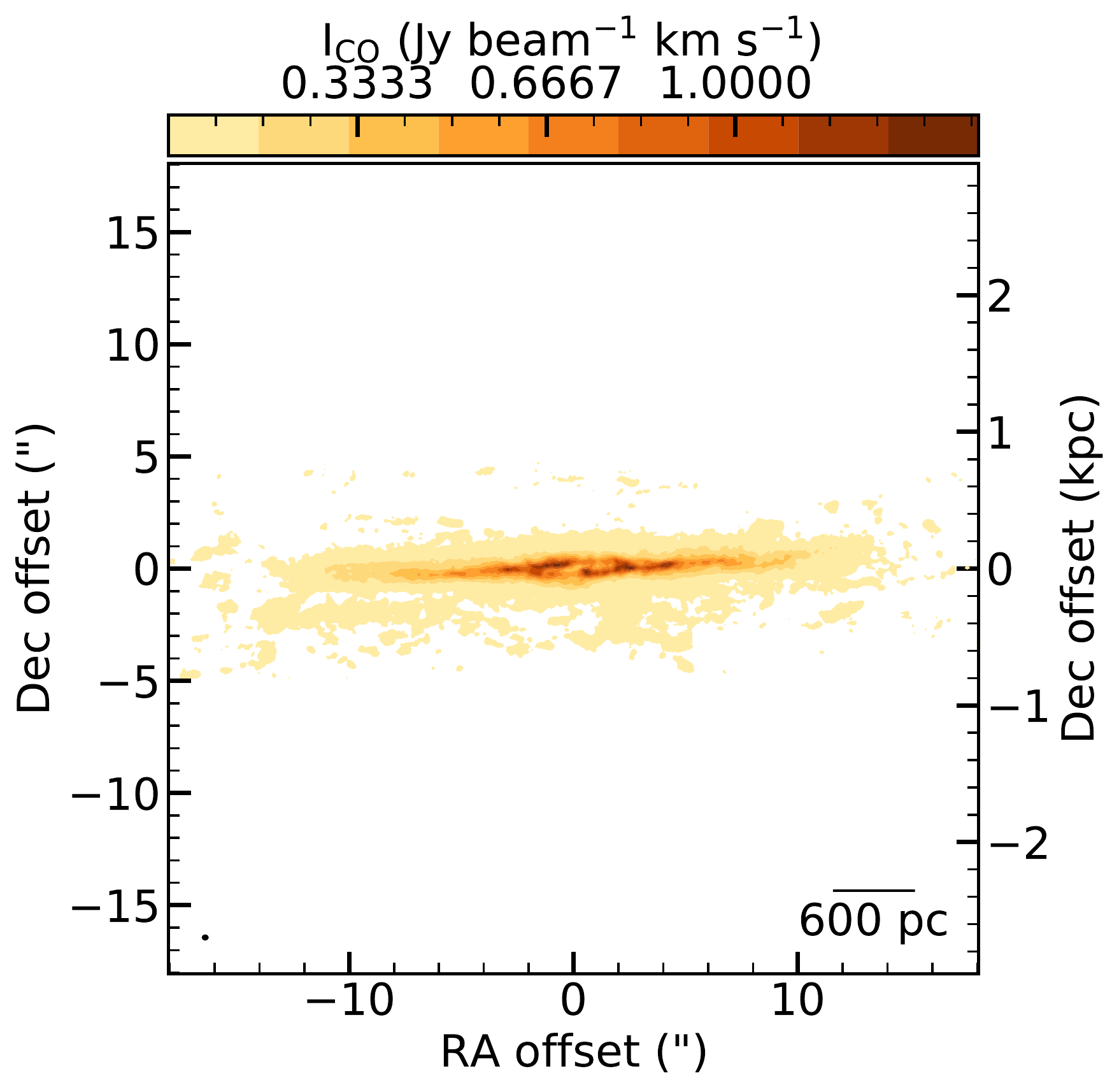}} &
\end{tabular}
\contcaption{} 
\end{figure*}
\begin{figure*}
\begin{tabular}{ccc}
\subfloat[MRK567]{\includegraphics[height=5cm,trim=0cm 0cm 0cm 0cm,clip]{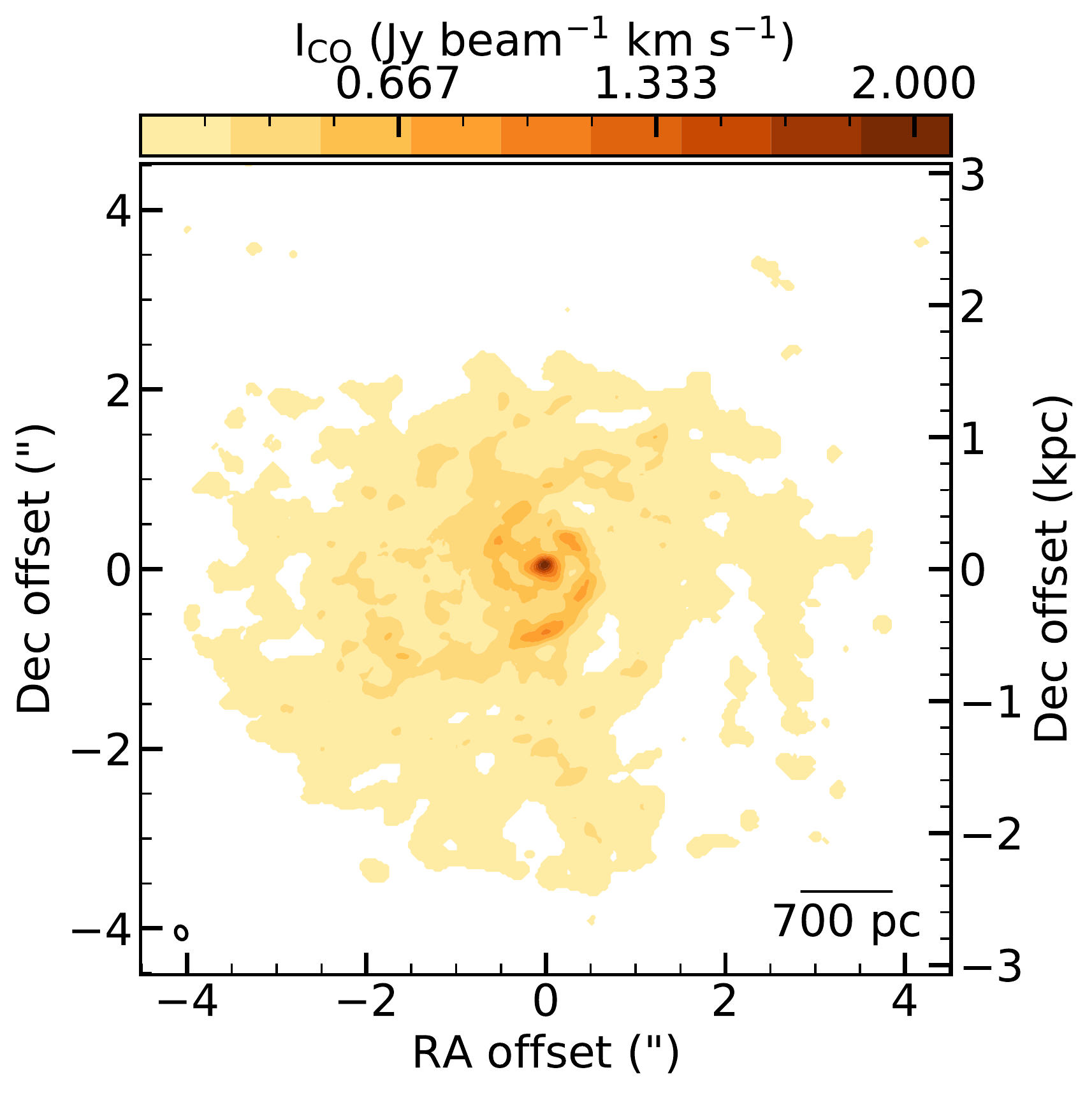}} &
\subfloat[NGC0449]{\includegraphics[height=5cm,trim=0cm 0cm 0cm 0cm,clip]{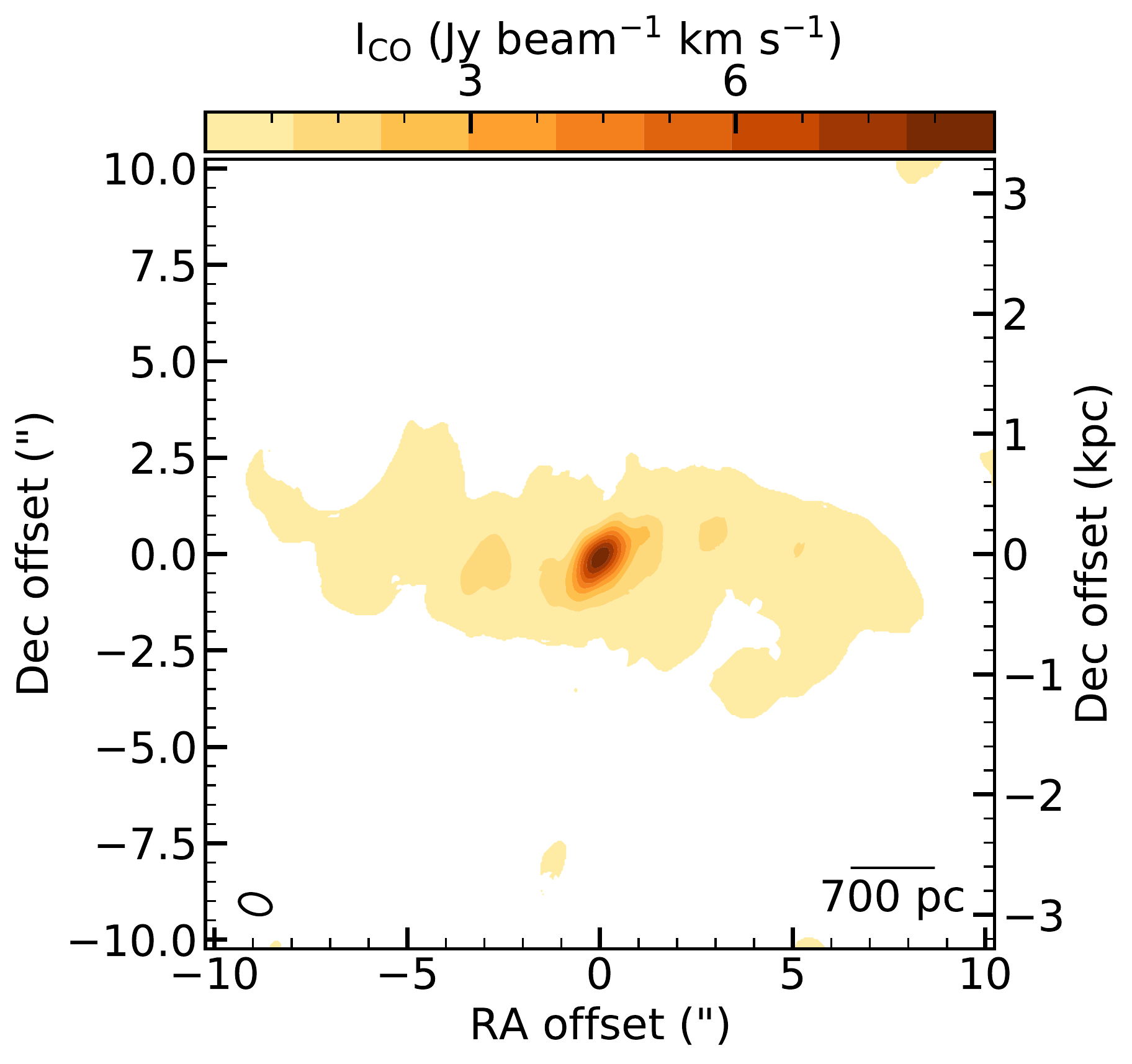}} &
\subfloat[NGC0612]{\includegraphics[height=5cm,trim=0cm 0cm 0cm 0cm,clip]{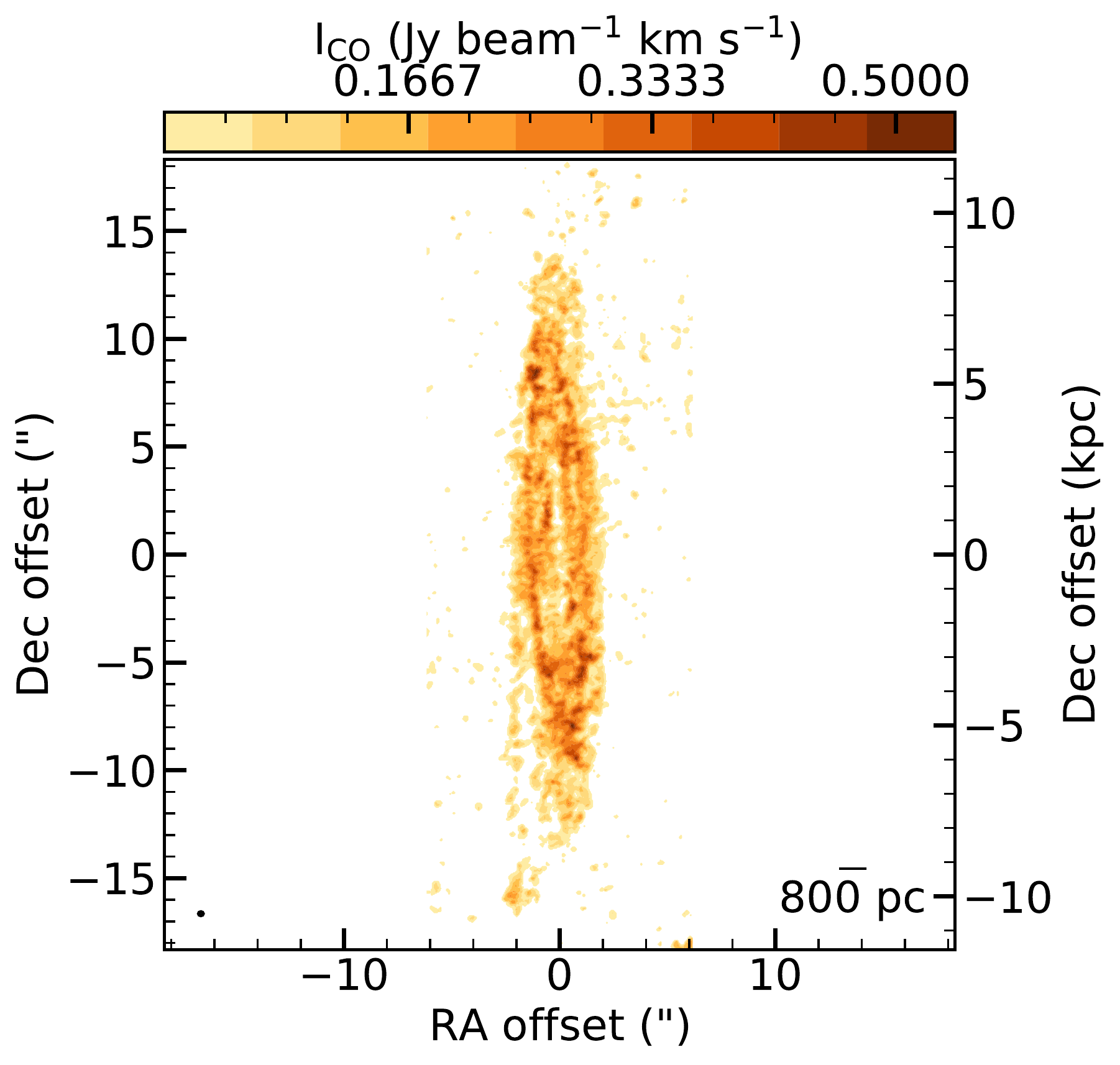}} \\
\subfloat[NGC3169]{\includegraphics[height=5cm,trim=0cm 0cm 0cm 0cm,clip]{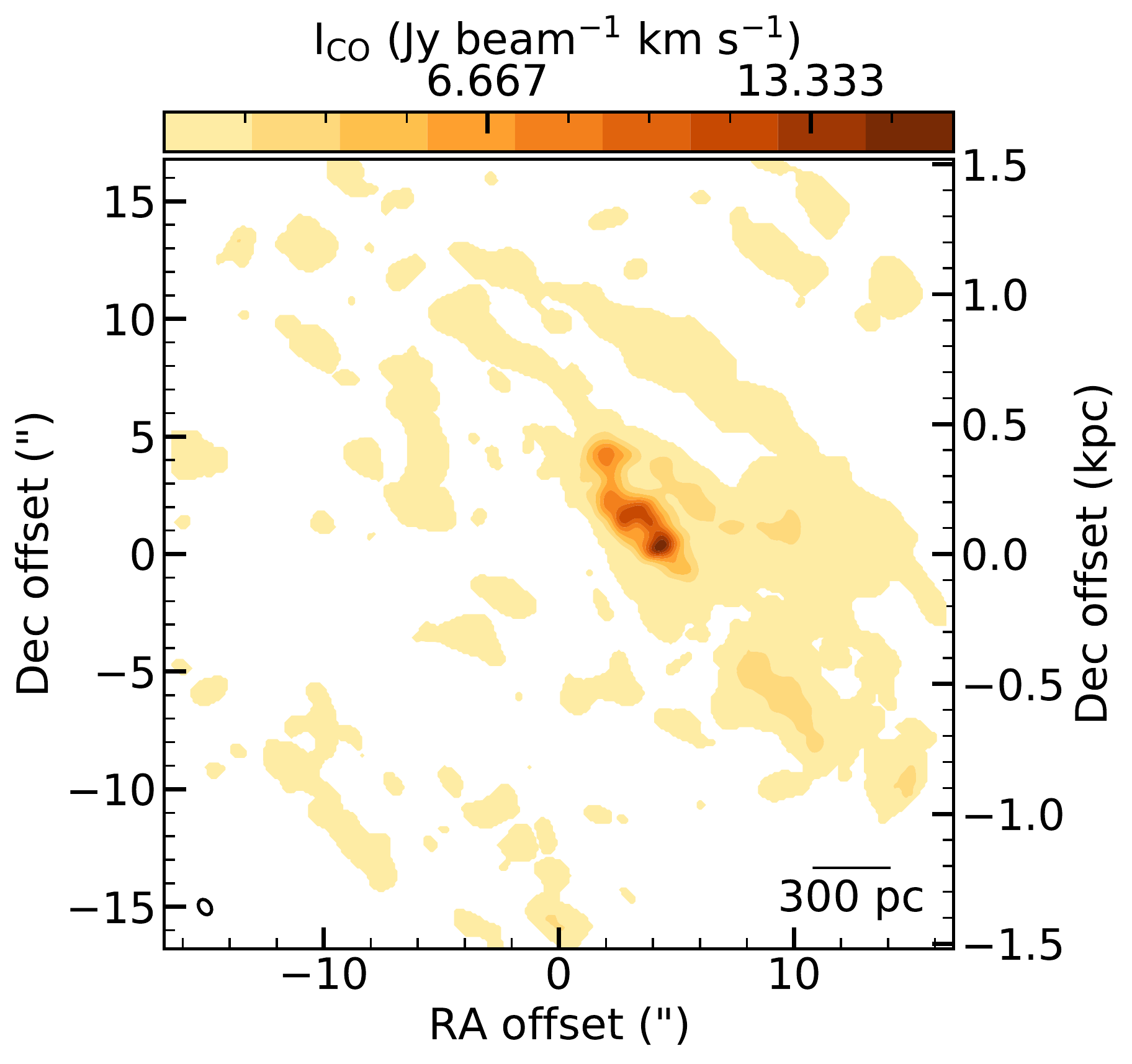}} &
\subfloat[NGC3368]{\includegraphics[height=5cm,trim=0cm 0cm 0cm 0cm,clip]{figs/NGC3368_moment0.pdf}} &
\subfloat[NGC4438]{\includegraphics[height=5cm,trim=0cm 0cm 0cm 0cm,clip]{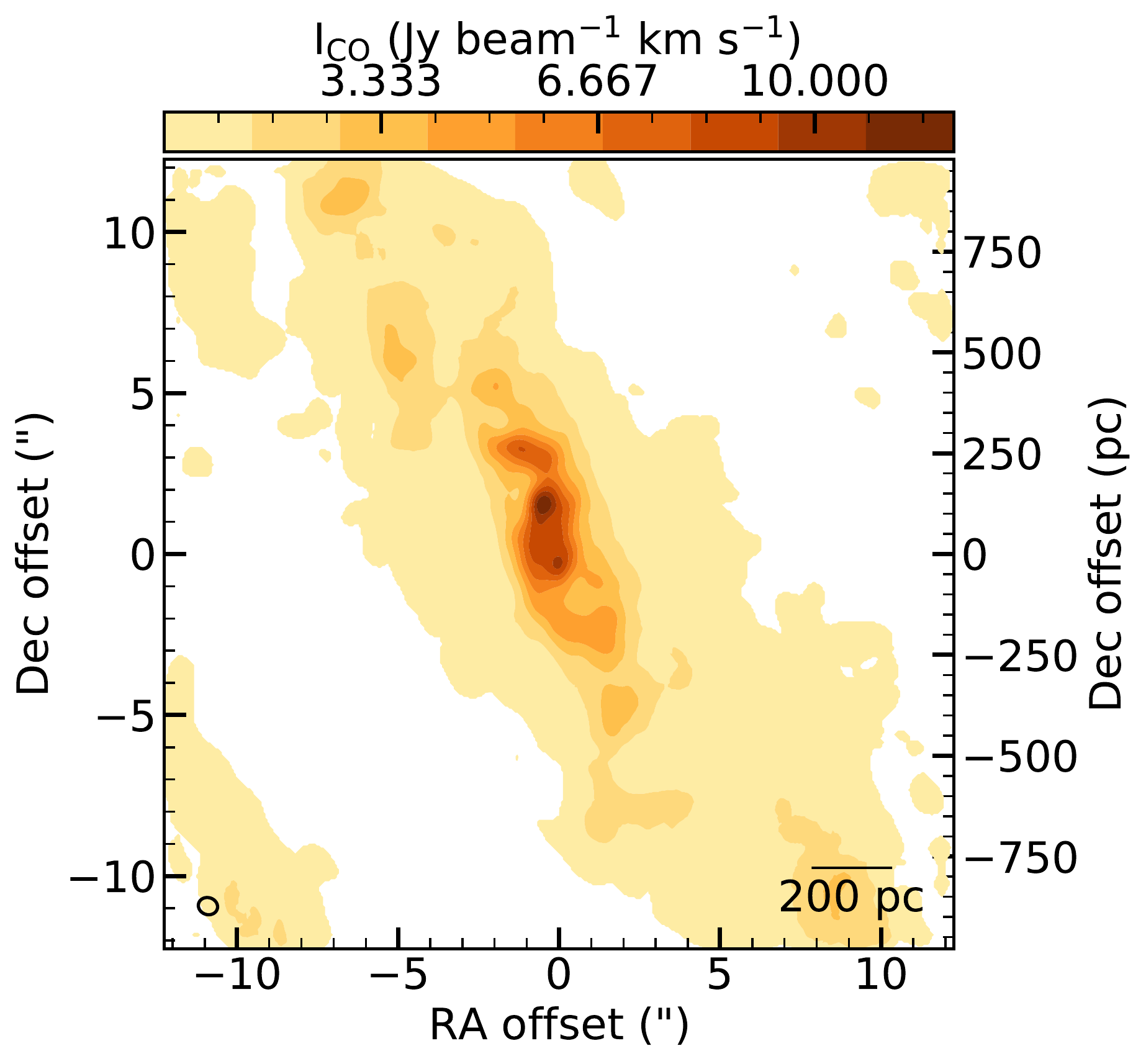}} \\
\subfloat[NGC4501]{\includegraphics[height=5cm,trim=0cm 0cm 0cm 0cm,clip]{figs/NGC4501_moment0.pdf}} &
\subfloat[NGC4826]{\includegraphics[height=5cm,trim=0cm 0cm 0cm 0cm,clip]{figs/NGC4826_moment0.pdf}} &
\subfloat[NGC5064]{\includegraphics[height=5cm,trim=0cm 0cm 0cm 0cm,clip]{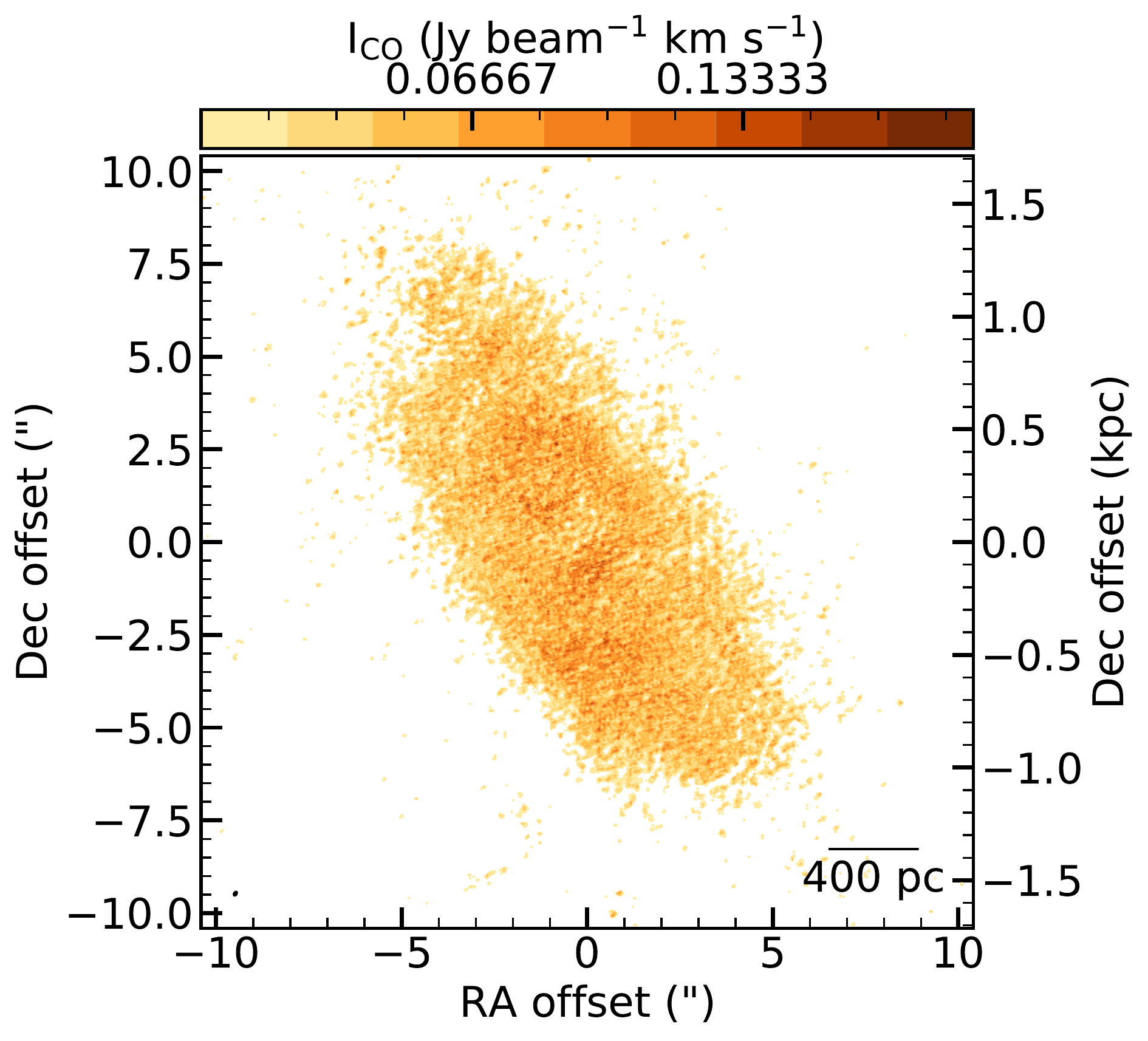}} \\
\end{tabular}
\caption{Integrated-intensity maps of the CO(2-1) or CO(3-2) transition for late-type galaxies in the WISDOM survey.} 
\label{mom0s_ltgs}
\end{figure*}
\begin{figure*}
\begin{tabular}{ccc}
\subfloat[NGC5765b]{\includegraphics[height=5cm,trim=0cm 0cm 0cm 0cm,clip]{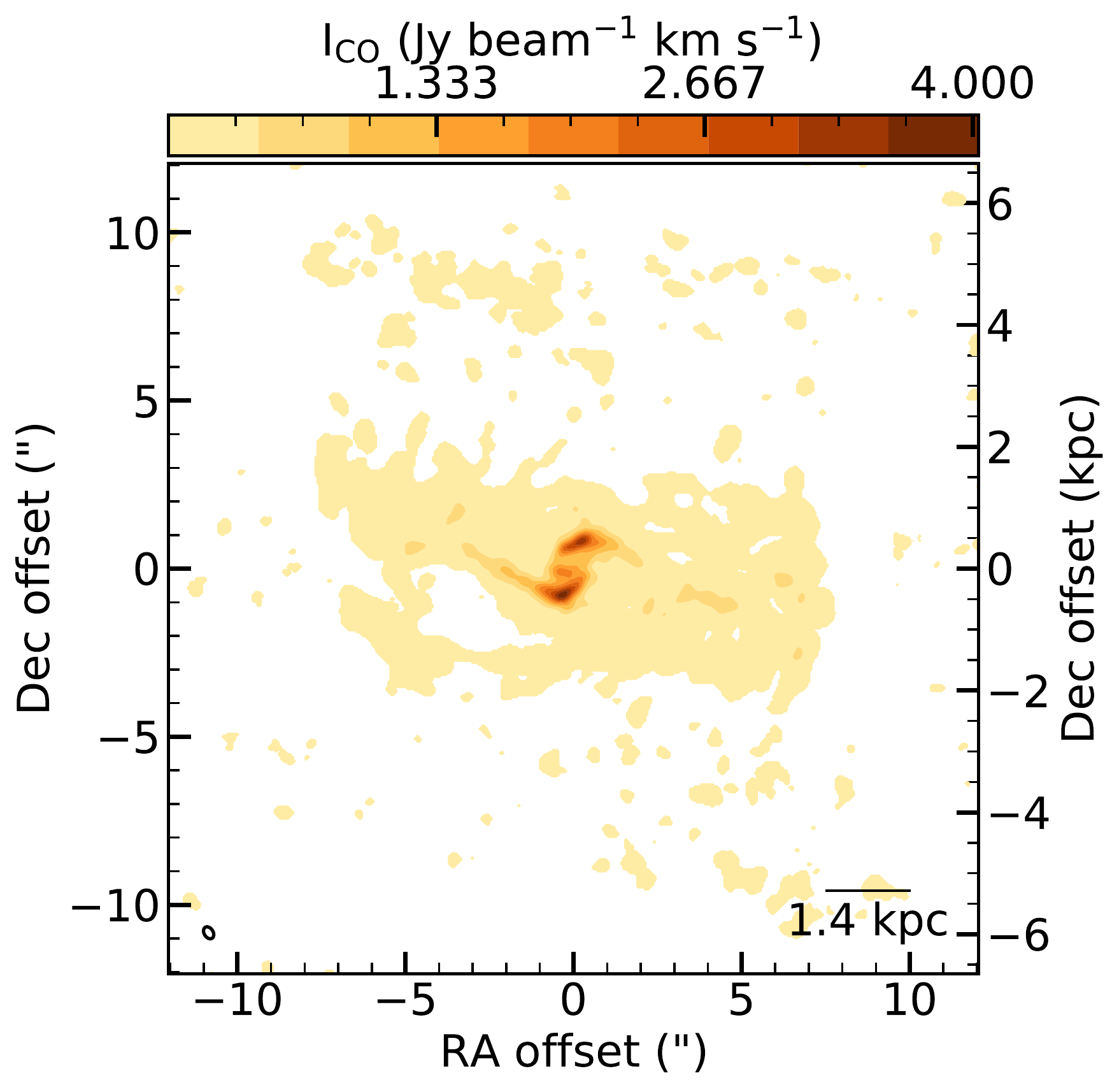}} &
\subfloat[NGC5806]{\includegraphics[height=5cm,trim=0cm 0cm 0cm 0cm,clip]{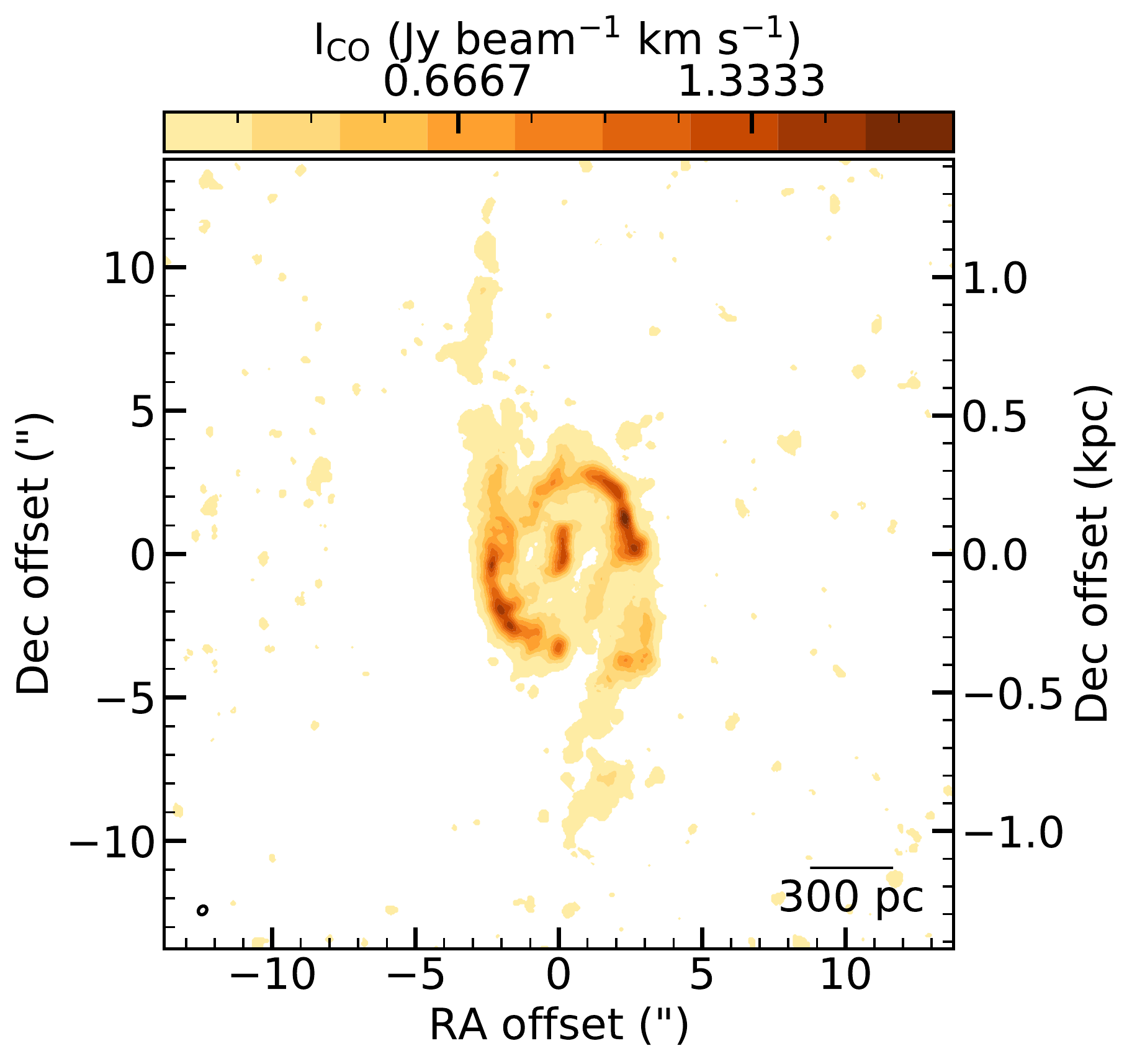}} &
\subfloat[NGC6753]{\includegraphics[height=5cm,trim=0cm 0cm 0cm 0cm,clip]{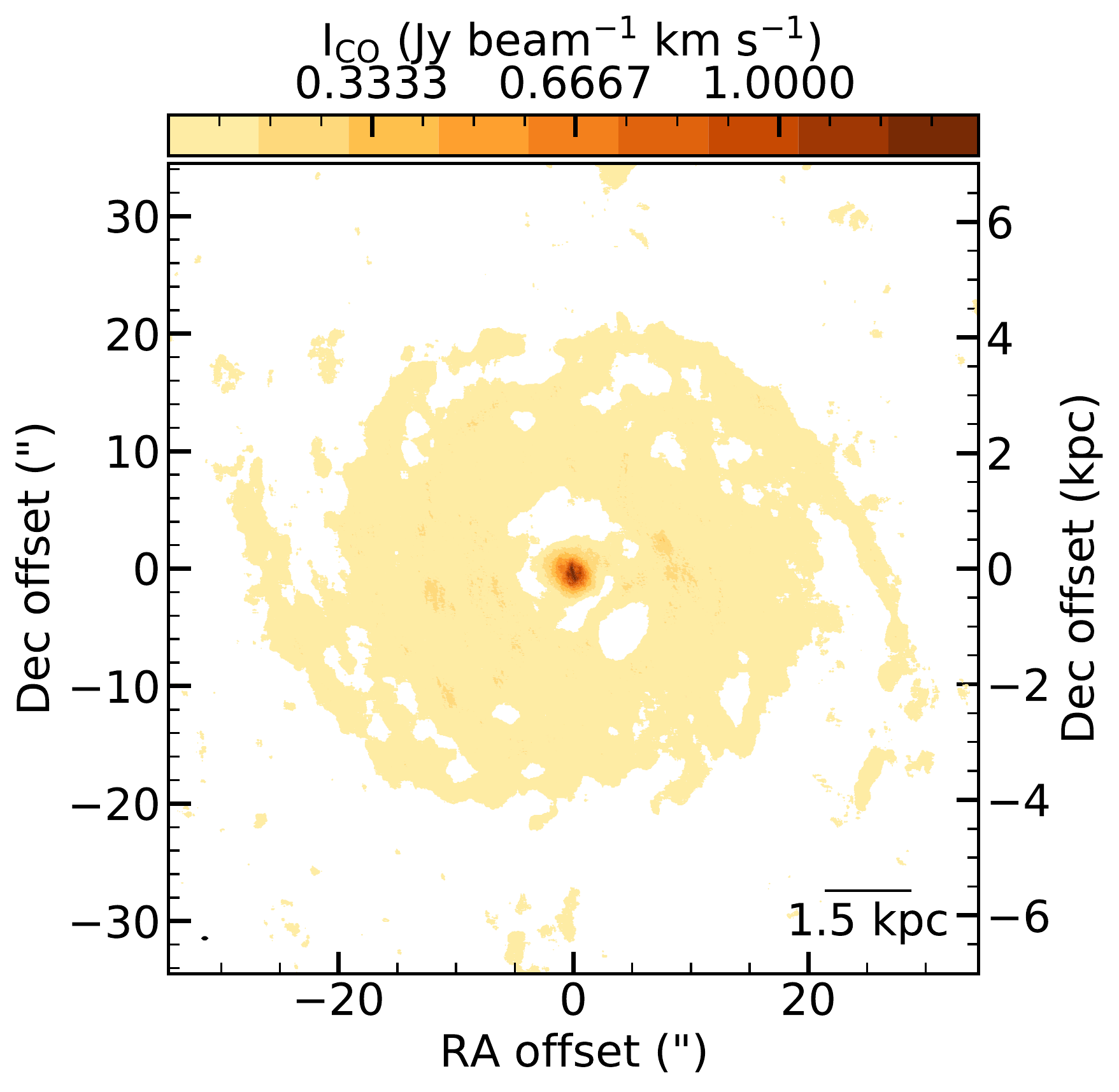}} \\
\end{tabular}
\contcaption{} 
\end{figure*}

\section{Non parametric morphology measurements}

\begin{table}
\caption{Non-parametric morphology measurements}
\begin{center}
\begin{tabular*}{0.45\textwidth}{@{\extracolsep{\fill}} l r r r r r r}
\hline
Name & A\,\, & $\Delta$A & S\,\, & $\Delta$S & Gini & $\Delta$Gini\\
 (1) & (2) &(3) &(4) & (5) &(6) &(7) \\
\hline
\\

WISDOM:\\
\hline
FRL49 & 0.30 & 0.04 & 0.17 & 0.01 & 0.41 & 0.12 \\
MRK567 & 0.76 & 0.05 & 0.23 & 0.01 & 0.50 & 0.05 \\
NGC0383 & 0.11 & 0.01 & 0.13 & 0.02 & 0.16 & 0.03 \\
NGC0449 & 0.60 & 0.02 & 0.40 & 0.01 & 0.64 & 0.01 \\
NGC0524 & 0.23 & 0.01 & 0.19 & 0.01 & 0.43 & 0.02 \\
NGC0612 & 0.63 & 0.03 & 0.52 & 0.01 & 0.47 & 0.02 \\
NGC0708 & 0.69 & 0.01 & 0.32 & 0.03 & 0.51 & 0.06 \\
NGC1387 & 0.15 & 0.02 & 0.10 & 0.01 & 0.27 & 0.02 \\
NGC1574 & 0.02 & 0.01 & 0.04 & 0.01 & 0.31 & 0.05 \\
NGC3169 & 0.99 & 0.03 & 0.33 & 0.01 & 0.81 & 0.03 \\
NGC3368 & 0.49 & 0.04 & 0.19 & 0.01 & 0.79 & 0.04 \\
NGC3607 & 0.32 & 0.01 & 0.24 & 0.02 & 0.57 & 0.01 \\
NGC4061 & 0.14 & 0.01 & 0.15 & 0.02 & 0.32 & 0.06 \\
NGC4429 & 0.15 & 0.01 & 0.09 & 0.02 & 0.29 & 0.06 \\
NGC4435 & 0.18 & 0.01 & 0.19 & 0.03 & 0.37 & 0.08 \\
NGC4438 & 0.30 & 0.04 & 0.17 & 0.01 & 0.60 & 0.04 \\
NGC4501 & 0.75 & 0.02 & 0.38 & 0.01 & 0.79 & 0.03 \\
NGC4697 & 0.13 & 0.01 & 0.20 & 0.02 & 0.51 & 0.04 \\
NGC4826 & 0.25 & 0.02 & 0.04 & 0.01 & 0.37 & 0.03 \\
NGC5064 & 0.26 & 0.03 & 0.14 & 0.02 & 0.26 & 0.07 \\
NGC5765b & 0.67 & 0.03 & 0.43 & 0.01 & 0.64 & 0.01 \\
NGC5806 & 0.36 & 0.03 & 0.32 & 0.01 & 0.53 & 0.09 \\
NGC6753 & 0.46 & 0.02 & 0.26 & 0.01 & 0.44 & 0.08 \\
NGC6958 & 0.11 & 0.01 & 0.10 & 0.01 & 0.42 & 0.09 \\
NGC7052 & 0.22 & 0.02 & 0.14 & 0.03 & 0.34 & 0.09 \\
NGC7172 & 0.21 & 0.01 & 0.18 & 0.02 & 0.64 & 0.03 \\
\\
Simulations:\\
\hline
noB & 1.57 & 0.08 & 0.52 & 0.027 & 0.81 & 0.03 \\
B\_M30\_R1 & 0.69 & 0.05 & 0.27 & 0.020 & 0.39 & 0.02 \\
B\_M30\_R2 & 1.24 & 0.08 & 0.44 & 0.037 & 0.67 & 0.03 \\
B\_M30\_R3 & 1.47 & 0.11 & 0.56 & 0.026 & 0.77 & 0.02 \\
B\_M60\_R1 & 0.40 & 0.04 & 0.20 & 0.013 & 0.24 & 0.02 \\
B\_M60\_R3 & 1.10 & 0.06 & 0.39 & 0.015 & 0.60 & 0.02 \\
B\_M60\_R2 & 0.75 & 0.07 & 0.30 & 0.027 & 0.42 & 0.04 \\
B\_M90\_R1 & 0.35 & 0.04 & 0.20 & 0.012 & 0.23 & 0.02 \\
B\_M90\_R2 & 0.46 & 0.05 & 0.21 & 0.016 & 0.25 & 0.02 \\
B\_M90\_R3 & 0.69 & 0.05 & 0.27 & 0.015 & 0.36 & 0.02 \\
\hline
\end{tabular*}\vspace{0.01cm}
\parbox[t]{0.45\textwidth}{ \textit{Notes:} Column 1 lists the galaxy/simulation name, and columns 2-7 the concentration, Asymmetry and Smoothness derived for each object, along with its estimated error. }
\end{center}
\label{castable}
\end{table}
 \begin{table}
\contcaption{}
\begin{center}
\begin{tabular*}{0.45\textwidth}{@{\extracolsep{\fill}} l r r r r r r}
\hline
Name & A\,\, & $\Delta$A & S\,\, & $\Delta$S & Gini & $\Delta$Gini\\
 (1) & (2) &(3) &(4) & (5) &(6) &(7) \\
\hline
\\
PHANGS:\\
\hline
IC1954 & 0.97 & 0.01 & 0.30 & 0.01 & 0.57 & 0.01 \\
IC5273 & 0.94 & 0.02 & 0.37 & 0.01 & 0.75 & 0.01 \\
NGC0628 & 0.70 & 0.02 & 0.32 & 0.01 & 0.65 & 0.01 \\
NGC0685 & 1.38 & 0.01 & 0.38 & 0.01 & 0.83 & 0.01 \\
NGC1087 & 0.83 & 0.03 & 0.27 & 0.01 & 0.67 & 0.01 \\
NGC1097 & 0.43 & 0.01 & 0.24 & 0.01 & 0.47 & 0.05 \\
NGC1300 & 0.61 & 0.02 & 0.19 & 0.01 & 0.61 & 0.10 \\
NGC1317 & 0.56 & 0.01 & 0.22 & 0.01 & 0.39 & 0.01 \\
NGC1365 & 0.71 & 0.02 & 0.17 & 0.01 & 0.52 & 0.04 \\
NGC1385 & 1.43 & 0.01 & 0.32 & 0.01 & 0.65 & 0.01 \\
NGC1433 & 0.63 & 0.01 & 0.23 & 0.01 & 0.47 & 0.06 \\
NGC1511 & 1.45 & 0.02 & 0.32 & 0.01 & 0.70 & 0.01 \\
NGC1512 & 0.66 & 0.02 & 0.36 & 0.02 & 0.49 & 0.05 \\
NGC1546 & 0.32 & 0.01 & 0.13 & 0.01 & 0.27 & 0.03 \\
NGC1559 & 1.31 & 0.01 & 0.38 & 0.01 & 0.70 & 0.01 \\
NGC1566 & 0.71 & 0.05 & 0.32 & 0.01 & 0.74 & 0.04 \\
NGC1637 & 0.58 & 0.03 & 0.33 & 0.01 & 0.71 & 0.01 \\
NGC1672 & 0.53 & 0.02 & 0.17 & 0.01 & 0.46 & 0.06 \\
NGC1792 & 0.66 & 0.04 & 0.18 & 0.01 & 0.42 & 0.02 \\
NGC2090 & 0.69 & 0.01 & 0.29 & 0.01 & 0.44 & 0.01 \\
NGC2566 & 0.41 & 0.01 & 0.16 & 0.01 & 0.75 & 0.06 \\
NGC2903 & 0.74 & 0.01 & 0.24 & 0.01 & 0.73 & 0.04 \\
NGC2997 & 0.62 & 0.02 & 0.20 & 0.01 & 0.64 & 0.04 \\
NGC3059 & 1.22 & 0.01 & 0.29 & 0.01 & 0.74 & 0.01 \\
NGC3137 & 1.26 & 0.01 & 0.40 & 0.01 & 0.64 & 0.02 \\
NGC3351 & 0.27 & 0.03 & 0.33 & 0.01 & 0.54 & 0.18 \\
NGC3511 & 0.52 & 0.03 & 0.24 & 0.01 & 0.48 & 0.02 \\
NGC3507 & 0.83 & 0.05 & 0.42 & 0.02 & 0.75 & 0.03 \\
NGC3521 & 0.41 & 0.03 & 0.18 & 0.01 & 0.29 & 0.01 \\
NGC3596 & 0.87 & 0.03 & 0.35 & 0.01 & 0.68 & 0.01 \\
NGC3621 & 0.92 & 0.03 & 0.27 & 0.01 & 0.47 & 0.01 \\
NGC3626 & 1.27 & 0.01 & 0.52 & 0.01 & 0.73 & 0.01 \\
NGC3627 & 0.69 & 0.04 & 0.38 & 0.01 & 0.80 & 0.01 \\
NGC4207 & 0.62 & 0.01 & 0.37 & 0.01 & 0.74 & 0.02 \\
NGC4254 & 0.70 & 0.01 & 0.24 & 0.01 & 0.45 & 0.01 \\
NGC4293 & 0.38 & 0.03 & 0.44 & 0.01 & 0.83 & 0.03 \\
NGC4298 & 0.63 & 0.01 & 0.24 & 0.01 & 0.44 & 0.01 \\
NGC4303 & 0.47 & 0.01 & 0.18 & 0.01 & 0.61 & 0.07 \\
NGC4321 & 0.56 & 0.01 & 0.29 & 0.01 & 0.59 & 0.08 \\
NGC4424 & 1.16 & 0.03 & 0.37 & 0.02 & 0.80 & 0.03 \\
NGC4457 & 0.95 & 0.01 & 0.31 & 0.01 & 0.75 & 0.01 \\
NGC4496A & 1.61 & 0.01 & 0.51 & 0.01 & 0.85 & 0.03 \\
NGC4535 & 0.41 & 0.05 & 0.21 & 0.01 & 0.82 & 0.08 \\
NGC4536 & 0.34 & 0.03 & 0.18 & 0.01 & 0.68 & 0.06 \\
NGC4540 & 1.31 & 0.01 & 0.47 & 0.01 & 0.74 & 0.01 \\
NGC4548 & 0.46 & 0.05 & 0.35 & 0.01 & 0.92 & 0.07 \\
NGC4569 & 0.71 & 0.01 & 0.22 & 0.01 & 0.73 & 0.04 \\
NGC4579 & 0.93 & 0.02 & 0.35 & 0.01 & 0.71 & 0.01 \\
NGC4654 & 0.37 & 0.02 & 0.13 & 0.01 & 0.45 & 0.03 \\
NGC4689 & 0.85 & 0.01 & 0.27 & 0.01 & 0.45 & 0.01 \\
NGC4694 & 1.50 & 0.03 & 0.35 & 0.01 & 0.91 & 0.02 \\
NGC4731 & 1.56 & 0.02 & 0.52 & 0.01 & 0.91 & 0.01 \\
NGC4781 & 0.97 & 0.01 & 0.27 & 0.01 & 0.57 & 0.03 \\
NGC4941 & 0.83 & 0.03 & 0.35 & 0.02 & 0.92 & 0.01 \\
NGC5134 & 1.62 & 0.01 & 0.51 & 0.01 & 0.85 & 0.01 \\
NGC5248 & 0.39 & 0.02 & 0.17 & 0.01 & 0.52 & 0.06 \\
NGC5530 & 1.00 & 0.01 & 0.37 & 0.01 & 0.60 & 0.01 \\
NGC5643 & 0.89 & 0.01 & 0.28 & 0.01 & 0.82 & 0.02 \\
NGC6300 & 0.78 & 0.01 & 0.36 & 0.01 & 0.87 & 0.01 \\
NGC7496 & 0.53 & 0.01 & 0.28 & 0.01 & 0.77 & 0.06 \\
\hline
\end{tabular*}\vspace{0.01cm}
\parbox[t]{0.45\textwidth}{ \textit{Notes:} Column 1 lists the galaxy/simulation name, and columns 2-7 the concentration, Asymmetry and Smoothness derived for each object, along with its estimated error. }
\end{center}
\end{table}
\end{document}